\documentclass[ showpacs,article,11pt,amsmath,amssymb]{revtex4-2}
\usepackage{graphicx}
\usepackage{bm}
\usepackage{epsf}
\usepackage[T2A]{fontenc}
\usepackage[utf8]{inputenc}
\usepackage[russian,english]{babel}
\newcommand{\R}{\mathcal{R}}
\newcommand{\K}{\mathcal{K}}
\newcommand{\Ro}{{|\cal R}_\omega|^2}

\newcommand{\Det}{{\mathfrak Det}_{\rm imp}}
\newcommand{\ee}{\epsilon}
\newcommand{\beq}{\begin{equation}}
\newcommand{\eeq}{\end{equation}}

\begin{document}
\title{Representation of the Luttinger Liquid with Single Point-like Impurity as a Field Theory for the Phase of Scattering.}
\author{V.V.Afonin}
\affiliation{Ioffe Physical-Technical Institute of the Russian Academy of Sciences, 194021, St.Petersburg, Russia.
(e-mail: vasili.afonin@mail.ioffe.ru)}
\begin{abstract}
A new approach describing Luttinger Liquid with point-like impurity as field theory for the phase of scattering is developed. It based on a matching of the electron wave functions at impurity position point. As a result of the approach, an expression for non-local action has been taken. The non-locality of the theory leads to convergence of the observed values in an ultraviolet region. It allows studying conductance of the channel up to electron-electron interaction strength of the order of unit. Expansion of the non-local action in small frequency powers makes possible to develop a new approach to the renormalization group analysis of the problem. This method differs from the “poor man's”\ approach widely used in solid-state physics. We have shown, in the Luttinger Liquid “poor man's”\ approach breaks already in two-loop approximation. We analyse the reason of this discrepancy. The qualitative description of the phenomenon is discussed in detail.
\end{abstract}
\maketitle
\section{Overview of the problem.}
\label{int}
So-called Luttinger Liquid (one-dimensional interacting electrons with linear
spectrum without back scattering; LL) has been studied for more than half a century. Unlike most problems, LL attracted researchers not because the way to solve it became clear, but because it turned out to be a more complicated problem than it seemed at first glance.
 Until the fifties, LL was considered as a paradoxical problem with trivial
solution: interaction does not change the transport properties of 1D channel. As it seems at first glance, this conclusion comes out of the  simple
considerations.
Absence of back scattering provides another conservation law:  chirality of the system must be conserved. (The chiral charge $j= \rho_R-\rho_L$ is the difference between the number of electrons moving to the right, R-electrons, and
to the left one, L-electrons.) After adding the  electric charge conservation law, we get two continuity equations for two quantities:
electric charge ($\rho= \rho_R+\rho_L$; $e_0=1$) and chiral one. These equations  coincide with  equations
 of the non-interacting system. It follows,  the electric current in one-dimensional channel is not changed due to electron-electron (e-e) interaction, in spite the  e-e interaction is strong.
By then, the latter was clear.(It follows that one-dimensional packets with a linear spectrum do not spread in one-dimensionality.)
Revision of the view point
came with publications \cite{T}-\cite{Hold},
where the authors  showed, the e-e interaction changes the value of electric
current in one-dimension channel with e-e interaction.
Afterwards it was clear, the LL has a direct outlet to more topical problems of today (to the helical and chiral liquids, see e.g. [5]).

Usually, changing of the transport properties of the LL associates  with anomalies in the correlators of  the  “density-density”\ type.
In the case of one component repulsive fermions, these anomalies are interpreted as a tendency to Peierls transition (\cite{E},\cite{Gim}), leading to the propagating charge density waves (see review \cite{Gr}). Of course, such a state can be considered as a candidate to ground state of the system. However, in 1D channel these states should have a high energy due to strong e-e interaction. In this situation, it seems natural to question about existence of the electro-neutral ground state. (Now, unless otherwise stated, we consider the problem of repulsive fermions.
To keep formulas simple, we will consider only single-component electrons; $\hbar =1$, all velocities are measured in the $v_F$, Fermi velocity, unit.)

The conservation laws  leading to the “paradoxical” picture arise from the symmetry properties of the Hamiltonian with e-e density-density type interaction without backscattering: $\hat\Psi_{R,L}\to \exp(i\Lambda) \hat\Psi_{R,L}$ -the gauge symmetry,
leading to electrical charge conservation; $\hat\Psi_{R,L}\to \exp(\pm
i\Lambda) \hat\Psi_{R,L}$ - the chiral symmetry   ($\hat\Psi_{R,L}$ - wave functions
of $R,L$  electrons). To retract the above-mentioned “paradox”  one should note, the wave functions of the excited states can be non-invariant in relation to these transformations in the case when the ground state wave function  of the system ($|\Omega\rangle$) has the lower symmetry in  compared with the original Hamiltonian. If one conservation law is not implemented, then the “paradox” will be cancelled. The non-invariance of the ground state under chiral transformations means,  it must consist of exciton-like (uncharged and moving in one direction) complexes $\hat a^{\dag}_R\hat b^{\dag}_L$ or larger complexes consisting of a number of such ones.
(Here $ \hat a^{\dag};\hat b^{\dag}$- are electron or hole creation operates). As regards to theirs statistical properties, they are Bose particles; i.e. they are accumulated in the ground state in a macroscopically large  (growing with channel length, $L$) number. Landau theorem prohibits the
phase transition of the second type in a one-dimension system \cite{Land}. The theorem is proved by the fact, the two-boson correlation function decreases exponentially on lengths greater than
$\zeta_c\sim v_c/T$ ($T$-temperature, and $v_c=\sqrt{1+V_0/\pi}$ is the
velocity of Bose particles; $ V_{e-e}(x-y)=V_0\delta(x-y))$ and by a power-law in the opposite limit case. However, the same
considerations lead to the conclusion, a channel of the     finite length
can have a macroscopically large number of coherent chiral pairs in the ground state at temperature $T\ll T_c=v_c/L$, i.e. $L\ll\zeta_c$.   (We do not make the limit $L\to \infty$ here, but consider the problem with many longitudinal quantization states, $N_e\sim p_FL\gg1.$  Therefore, in the leading order in $1/N_e$, Riemann sums over $p_n$ can be replaced by integrals. The exceptions are quantities divergenting in the thermodynamic limit: such as energy shift of the whole system due to e-e interaction, etc.   As result, the channel length will enter to the parameter for $T_c$, and calculation of observed values can be made as in unlimited case. It is important here,  the limit $\omega \to 0$ (the transition from linear response to conductance) is $\omega \ll T_c\sim v_c/L$ \cite{Af}.

LL is the exactly solvable problem in a sense of, it is possible to calculate any n-particle Green function. However, to obtain clarity regarding the ground state (GS), it is necessary to correctly interpret (in a physical sense)expressions of these correlators.
Thus, to prove existence of
a symmetry-broken phase, calculation of GS-wave function is required.  This problem is not exactly solvable,  an analytical solution can  be obtained  in the leading order in $v_c\gg 1$ only.
In a one-dimensional system, the GS-wave function with broken symmetry always depends on one more temperature, the degeneracy  temperature, $T_d\sim v_F/L.$ Above this temperature $\langle\Omega| a^{\dag}_R\hat b^{\dag}_L|\Omega\rangle\ne 0 $, and below $\langle\Omega| a^{\dag}_R\hat b^{\dag}_L|\Omega\rangle = 0.$ The reason is:  non-zero anomalous average requires, the GS-wave function has to be a packet of the states with different chirality. The characteristic difference between the energies of these states is of the order of $T_d\ll T_c$. At $T_d\ll T\ll T_c$ the
wave function of the GS with fixed phase of condensate was calculated analytically \cite{Ph} and equals to:
\begin{eqnarray} |\Omega_{\theta} \rangle = \sqrt
{Z}\exp \left[\int dx \exp(i\theta)\hat a^{\dag}_R \left( x  \right) \hat
b^{\dag }_L\left( x  \right) + \int dy\exp(-i\theta)
 \hat a^{\dag }_L\left( y  \right)
\hat b^{\dag }_R \left( y  \right) \right] | F \rangle,
\label{rt5}
\end{eqnarray}
here $|F>$ is the filled Fermi sphere,  $Z$ - normalization coefficient. It can be shown,  in the case of lower temperature $T\ll T_d$ one should keep  from all expansion  of exponent  function in  Eq.(\ref{rt5}) only the summands with same chirality. The state with the lowest energy corresponds to the state with zero chirality. However, an external electrical circuit may require implementation of a state with a non-zero chirality.
A direct analytical calculation of $|\Omega_{\theta} \rangle$ shows,  at $T\gg T_c$ only the pairs located at a distance less than $\zeta_c$ remain correlated. So, the wave function (\ref{rt5}) for a channel of finite length does not contradict Landau’s theorem. \footnote{More complex electro-neutral complexes consisting of two electrons and two holes in GS-wave function are forbidden by the Pauli principle for one-component fermions (in the limit of infinite strong interaction). In a system of interacting two-component fermions, the GS- wave function contains a macroscopic number of electro-neutral complexes consisting of two electrons with opposite spins and two holes. An increase in the number of components leads to an increase in the number of particles combined into interacted complexes \cite{NT}.}

For $T\ll T_c$ the GS-wave function Eq.(\ref{rt5}) corresponds  to  long-range ordering phase with a finite density of the chiral pairs. It is not possible to calculate the GS-wave function for the finite $v_c$. However, it is possible to calculate the correlator $\langle\Omega|\hat b_L(x)\hat a_R(x)\hat a^{\dag}_R(y)\hat b^{\dag}_L(y)|\Omega \rangle$, decreasing as a power of $\zeta=|x-y|$. It demonstrates, the number of correlated chiral pairs in the GS increases as $N\sim L^{\beta}; \beta\le1$, i.e. the number of chiral pairs is  macroscopically large. That means, in real systems one has the long-range ordering  phase, the phase by Berezinskii-Kosterlitz-Thouless (BKT) \cite{BKT},\cite{KT}.
This allows us to take a different view on the statement: “LL is a non-Fermi liquid” (see \cite{Gim} and references there).
Indeed, the basic assumption of the Fermi Liquid theory is: “as a result of adiabatic switching on an interaction, a GS-wave function of a non-interacting system moves into a GS-wave function of the interacting electron system” \cite{AGD}. Therefore, quasiparticles should be defined over $|\Omega\rangle$ wave function, not over $|F>$ one.
 The requirement of the transition from the GS of a non-interacting system to the interacting one is a usual condition in quantum mechanics: the perturbation theory can be formulated only over the stable GS-wave function,  taking it as a  zeroth-order approximation. In the opposite case, perturbation theory does not occur. Transition to  quasi-particles description is a formulation  of this perturbation procedure differently.
In the case of point-like e-e interaction, the explicit expression of normal quasi-particles in LL are presented in \cite{next}. They represent non-interacting fermions with the electric charge equals $e^*=1/\sqrt v_c$  moving with velocity $v_c$ (compare with \cite{Pham}). The difference between quasi-particle's charge and free fermion one comes out from polarization of the GS. The quasi-particle, moving to the right, is a non-linear package, consisting of right electrons and left holes (taken with unequal weight, since $e^*\ne 0$) and orthogonal to $|\Omega_{\theta}\rangle$.

One of the most interesting effects discovered in the LL problem with repulsion is the cutting of a one-dimensional channel  with respect to direct current after implantation of a weakly reflective point-like impurity into the channel \cite{FK}-\cite{BeteF}. Description of the GS wave function as a state with chiral condensate makes this effect clear on a qualitative level too. Indeed, let's take  into account, the point-like impurity distorts the condensate wave function. It becomes non-orthogonal to the quasi-particle wave function. As a result, a new channel of electrons “scattering”  appears (similar to Andreev’s reflection in superconductivity \cite{sup}). The process relates to the transition the normal
excitations to the condensate. It becomes possible only due to the non-orthogonality of the new GS-wave function (the GS of the system with an impurity) and the wave function of quasi-particles. For this to happen, the quasi-electron, moving toward impurity from the right, must polarize the electron liquid (in the region of non-orthogonality) and pairs with the left hole from the polarization cloud. This process must be accompanied by  creation of a left-electron due to  conservation of electric charge: $\hat
a_R^{\dag};\hat a_L^{\dag}+\hat b_L^{\dag}\to \hat a_R^{\dag}\hat  b_L^{\dag}+\hat a_L^{\dag}$. It is important, the probability of this transition is proportional to $N\sim L^{\beta}$, while the channel of “real”  impurity scattering of the right electron will not have this factor. However, the  probability  of all possible scattering processes has to be equal to unit.
 Therefore, we can neglect  the  channel related to “real” scattering of quasi-particles on impurity (in parameter $1/N$). The last is the single channel containing the transition wave.  So, transition of right-electrons to condensate will look like  their perfect reflection by impurity.

 The physical effect discussed above is confirmed by the exact solution of the problem for $v_c=2$. This solution can be formulated in terms of two Majorana's particles \cite{ExRep}. One of these particles enters into the scattering Hamiltonian, which does not conserve the electric charge. The conservation law of electric charge is satisfied in the entire system due to the second particle. It does not enter into the scattering Hamiltonian and moves in the opposite direction. This particle is registered as an electron reflected by impurity. The only possible interpretation of this solution is the appearance of a second excitation due to  creation of an additional exciton-like pair in $|\Omega_{}\rangle$.

For formulation a quantitative theory of LL with impurity, Hamiltonian of the electron should be discussed. At present, it is generally accepted to describe e-i scattering by an one-dimensional modification of the tunnelling Hamiltonian \cite{B}, which contains only amplitude of $R\leftrightarrow L$ transitions of electrons in the impurity localization point \cite{FK}. In the case of  point-like isotropic impurity, the Kane-Fisher Hamiltonian (KF-Hamiltonian) can be represented as \begin{equation}
 H_{KF}= V_{imp}[\hat\Psi^{\dag}_R(0)\hat\Psi_L(0)+h.c.],
\label{KF}
\end{equation} here $\hat\Psi^{\dag}_{R(L)}(0)$ are the $R(L)$-electron creation operators.  Thus, one omits a transmitted wave coming from the impurity. This simplification can be correct if, only the fact of chirality violation is important.
Expression (\ref{KF}) should be added to the ordinary Hamiltonian of e-e interaction:
\begin{equation}
 H_{e-e}= \int dx [\hat\Psi^{\dag}_R(x)(-i\partial_x)\hat\Psi_R(x)-R\to L]+\frac{1}{2}
\int dxdy\hat\rho(x)V_{e-e}(x-y)\hat\rho(y).
\label{KFfree}
\end{equation}
Unfortunately,  the received Hamiltonian cannot describe  scattering an electron by the point-like isotropic impurities correctly. Indeed, one should not solve a complicated problem  to see this. To that it is sufficient to note, the isotropy of the scattering should lead to the electron wave function parity,
 i.e. $\Psi_{R,L}(-x)=\Psi_{R,L}(x)$. At the same time, integration of the Schr\"odinger equation, corresponding to the KF-hamiltonian without interaction, around the  impurity position point requires a jump in these functions at that point:
\begin{equation}
i[\Psi_R(\epsilon)-\Psi_R(-\epsilon)]= \int^{\epsilon}
_{-\epsilon} dy V_{imp}\Psi_L(y)\delta(y).
\label{ChJ}\end{equation}
This means, the local Hamiltonian without transmitted wave cannot correctly describe a conducting channel at small distance from impurity. (The smooth potential of the e-imp scattering would provide equality to zero the r.h. part of Eq.(\ref{ChJ}). However, a smooth potential reduces to a qualitatively different problem in comparison with  we are considering.) As a result of this jump, we will get the ultraviolet (UV) unphysical divergences in an observable value  with incorrect symmetry property. The last is important for the renormalization group (RG) analysis. (This property brings to the renormalizability of the problem; see Section \ref{twoloop}.)  Later we will make sure,  interaction of an electron  with a point-like isotropic impurity will be determined by an odd  under spatial inversion quantity: a jump of  the electron  density at the impurity position point. However, this interaction cannot be represented as a local term in the Hamiltonian. The appearance of a new quantity in the theory is easily understood from the hydrodynamic analogy:
the effect existing in the liquid flowing around the hurdle. According to it, the hump before the hurdle and hollow behind it are formed. The characteristic scale of this construction is about the scale of the hurdle. In the case of a point-like impurity, this scale tends to zero, and the structure looks like a double layer. To have a finite value of the double layer, one should have a UV divergence in an expression of electron density.  The derived  expression of the charge jump will be odd under the space inversion, as required Eq.(\ref{ChJ}).
However, the electrical field, creating by the double charge layer, is  slowly-decreasing and  significant for the one-dimensional problem.
Besides, there is a laminar wake is described by 1D theory directly. Both effects define abnormal frequency dependence of conductance.  Therefore, as a first step, a consistent derivation of a long-wave Hamiltonian  is necessary.

The common way to solve a problem with point-like impurity is to consider impurity as a boundary condition for the Schr\"odinder equation. Impurity may be considered  as a point-like  if $p_Fa_i\ll 1$, where $a_i$ is the impurity scale.  At the same time, we can divide the whole electron wave function  into the left and right electrons only on the scale greater  than $1/p_F$. Therefore, before linearizing the Schrödinder equation, it is necessary to match the wave functions of the incident, transition and reflected electrons.
 To apply this approach to the e-e interaction problem, we should recall the Hubbard's trick \cite{Hub}.  It allows us to transfer the problem of e-e interaction to the problem of a non-interacting electronic system placed in a slowly varying external field $U(x,t)$ (with averaging of the resulting expressions over the Hubbard fields). This approach makes it possible to match the wave functions. The resulting complete set of solutions at the scale greater than $1/p_F$ will depend on the boundary conditions, as well as on the fields $U(x,t)$ entered in the phase of scattering. After this, one can transform the averaging over the Hubbard fields into an averaging over the scattering phase, $\alpha$.  As a result, the phase becomes a field variable of the Hamiltonian. This method simplifies the problem in compared to the direct description by left and right electrons. (Instead of calculation a lot of diagram to obtain conductivity beyond the leading logarithm approximation \cite{aristov2}, you can consider several ones; see Section \ref{RGgrp}.)
 As mentioned above, existing the double charge layer in the interacting electron system can be obtained  only together with ultraviolet divergence in the electron density. (A finite  charge of the double layer is obtained only as uncovering of uncertainty $a_i(\rho^{UV}_R+\rho^{UV}_L)$, where $\rho^{UV}_{R,(L)}$ is the part of the $R(L)$-electron density diverged in the ultraviolet region.) The uncertainty has to be removed before deriving the long-range Hamiltonian. To this, we should regularize expressions of the electron density. It is impossible to write an analytical theory on the scales $\le a_i$ for the interacting  1D fermions.   Nevertheless,  requirements of gauge invariance of the problem and electric charge conservation law make possible to define the value of the charge jump unique. (Discussion of the question is in the Section \ref{ChDens} and Appendix \ref{Adler} of the paper.)
 At this step, absence of ultraviolet divergence with incorrect parity, coming out from the initial Hamiltonian, becomes important. Later, one can  pass from second order Schr\"odinder equation to the first order one.  This step allows us to solve the Schr\"odinder equation with external field $U(x,t)$ and point-like impurity exactly and proceed to the construction  of the averaging procedure.
 As a result, we will get the effective 1D  Hamiltonian.  This Hamiltonian will be non-local, but the observed quantities will not have the ultraviolet divergences (see Section \ref{TH}). It allowed us to reject approximation of weakly interacting electrons and extend results up to the interaction constant of the order  of unit (see Subsection \ref{AttrP1}).

The common way to investigate the  system with
long-range order  is the renormalization RG-approach. In our problem, the first step on this way  was done in the paper \cite{RG}, where  expression of the conductance had been calculated in the leading log-approximation. The authors of the papers have used  the so-called "poor man's"\ RG-approach. It is a simplified version of the original Gell-Mann - Low approach (GL) \cite{And} (the modern review - \cite{Collins}). In "poor man's" approach one assumes, the renormalized RG-charge coincides with the observed quantity and, so far as Gell-Mann - Low equation defines the renormalized charge, this quantity is defined by  GL-equation too. Later, the two- and three-loop contributions within the "poor man's"\ RG framework  have been calculated  \cite{aristov2}, \cite{aristov}. If the assumptions of the “poor” RG approach were correct, then a significant simplification of the calculations would take place. In this case, to derive the
Gell-Mann — Low  equation in a given order over of e-e interaction, it would be sufficient to calculate  in a quantity observed only the  logarithmic summands. (The higher powers of the logarithmic expansion would be reproduced by the Gell-Mann-Low equations.) Unfortunately, these assumptions cannot be correct
in all orders on e-e interaction. The point is, starting from a certain order, the Gell-Mann-Low equations always depend on the regularization scheme (i.e., on the calculation method). This is possible for an unobservable RG charge, but is unacceptable for an observable quantity. The only question is: in what order this will happen. The "poor man's approach"  widely  used in solid state physics, but the domain of its applicability had not been discussed in the literature.\footnote{In this paper, we do not analyse all RG-approaches used for this problem. It was important for me to use this model as an example to explain  the impossibility of applying the “poor man's approach” beyond the leading logarithm to any problem. Therefore, I referenced the very first works on this topic and those that used the “poor man's approach” beyond the leading logarithm.}
The answer to this question depends on the kind of the logarithmically divergent loop. (More precisely, on the number of vertices in the loop.) Usually, it takes place in the three-loop approximation, but in our problem - in the second loop  (see Subsections \ref{Robs},\ref{AndSec}). Therefore, if one limits itself by the leading logarithmic approximation,  it is possible to use the "poor man's"\ approach. Otherwise, it is necessary to check dependence of an observed value on the regularization scheme.
However, in the leading-log approximation, any logarithmically divergent theory looks like a renormalizable one. The renormalizability of logarithmic theories arises only as a result of sufficiently delicate cancellations of the divergences in expressions for observed quantities. Only they allow us to introduce a Lagrangian with renormalized charges that do not depend on the “external” frequency of the diagrams.  (see, Section \ref{renorm2}). These cancellations occur only beyond leading-log approximation and only thanks to them the observed values become independent of the form of a Lagrangian in the UV-region, the method of calculations, etc.

In the paper \cite{aristov2}, expression for conductance was calculated from the Callan-Symamanzik (CS) approach. Here, the authors have adapted the “poor man's” approach to this RG scheme. It resulted in  analogous dependence of the quantity observed from the subtraction scheme used to calculate the counter-terms. The authors interpreted this as a physical phenomenon indicating that the problem is not universal.   We believe, such dependence is forbidden in the CS approach too (see, Subsection \ref{AndSec}). The reason for its appearance is the same: identification   of the nonobservable renormalized charge with the observable quantity. We will show that the rejection of this identification leads to observable values  independent of the calculation method (as it should be in a renormalizable theory).

The paper is organized as follows:
In Section \ref{Imp} we adapted Hubbard's trick to the  problem with impurity.  Here, we obtained an expression of the Green's function of the system convenient  for the subsequent calculation  of the charge jump and laminar wake.  A method of obtaining the complete set of wave functions required to calculate the charge jump is shown in Appendix \ref{funckSet}.  The charge jump is calculated in Section \ref{ChDens}.
At the next step, we convert  the problem of a non-interacting electronic system placed in an external field to a system with e-e interaction.  Transition to the equivalent field theory is developed in Section \ref{TH}.
In section \ref{AttrP1}, as an example of using our approach,
we have calculated the electron reflection coefficient by an impurity in the lowest order on the bare reflection coefficient. Here we showed,  the absence of UV divergences changes the frequency asymptotic of the conductance for the strong e-e interaction case.
In Section \ref{RGgrp} we developed a RG approach in terms $\alpha$-variable and shown universality of the observable quantities.

The  brief  summaries of the most significant final expressions concerning the effective field theory of the LL have been published in \cite{EqTh} and \cite{AP}. However, this format did not allow a discussion of the qualitative description in  interacting 1D electron system, and  made it impossible to discuss the reason for a bit non-standard “technical”\ steps, have been taken to the correct description LL with impurity. Therefore, in the main part of the paper I limited oneself to the discussion the general results, physical picture and the reasons why we use a calculation method. Any non-trivial “technical”\ calculations that did not discuss earlier are moved to the Appendixes.
As an exception, I have not moved the charge jump calculation from the \ref{Imp} and \ref{ChDens} there.
This is because the Feynman boundary condition, commonly used in solid-state physics, is not applicable to our problem.  Usually, the correct boundary condition (Dirac one) is a consequence of the Lorentz invariance of a theory.  This cannot be an argument in a non-relativistic problem.  In  addition, in our case the difference between these conditions is more deep: owing to the time-dependency of the Habbard fields  the first condition leads to the heating up of the electron system (after averaging over the set of the fields), while the second one conserves the total energy of the system (as it should be for the e-e interaction).
So, application the first boundary conditions in our problem will lead to incorrect expressions for the observed quantities.
The remainder part of the Supporting Information proves assertions stated in the  main body of the paper. The qualitative description of the phenomenon is considered in the Section “Overview of the problem.”

 \section{Short range impurity as a boundary condition.}
\label{Imp}
There are two main approaches suitable for obtaining  Hamiltonian of Luttinger liquid. The first approach
 is bosonization
procedure. It permits to reduce the Luttinger Hamiltonian without impurity
to a diagonal form. The approach is  failure for the LL with impurity, because  for the case
the Hamiltonian  cannot be diagonalized. The
second one, base on the Hubbard trick \cite{Hub} (the short overview is in Introduction). Therefore, as a first step, we will consider the system of the non-interacting electrons
in external field  before linearization the electron Hamiltonian. To construct the electron-impurity part of the Hamiltonian,  we will use definition  of the "energy shift"\  of  electron system
under influence  of the external field:
$
\delta\mathcal{H}(x,t)/\delta U(x,t)  = \delta\rho(U,x,t)$.
Here $\delta\rho(x,t)$ is the non-linear changing of electron density, and
$\delta\mathcal{H}(x,t)$ is the field-dependent part of the Hamiltonian. So, if we calculate the
electron density, we can construct the Hamiltonian.  To
this, we will calculate the Green function of the system ($G(x,t)$).  Generally speaking, under influence of the time-dependent field $U$ an electron system can transmit  to the  excited state. However, after integration over all set $U$, the
external field will be converted to the e-e interaction. The last conserves the total energy of electron system. Thus, all excited states should not  give  a contribution to
the result because the initial state of our system is the GS. This allows us to limit ourselves to calculating the Feynman's Green function. It describes  transition of a system from
GS (at $t\to -\infty$) to the GS (at $t\to \infty$) and
obeys the inhomogeneous linearized Schr\"odinger equation everywhere,
except the point $x=0$ (outside of  impurity):
\begin{equation}
\left(i\frac{\partial}{\partial t}\pm i\frac{\partial}{\partial
x}-U(x,t)\right)G_{R(L)}(xt,x't')=i\delta^{(2)}(x-x'),
\label{Dirac-inhomogeneous}
\end{equation}
and complex conjugated equation in variables $x^{'},t^{'}$.

To construct the Green's function, it is necessary to have a complete set of solutions of the homogeneous one-dimensional Schrödinger equation, satisfying the Feynman boundary conditions and matching at the point $x=0$. (We will denote it as $\psi_{\pm\varepsilon ,\alpha}(x,t)$; $\alpha =
R,L;\varepsilon>0 $ is the electron energy, with is well-definite  at
$t\to\pm\infty$.) The Feynman boundary conditions are
\begin{enumerate}
\item
at $t\to -\infty$ one allows for the $\psi_{-\varepsilon ,\alpha}$  only
hole-like solutions ($\propto \exp(i\varepsilon t))$, while
\item
at $t\to\infty$  only electron-like solutions for $\psi_{\varepsilon
,\alpha}$ exist ($\propto \exp(-i\varepsilon t))$ .
\end{enumerate}
 The reason the boundary conditions for the electron-like and hole-like states are specified at different times is that the Eq.(\ref{Dirac-inhomogeneous}) is a first-order differential equation with respect to $t$. Therefore, we cannot put two boundary conditions (at $t\to \pm\infty$) for each function. Instead, we can specify one condition for each wave function, but at different times. In Feynman's boundary conditions, we account for: at $t\to\infty$ electron-like states satisfy the condition
$\theta(\epsilon)\hat c_{\epsilon}\psi_{\varepsilon
,\alpha}|F>=0$.  (Here $\hat c_{\epsilon}$ is the electron annihilation operator. It  is defined over an empty state: $\hat c_{\epsilon}|0>=0$.) Therefore, the electron part of $\psi_{\varepsilon
,\alpha}$ can be any. It cannot create an excited state,  etc. (See Appendix \ref{funckSet} for details.)
The Feynman boundary conditions is no more than the assertion:
the Feynman Green function connects the incident and transition waves in a
scattering problem: $S_{f,i}=\int
dxdx'\psi_f(x,t)G(x,t;x't')\widetilde{\psi}_i(x',t')|_ {t\to\infty;t'\to -\infty}.$

In addition, it is necessary to have a {\em Dirac conjugate} set,
$\widetilde{\psi}_{\pm\varepsilon ,\alpha}(x,t)$. In problem with time-dependent fields, the Dirac
conjugated boundary condition does not coincide with hermitian conjugation of the Feynman  one.
They satisfy complex conjugated  Schr\"odinger equation {\em  plus the time reversion}:
\begin{enumerate}
\item
at $t\to \infty$ only electron-like solutions are allowed: $\widetilde{\psi}_{-\varepsilon ,\alpha}$ ($\propto \exp(-i\varepsilon t))$, and
\item
at $t\to-\infty$  only hole-like solutions $\widetilde{\psi}_{\varepsilon
,\alpha}$ exist ($\propto \exp(i\varepsilon t))$ .
\end{enumerate}
Let us discuss the cause for using the Dirac conjugated boundary condition for the wave function in the problem. Later we will make sure, before averaging over all set of the Habbard's fields among solutions of Eq.(\ref{Dirac-inhomogeneous}) the soliton-like (undamped in time) solutions exist. Solitons go away from the impurity to the edges of the channel (at $t\to\pm \infty;$
$x\to\mp\infty$, and the difference of $|x\pm t|$ is finite). It is natural,  the state is described of these solutions is the exited state of the system. Would the time-dependent fields $U(x,t)$ the real fields,  they  heat up the system,  taking the energy from an electric circuit.  However, in   calculation end,  these fields should describe the e-e interaction.
The last conserves the total energy of the whole system.  This fact should be taken into account from the very beginning.   The problem has to be formulated in  a way eliminated heating of the electronic system.  This condition should  take into account as the supplementary one.

Usually used complex conjugated Feynman boundary conditions for a wave function do not exclude the soliton-like solutions, while adding  the time-reversion  condition  exclude these solutions (see below). One must think about serious consequences of excluding some solutions from a complete set. Condition for the completeness of a set of functions (Eq.\ref{compl}) and  method for constructing Green's functions (Eq.\ref{Green-general}) should be different for different sets corresponding to different boundary conditions, { \em if one of the conditions excludes a part of solutions}. For complex conjugated Feynman's boundary conditions,  these expressions will be diagonal  in the label of linearly independent solutions. After  exclusion of the soliton-like solutions, these expressions will be non-diagonal.
To verify correctness of the statements are formulated above, we begin
our discussion from the case $U=0$. For this case, the
set of solutions may be get in various ways. Usually, one takes solutions
corresponding to two waves: incident to the impurity from the right or left
and two transmitted and reflected ones. They are corresponded to Feynman's boundary condition. However, solutions of
Eq.(\ref{Dirac-inhomogeneous}) with $U\ne 0$,  cannot correspond to this set.
To see this, let's note that without impurity solution of the Schr\"odinger equation  are
\begin{equation}
\psi_{R,L}(x,t)=\chi_{R,L}(x,t)e^{i\gamma_{R,L}(x,t)},\quad{\rm with}\quad \gamma_{R,L}(x,t)=-\int d^2x' G^{(0)}_{R,L}(xt,x't')U(x't'),
\end{equation}
(here $\chi_{R,L}(x,t)$ obey the free Schr\"odinger equation, and
$G^{(0)}_{R}(xt,x't')$ is the free {\em Feynman} Green function). They obey the Schr\"odinger
boundary condition  (because Feynman Green function does) and
$\chi_{R,L}(x,t)$ are taken properly. Let's come back to the problem with
impurity. An attempt to substitute the $\chi_{R,L}$ by incident, reflected
and transmitted waves breaks the matching conditions at the impurity position point, because
$\gamma_{R}(0,t)\ne \gamma_{L}(0,t).$ One can correct the fact by adding  the
phase shift
$ \alpha (t) = \gamma_R(t) -\gamma_L(t)
$ depends on $t+x$ for the left wave function (and on $t-x$ to the right one),
because $e^{i\alpha (t\pm x)} $ are solutions of the homogeneous Schr\"odinger
equation without impurity.  In the case, we would have:
$$
\psi_{\varepsilon ,R}(x,t)={\exp(-i\varepsilon t + i\varepsilon x
+i\gamma_R(x,t))}\left[\theta (-x) + \mathcal{K} \theta (x)
\right];$$ $$
\psi_{\varepsilon ,L}(x,t)={\exp(-i\varepsilon t - i\varepsilon x +
i\gamma_L(x,t)+i\alpha (t+x))}\mathcal{R}\theta (-x),$$ here
$\mathcal{K}(\mathcal{R})$ is a well-known bare transition  (reflection)
coefficient satisfying to the conditions:
\begin{equation}
|\mathcal{R}|^2 +|\mathcal{K}|^2=1
\quad\mathcal{R}\mathcal{K}^*+\mathcal{R}^*\mathcal{K}=0.
\label{KR}
\end{equation}
The solution obeys the matching condition at the point $x=0$ and the
boundary condition for $t\to\infty;x$ is finite, but for the case
$t\to\infty;x\to-\infty;t+x$ is finite, the solution has negative and positive
frequency  part simultaneously (because the first argument of the $G^{(0)}$ is
finite). It means, the solution describes the soliton-type excitations in
the final state. If one write the $\theta (x)$
in $\psi_{\varepsilon, L}$, the solution will be obeying the hermitian conjugated  boundary condition, but it will be
forbidden by the matching one.
The correct set of solutions satisfies the boundary conditions with the Dirac conjugation. They are calculated in Appendix\ref{funckSet}. Here, we have matched solutions of the non-linearized Schr\"odinger equation at the impurity position point. These wave functions can be
represented in the  "spinor"\ form, where the upper term is the wave function of an R-particle and the lower term is an L one.
\begin{equation}
\hat{\psi}_{\varepsilon}^{(1)}(x,t)=
\begin{array}
[c]{c}%
\left[  {\mathcal{K}}^{\ast}\Theta(-x)+\Theta(x)\right]  e^{-i\varepsilon
(t-x)}e^{i\gamma_{R}(x,t)}\\
{\mathcal{R}}^{\ast}\Theta(x)e^{i\gamma_{L}(x,t)+
i\alpha(t+x)}%
e^{-i\varepsilon(t+x)}%
\end{array}
\hat{\psi}_{\varepsilon}^{(2)}(x,t)=
\begin{array}
[c]{c}%
{\mathcal{R}}^{\ast}e^{-i\varepsilon(t-x)}\Theta(-x)e^{i\gamma
_{R}(x,t)-i\alpha(t-x)}\\
\left[  \Theta(-x)+{\mathcal{K}}^{\ast}\Theta(x)\right]  e^{-i\varepsilon
(t+x)}e^{i\gamma_{L}(x,t)}%
\end{array}
\label{set1}
\end{equation}
Here $i=1,2$ is a running number of linearly independent solution (all
$\varepsilon>0$).

For the case $U= 0$ they are a linear combination of the
left- and right-incident waves. At $U\ne 0$ only these
solutions satisfy the boundary conditions. The solutions with negative energy can be written in the form
\begin{equation}
\hat{\psi}_{-\varepsilon}^{(1)}(x,t)=
\begin{array}
[c]{c}%
\left[  \Theta(-x)+\Theta(x){\mathcal{K}}\right]  e^{i\varepsilon
(t-x)}e^{i\gamma_{R}(x,t)}\\
{\mathcal{R}}\Theta(-x)e^{i\varepsilon(t+x)}
e^{i\gamma_{L}(x,t)}%
e^{+i\alpha(t+x)}%
\end{array}
\quad
\hat{\psi}_{-\varepsilon}^{(2)}(x,t)=
\begin{array}
[c]{c}%
\Theta(x){\mathcal{R}}e^{i\varepsilon(t-x)}
e^{i\gamma_{R}(x,t)}%
e^{-i\alpha(t-x)}\\
\left[  {\mathcal{K}}\Theta(-x)+\Theta(x)\right]  e^{i\varepsilon
(t+x)}e^{i\gamma_{L}(x,t)}%
\end{array}
$$
{\em Dirac conjugated} solutions have an analogous form:
$$
\widetilde{\hat{\psi}}_{-\varepsilon}^{(1)}(x,t)=
\begin{array}
[c]{c}%
\left[  {\mathcal{K}}\Theta(-x)+\Theta(x)\right]  e^{-i\varepsilon
(t-x)}e^{-i\gamma_{R}(x,t)}\\
{\mathcal{R}}\Theta(x)e^{-i\gamma_{L}(x,t)
-i\alpha(t+x)}
e^{-i\varepsilon (t+x)}%
\end{array}
\quad\widetilde{\hat{\psi}}_
{-\varepsilon}^{(2)}(x,t)=
\begin{array}
[c]{c}%
{\mathcal{R}}\Theta(-x)e^{-i\gamma_{R}(x,t)
+i\alpha(t-x)}
e^{-i\varepsilon (t-x)}\\
\left[  \Theta(-x)+{\mathcal{K}}\Theta(x)\right]  e^{-i\varepsilon
(t+x)}e^{-i\gamma_{L}(x,t)}%
\end{array}
$$
$$\widetilde{\hat{\psi}}
_{\varepsilon}^{(1)}(x,t)=
\begin{array}
[c]{c}%
\left[  \Theta(-x)+\Theta(x){\mathcal{K}}^{\ast}\right]  e^{i\varepsilon
(t-x)}e^{-i\gamma_{R}(x,t)}\\
{\mathcal{R}}^{\ast}\Theta(-x)e^{i\varepsilon(t+x)}
e^{-i\gamma_{L} (x,t)}e^{-i\alpha(t+x)}%
\end{array}
\quad\widetilde{\hat{\psi}}
_{\varepsilon}^{(2)}(x,t)=
\begin{array}
[c]{c}%
\Theta(x){\mathcal{R}}^{\ast}e^{i\varepsilon(t-x)}
e^{-i
\gamma_{R} (x,t)}e^{+i\alpha(t-x)}\\
\left[  {\mathcal{K}}^{\ast}\Theta(-x)+\Theta(x)\right]  e^{i\varepsilon
(t+x)}e^{-i\gamma_{L}(x,t)}%
\end{array}
\label{Dirac11}
\end{equation}
At this set of solutions the argument of a phase $\alpha(t\pm x)$ does not vanish.
 We see, the functions ($\widetilde{\psi}$ and $\psi$) would be complex conjugated, if $\gamma$ \emph{would be real.} However, Green function with arbitrary $\gamma$ has an imaginary part.
 Its contribution to the wave function would correspond to the excitations moving to the contacts (now they  do not allowed by the boundary conditions). It is important, the scattering phase  depends on $U$, which means that after averaging over $U$, the resulting vertex of the e-i scattering will depend on the e-e interaction. Therefore, it cannot be described by a local three-fermion vertex in a one-dimensional region.

 The set of functions (\ref{set1}-9)  is not complete in sense of the standard scalar product (with complex-conjugated wave functions and diagonal in upper indices). That is absolutely understandable: part of solutions of Eq.(\ref{Dirac-inhomogeneous})
were discarded.
Therefore, we will seek expression of the Feynman's Green function with external field $U(x,t)$ in a more generally  form:
\begin{equation}
G_{\alpha,\beta}(xt,x't')=\sum_{i,k=1,2}\int_0^
\infty\frac{d\ee d\ee'}{(2\pi)^2}\left[S^{(i,k)}(\ee,\ee')\theta(t-t')
\psi^{i}
_{\alpha,\ee}(x,t)\widetilde{\psi}^{k}_{\beta,
\ee'}(x',t')
\right.-
\label{Green-general}
\end{equation}
$$\left.- S^{(i,k)}(-\ee,-\ee')\theta(t'-t)\psi^i_{\alpha
,-\ee}(x,t)\widetilde{\psi}^k_{\beta ,-\ee'}(x',t')\right],$$
non--diagonal in upper indexes. It gives a set of equations that define the functions $S^{(i,k)}$, because
expression (\ref{Green-general}) will be a Green function of the Eq.(\ref{Dirac-inhomogeneous}) only if the kernels $S^{(i,k)}$ obey the expression:
\begin{equation}
\delta_{\alpha,\beta}\delta(x-x')=\sum_{i,k=1,2}\int
\limits_0^\infty
\frac{d\ee d\ee'}{(2\pi)^2}\left[S^{(i,k)}(\ee,\ee')\psi^{i}
_{\alpha,\ee}(x,t)\widetilde{\psi}^{k}_{\beta,\ee'}(x',t)
+\ee,\ee'\to -\ee,-\ee'
\right]
\label{compl}
\end{equation}

\section{Charge density in the time-dependent  external field.}
\label{ChDens}
To construct Feynman Green function in the external field with short-range impurity, one has to solve Eqs.(\ref{compl}). First of all, we note the
important fact:
\begin{equation}
\int dx\hat{\psi}_{-\varepsilon_{1}}^{(i)}(x,t)\widetilde{\hat{\psi}}
_{\varepsilon_{2}%
}^{(k)}(x,t)=\int
dx\hat{\psi}_{\varepsilon_{1}}^{(i)}(x,t)\widetilde{\hat{\psi}
}_{-\varepsilon_{2}}^{(k)}(x,t)=0
\label{orto1}
\end{equation}
One can check it directly, taking into account  condition (\ref{KR}).
Besides, let us introduce two overlap integrals matrices
\begin{equation}
T_{ik}^{(\pm)}(\varepsilon_{1},\varepsilon_{2})=\int dx\widetilde{\hat{\psi}
}_{\pm\varepsilon_{1}}^{(i)}(x,t)\hat{\psi}_
{\pm\varepsilon_{2}}^{(k)}
(x,t)
\label{overlap1}
\end{equation}
Elements of the matrix for negative energies are:\begin{equation}
T_{11}^{(-)}(\varepsilon_{1},\varepsilon_{2})=
2\pi{\mathcal{K}}%
\delta(\varepsilon_{2}-\varepsilon_{1});\qquad T_{12}^{(-)}={\mathcal{R}}
\varphi_{-}(\varepsilon_{2}-\varepsilon_{1})$$ $$
T_{21}^{(-)}(\varepsilon_{1},\varepsilon_{2})={\mathcal{R}}
\varphi
_{+}(\varepsilon_{2}-\varepsilon_{1});\qquad T_{22}^{(-)}(\varepsilon
_{1},\varepsilon_{2})=2\pi{\mathcal{K}}\delta
(\varepsilon_{2}
-\varepsilon_{1}).
\label{Tminus}
\end{equation}
Here we have introduced two Fourier transforms:
\begin{equation}
\varphi_{\pm}(\varepsilon)=\int\! dz\; e^{i\varepsilon z\pm i\alpha(z)}.%
\label{defphi}
\end{equation}

 The quantities for $\varepsilon >0$ enter into the equations (\ref{compl}) for
$S^{(i,k)}(\varepsilon ,\varepsilon')$ but we will see later, to find the density it is sufficient to know only $S^{(i,k)}(-\varepsilon ,-\varepsilon')$. Therefore, we  restrict ourselves by the case $\varepsilon<0$.
After applying operation $\sum_{\alpha ,\beta }\int dxdx'\widetilde{\psi}^{i}
_{\alpha,-\ee}(x,t)\psi^{k}_{\beta,-\ee'}(x',t)$ to the equations (\ref{compl}), we have:
$$
\sum_{i,k}\int\frac{d\varepsilon d\varepsilon '}{(2\pi
i)^2}T^{-}_{j,i}(\varepsilon_1,\varepsilon )S^{i,k}(-\varepsilon, -\varepsilon
')T^-_{k,m}(\varepsilon ',\varepsilon_2 ) =
T^-_{j,m}(\varepsilon_1,\varepsilon_2).
$$ These expressions are valid provided
\begin{equation}
\sum_{i}\int\frac{d\varepsilon_1 }{2\pi
i}T^{-}_{j,i}(\varepsilon,\varepsilon_1 )S^{i,k}(-\varepsilon_1, -\varepsilon
')=2\pi\delta_{j,k} \delta (\varepsilon -\varepsilon ').
\label{invers}
\end{equation}
It means, $S$ and $T$ are the inverse operators. One can represent Eqs.(\ref{compl}) in an explicit form:
$$
{\mathcal{K}}S^{(1,1)}(-\varepsilon_1, -\varepsilon ') +
{\mathcal{R}}\int_0^{\infty} \frac{d\varepsilon}{2\pi}d\tau\exp{[-i\tau
(\varepsilon_1-\varepsilon)-i\alpha(\tau)]}S^{(2,1)}(-\varepsilon,
-\varepsilon ')=2\pi\delta(\varepsilon_1-\varepsilon ')
$$\begin{equation}
{\mathcal{K}}S^{(2,1)}(-\varepsilon_1, -\varepsilon ') +
{\mathcal{R}}\int_0^{\infty} \frac{d\varepsilon}{2\pi}d\tau\exp{[-i\tau
(\varepsilon_1-\varepsilon)+i\alpha(\tau)]}S^{(1,1)}
(-\varepsilon, -\varepsilon ')=0
\label{invers1}
\end{equation}
There is a system of Wiener-Hopf equations. We have to solve  Eqs.
(\ref{invers1}) with arbitrary $\alpha (\tau)$ explicitly to
calculate  the functional integral. Eqs. for $S^{(2,2)}$
and $S^{(1,2)}$ can be obtained from Eqs.(\ref{invers1}) by replacement
$\alpha\to-\alpha$.

We will see later, expressions
 for the currents  will have the ultraviolet divergences.
As a result, asymptotic behaviour of ${S}_{}$ (we will indicate it as ${S}_{as}$) in the region of very high energies ($\varepsilon,
\varepsilon '\gg \widetilde{\partial_t\alpha (\tau)}$) requires. (Here the quantity $\widetilde{\partial_t\alpha (\tau)}$
is a typical value of the $\partial_t\alpha (\tau)).$ In these regions the function
$\hat{T}(\varepsilon_1,\varepsilon)$ depends on $\varepsilon_1-\varepsilon$ only and it should decrease at
$|\varepsilon_1-\varepsilon|\to\infty$. Therefore, one can expand the integration range
  in the Eq.(\ref{invers1}) for the
$S_{as}(\varepsilon_1,\varepsilon )$ up to $-\infty$. In this
energy region $S_{as}(\varepsilon_1,\varepsilon )$ depends only on
$\varepsilon_1 -\varepsilon$ too, and we obtain equation with difference kernel.
It can be reduced to the matrix equation:
$$
 \sum_i T^{-}_{j,i}(\tau)S^{i,k}_{as}(\tau)=\delta_{j,k}.
 $$ So, at large $\varepsilon_1, \varepsilon_2$ the matrix
 $S^{i,k}_{as}(\varepsilon_1-\varepsilon)$ can be expressed in the form:
 \begin{equation}
S^{ik}_{as}(\varepsilon_{1}-\varepsilon_{2})=\left(
\begin{array}
[c]{cc}%
2\pi{\mathcal{K}}^{\ast}\delta(\varepsilon_{1}-\varepsilon_{2}) &
{\mathcal{R}}^{\ast}\varphi_{-}(\varepsilon_{2}-\varepsilon_{1})\\
{\mathcal{R}}^{\ast}\varphi_{+}(\varepsilon_{2}-\varepsilon_{1}) &
2\pi{\mathcal{K}}^{\ast}\delta(\varepsilon_{1}-
\varepsilon_{2})
\end{array}
\right)
\label{Sasympt}
\end{equation}
The difference $S^{i,k}(\varepsilon_1-\varepsilon_2) -
S^{i,k}_{as}(\varepsilon_1-\varepsilon_2)$ decreases for large $\ee_{1,2}$.
Let us introduce special notation for this
difference:
$
\widehat{S}^{ik}(\varepsilon_{1},\varepsilon_{2})=
S^{ik}(-\varepsilon_{1},-\varepsilon_{2})-S^{ik}_{(as)}
(\varepsilon_{1}-\varepsilon_{2}).
$
One can find  $\widehat{S}_{ik}(\varepsilon_{1},\varepsilon_{2})$ simply as a
series in reflection coefficient,  assuming the coefficient is small.
Also, we will need the following function of one variable:
\begin{equation}
\Pi_{ik}(t)=\int_0^\infty\frac{d\ee_1d\ee_2}{(2\pi)^2}
\widehat{S}_{ik}(\ee_1,\ee_2)e^{i(\ee_1-\ee_2)t}.
\label{defPi}
\end{equation}
As for $S^{ik}_{as}$, it should be "calculated"\ exactly.
Now one can  construct an expression for electron density  in the arbitrary
scalar external field and with presence of impurity.

It is  well-known, the currents in 1D theory are the subject of ultraviolet divergences, which lead, in particular, to the Adler anomaly (\cite{AD}). It is believed, the divergence occurs because in our approach the  filled Fermi sphere is unlimited from below. For this reason, the charge density is effectively undefined and must be regularized. For us,  the physical conditions for regularization  will be  the gauge invariance of the expressions  for the electron density and appearance of the $\delta$-functional source in the Adler anomaly associated with direct violation of the chiral symmetry. (Provided,  continuity equation of the electric charge is conserved).
In our case, the most convenient regularization way is the splitting arguments of the current. We define regularized $R$ and $L$-densities as:
\begin{equation}
\rho_{R}(x,t)=-<G_{RR}(x-\delta x,t-\delta t||x+\delta x,t+\delta
t)e^{2i\delta
tU(x,t)}>_{\delta t\rightarrow+0; \delta x \rightarrow 0}.
\label{RL-splitting}
\end{equation}
Splitting $\delta x$\ and $\delta t$\ are introduced  to regularize
the singularity. One has to keep $\delta t>0$, that provide the correct
order of operators. The factor $e^{2i\delta tU(0,t)}$  guaranties the gauge
invariance of the current, see Appendix \ref{Adler}. (Detailed  discussion about physical meaning and nature of the regularization is given in \cite{reg}.) Mainly this problem concerns
impurity-independent part of the currents ($\rho _{R,L}(x,t)_{bal})$ and has
been discussed in a context of LL problem without impurity, while
the impurity-depended part $(\delta\rho _{R,L}(x,t)$) contains
indeterminateness and should be uncovered by the same splitting. (One has to use
here the same regularization method as in the impurity-independent part.)

To calculate the charge density, we express it in terms of matrix
$S_{ik}(-\varepsilon_{1},-\varepsilon_{2})$. Let us label by symbols $<$ and
$>$\ the values of densities at $x<0$\ and $x>0,$ respectively. So,
\[
\rho_{R}^{<}(x,t)=\int_{0}^{\infty}\frac{d\varepsilon_{1}
d\varepsilon_{2}%
}{(2\pi)^{2}}e^{2i\delta tU(x,t)}e^{-i\gamma_{R}(x+\delta x,t+\delta
t)+i\gamma_{R}(x-\delta x,t-\delta t)}e^{i(\varepsilon_{1}-\varepsilon
_{2})(t-x)}\times
\]
\begin{equation}
\times e^{-i(\varepsilon_{1}+\varepsilon_{2})(\delta t-\delta x)}\left[
{\mathcal{K}}S_{11}(-\varepsilon_{1},-\varepsilon
_{2})+{\mathcal{R}}e^{i\alpha(t-x+\delta t-\delta x)}S_{12}%
(-\varepsilon_{1},-\varepsilon_{2})\right]
\label{rhoR1}
\end{equation}
at negative $x$ and
\begin{equation}
\rho_{R}^{>}(x,t)=\int_{0}^{\infty}\frac{d\varepsilon_{1}
d\varepsilon_{2}%
}{(2\pi)^{2}}e^{2i\delta tU(x,t)}e^{-i\gamma_{R}(x+\delta x,t+\delta
t)+i\gamma_{R}(x-\delta x,t-\delta t)}e^{i(\varepsilon_{1}-\varepsilon
_{2})(t-x)}\times
\end{equation}
\begin{equation}
\times e^{-i(\varepsilon_{1}+\varepsilon_{2})(\delta t-\delta x)}\left[
{\mathcal{K}}S_{11}(-\varepsilon_{1},-\varepsilon
_{2})+{\mathcal{R}}e^{-i\alpha(t-x-\delta t+\delta x)}S_{21}%
(-\varepsilon_{1},-\varepsilon_{2})\right]\label{rhoR2}
\end{equation}
for positive one. We also need expressions for the densities of left-handed particles:
\[
\rho_{L}^{<}(x,t)=\int_{0}^{\infty}\frac{d\varepsilon_{1}d\varepsilon_{2}%
}{(2\pi)^{2}}e^{2i\delta tU(x,t)}e^{-i\gamma_{L}(x+\delta x,t+\delta
t)+i\gamma_{L}(x-\delta x,t-\delta t)}e^{i(\varepsilon_{1}-\varepsilon
_{2})(t+x)}\times
\]
\begin{equation}
\times e^{-i(\varepsilon_{1}+\varepsilon_{2})(\delta t+\delta x)}\left[
{\mathcal{K}}S_{22}(-\varepsilon_{1},-\varepsilon
_{2})+{\mathcal{R}}e^{i\alpha(t+x-\delta t-\delta x)}S_{12}%
(-\varepsilon_{1},-\varepsilon_{2})\right]\label{rhoL1}
\end{equation}
for negative $x$ and
\[
\rho_{L}^{>}(x,t)=\int_{0}^{\infty}\frac{d\varepsilon_{1}d\varepsilon_{2}%
}{(2\pi)^{2}}e^{2i\delta tU(x,t)}e^{-i\gamma_{L}(x+\delta x,t+\delta
t)+i\gamma_{L}(x-\delta x,t-\delta t)}e^{i(\varepsilon_{1}-\varepsilon
_{2})(t+x)}\times
 \]
\begin{equation}
\times e^{-i(\varepsilon_{1}+\varepsilon_{2})(\delta t+\delta x)}\left[
{\mathcal{K}}S_{22}(-\varepsilon_{1},-\varepsilon
_{2})+{\mathcal{R}}e^{-i\alpha(t+x+\delta t+\delta x)}S_{21}%
(-\varepsilon_{1},-\varepsilon_{2})\right]
\label{rhoL2}
\end{equation}
for positive one.

Ultraviolet part of impurity-depended concentration comes from
Eqs.(\ref{rhoR1} - \ref{rhoL2}) after substitution into equations  the
asymptotic value of $S^{i,k}$. It will be an exact relation (not expansion
in power of small $\mathcal{R})$. After integration over $\varepsilon_{1,2}$
one has
\begin{equation}
\delta\varrho_{R}^{<}(x,t)_{UV}=-\frac{|{\mathcal R}|^2}{(2\pi)^{2}}\int\!
d\tau\; \frac
{1-e^{i\alpha
(x_-+\delta x_-)-i\alpha(\tau)}}{(\tau-x_-+\delta x_--i\delta)(\tau-x_--\delta
x_-+i\delta)}%
\label{adler-3}
\end{equation}
(here $x_{\pm}=t\pm x$). The expression is finite at $\delta x_{\pm}\to 0 $.
We proceed to the Fourier transform of the exponent  to calculate the integral:
\[
\delta\varrho_{R}^{<}(x,t)_{UV}=-\frac{|{\mathcal
R}|^2}{(2\pi)^{2}}\int\!\frac{d\omega d\tau}{2\pi}\;
\frac{e^{+i\alpha(x_-+\delta
x_-)}\varphi_-(\omega)\left(e^{-i\omega(x_-+\delta x_-)}-e^{-i\omega
\tau}\right)
}{(\tau-x_-+\delta x_--i\delta)(\tau-x_--\delta x_-+i\delta)},
\]
where the functions $\varphi_{\pm}$ are defined in Eq.(\ref{defphi}). There is the only
pole in the integrand located in the upper semi-plane which operates when
$\omega<0$.  After calculation the integral in $\tau$ and taking  the  limit $\delta
x_-\to 0$ we arrive at:
\begin{equation}
\delta\varrho_{R}^{<}(x,t)_{UV}=\frac{|{\mathcal
R}|^2}{2\pi}\int\!\frac{d\omega}{2\pi}\;
\theta(-\omega)\omega\varphi_-(\omega) e^{i\alpha(x_-)-i\omega x_-}=\frac{|{\mathcal R}|^2}{(2\pi)^2}\int\!\
d\tau\;
\frac{e^{-i\alpha(\tau)+i\alpha(x_-)}}{(\tau-x_--i\delta)^2}.
\end{equation}
Let us consider the $\rho_R$ at $x>0$. According to Eq.(\ref{rhoR2})
expression for $\varrho^>_R(x,t)$ differs from the considered case by the changes
 $\alpha\to -\alpha$ and by the sign in the argument splitting in $\alpha$:
\begin{equation}
\delta\varrho_{R}^{>}(x,t)_{UV}=-\frac{|{\mathcal R}|^2}{(2\pi)^{2}}\int\!
d\tau\; \frac
{1-e^{-i\alpha(x_--\delta x_-)+i\alpha(\tau)}}{(\tau-x_-+\delta
x_--i\delta)(\tau-x_--\delta x_-+i\delta)}=
\frac{|{\mathcal R}|^2}{(2\pi)^2}\int\!\
d\tau\;
\frac{e^{i\alpha(\tau)-i\alpha(x_-)}}{(\tau-x_-+i\delta)^2}%
\end{equation}
Let us proceed the same procedure with densities of the left particles. From
(\ref{rhoL1}) we obtain:
\[
\delta\varrho_{L}^{<}(x,t)_{UV}=-\frac{|{\mathcal{R}}|^{2}}
{(2\pi)^2}\int d\tau
\frac{1-e^{-i\alpha(\tau)+i\alpha(x_+-\delta x_+)}}{(\tau-x_++\delta
x_+-i\delta)(\tau-x_+-\delta x_++i\delta)}.
\]
The expression does not  coincide with $\varrho^<_R(x,t)$ after substitute
$x_-\to x_+$ (see the signs of $\alpha$.) In the limit $\delta x_{\pm}
\to 0$ the calculation gives finally:
\begin{equation}
\delta\varrho_R(x,t)_{UV}=\frac{|{\cal R}|^2}{(2\pi)^2}\int\! d\tau \left[
\frac{e^{-i\alpha(\tau)+i\alpha(x_-)}}{(\tau-x_--i\delta)^2}
\theta(-x)+
\frac{e^{i\alpha(\tau)-i\alpha(x_-)}}{(\tau-x_-+i\delta)^2}
\theta(x)
\right];
\label{UV-rhoR}
\end{equation}
\begin{equation}
\delta\varrho_L(x,t)_{UV}=\frac{|{\cal R}|^2}{(2\pi)^2}\int\! d\tau \left[
\frac{e^{i\alpha(\tau)-i\alpha(x_+)}}{(\tau-x_+-i\delta)^2}
\theta(x)+
\frac{e^{-i\alpha(\tau)+i\alpha(x_+)}}{(\tau-x_++i\delta)^2}
\theta(-x)
\right].
\label{UV-rhoL}
\end{equation}
Impurity-depending part of the density consists of the two parts: regular part
$(\varrho_{R,L}(x,t))_{reg})$ and ultraviolet one.  The regular part can be
obtain from Eqs.(\ref{rhoR1} - \ref{rhoL2}) (without splitting) by
substitution $S \to \widehat{S}$. In this way one has
\begin{equation}
\varrho_{R}(x,t)_{reg}={\cal K}\Pi_{11}(x_-)+{\cal R}
\left[
\theta(-x)\Pi_{12}(x_-)e^{i\alpha(x_-)}+
\theta(x)\Pi_{21}(x_-)e^{-i\alpha(x_-)}
\right]
\label{reg1}
\end{equation}
\begin{equation}
\varrho_{L}(x,t)_{reg}={\cal K}\Pi_{22}(x_+)+
{\cal R}\left[
\theta(-x)\Pi_{12}(x_+)e^{i\alpha(x_+)}
+
\theta(x)\Pi_{21}(x_+)e^{-i\alpha(x_+)}
\right] .
\label{reg2}
\end{equation}
Equations  (\ref{UV-rhoR} - \ref{reg2}) show,
the current
($\varrho_{R}-\varrho_{L}$) is continuous at the point $x=0$, as it should be.
On the contrary, the total density  undergoes the jump ($ {\mathfrak
D}(\omega)$).  The jump  plays a central role in the problem. This is the single unknown quantity demanding a calculation to obtain the current. To prove these assertions one observes, the
current and electron charge density should satisfy the conservation law of the electron current.
The "ballistic" electron current is completely determined by the Adler anomaly (see Appendix \ref{Adler}) and
satisfies the conservation law. This means, the impurity-dependent part of the current  separately satisfies  the  conservation  law, and the current $\delta j =\delta\rho_R-\delta\rho_L $ should  be continuous in the point $x=0$. All impurity-depending currents are functions of $t\pm x$. As a result, one has
\begin{equation}
\delta\rho_{R}(k,\omega)=\frac{\varrho_R^<(\omega)}
{i(\omega-k-i\delta)}-\frac{\varrho_R^>(\omega)}
{i(\omega-k+i\delta)};\quad
\delta\rho_{L}(k,\omega)=-\frac{\varrho_L^<(\omega)}
{i(\omega+k+i\delta)}+\frac{\varrho_L^>(\omega)}
{i(\omega+k-i\delta)}.
\label{}
\end{equation}
The relations contain the terms proportional to
$\delta (\omega \pm k)$  that have to be forbidden. They
describe  soliton-like excitations outgoing to the ends of the channel at $t\to\pm\infty$. It contradicts the boundary condition.
These terms disappear if the following conditions are fulfilled:
\begin{equation}
\rho^{<}_R(\omega)=0 \;{\rm at}\; \omega>0 ,\quad \rho^{>}_R(\omega)=0 \;{\rm
at} \;\omega<0 \quad {\rm and}\quad
\rho^{>}_L(\omega)=0\; {\rm at}\; \omega>0 , \quad \rho^{<}_L(\omega)=0\;
{\rm at}\; \omega<0
\label{Feynman-current}
\end{equation}
In view Eq.(\ref{Feynman-current}) and continuity the $\delta j(x=0,\omega)$ one can represent the impurity-dependent part of concentration in the simple form:
\begin{equation}
\delta\rho(k,\omega)= \frac{ik}{(\omega^2-k^2+i\delta)}{\mathfrak D}(\omega),
\quad{\rm here}\quad
{\mathfrak D}(\omega)=(\varrho^>_R(\omega,0)+\varrho^>_L(\omega,0))-
(\varrho^<_R(\omega,0)+\varrho^<_L(\omega,0))
\label{final-rho}
\end{equation} is the total charge jump.
Expression for the current  can be found from the conservation law:
\begin{equation}
\delta j(k,\omega)= \frac{i\omega}{(\omega^2-k^2+i\delta)}{\mathfrak
D}(\omega).
\label{final-j}
\end{equation}
From these equations one can see, the impurity-dependent parts of the "currents"\  conserve the chirality everywhere except the point $x=0$.  Corresponded conservation law is:
\begin{equation}
\partial_t \delta j+\partial_x \delta\rho =\mathfrak{D}(t)\delta (x),
\label{}
\end{equation}
i.e., the point-like impurity reduces to non-conservation point-like source of the chiral current, as it should be. The source should be added to the expression  for the Adler anomaly without impurity:
\begin{equation}
\partial_t j+\partial_x \rho=-\frac{\partial_xU}{\pi}+{\mathfrak
D}(t)\delta(x).
\label{A2}
\end{equation}
Solution of the conservation laws can be written in the form
\begin{equation}
\left(
\begin{array}{c}
\rho[U]\\
j[U]
\end{array}
\right)=
\left(
\begin{array}{c}
ik\\
i\omega
\end{array}
\right)
\frac{ikU(k,\omega)/\pi+{\mathfrak D}([\alpha],\omega)
}{(\omega^2-k^2+i\delta)}.
\label{jf1}
\end{equation}
 Thus, the non-trivial parts of the current are dependent only on the total value of the charge jump. It consists of two parts: the ultraviolet part (determined by Eqs.\ref{UV-rhoR},\ref{UV-rhoL}) and the regular one. One can see
from Eqs.(\ref{reg1},\ref{reg2}), the regular part
of the charge jump at the point $x=0$  equals to:
\begin{equation}
{\mathfrak D}_{reg}([\alpha],t)=2{\cal R}\left[
e^{-i\alpha(t)}\Pi_{21}(t)-e^{i\alpha(t)}\Pi_{12}(t)
\right].
\label{jump-reg}
\end{equation}
Note, only  the off-diagonal components of $\Pi(t)$ enter to the regular part
of  the charge jump.  We will show later that at small reflection coefficients they contribute only on the order of $|{\cal R}|^4$. Therefore, in the lowest
order in $R$ only ultraviolet part of a concentration produces the impurity-depended  current. The convergent series for ${\mathfrak
D}_{reg}([\alpha],t)$ are calculated in Appendix \ref{K}.
 One can see from Eq.(\ref{jf1}), the non-trivial part of the current does
 not depend on $U(t,x)$ directly, but on the function $\alpha([U],t)$ equals
\begin{equation}
\alpha([U],t)=\gamma_{R}(0,t)-\gamma_{L}(0,t)=
-\int\frac{d\omega dk}
{(2\pi)^2}\frac{2ikU(k,\omega)}{\omega^2-k^2
+i\delta}e^{-i\omega t}.
\label{defalpha}
\end{equation}
In this case, averaging over all realizations of the Hubbard fields can be represented as  an  averaging over phase shift $\alpha$. The first term in the r.h. of the Eq.(\ref{jf1}) (directly depended on $U(x,t)$) represents the ballistic current. It can be easily calculated.

\section{Equivalent field theory.}
\label{TH}
To construct an effective Hamiltonian of interacting electrons scattered by a point-like impurity, it is necessary to rewrite the action of the system (${\cal S}_{}$) in terms of the $\alpha$-variable.
We want to begin from useful for further consideration expressions.
The action consists of two parts: the
ballistic and impurity ones. The phenomenological definition of density
variation $
\delta\mathcal{H}(x,t)/\delta U(x,t)  = \delta\rho([U],x,t)$ makes possible to
calculate  variation of the action ($ i{\cal
S}_{}\equiv\log\mathfrak{Det}_{}[U] $) under influence of an external
field.
Taking into account that
\begin{equation}
 \log \mathfrak{Det}[U] =-i\int_{0}^{1}d\lambda\int d^{2}xU(x,t)\rho
\lbrack\lambda U](x,t),
\label{inlambda}
\end{equation}
 one can calculate the impurity part of the  action.  (Integration in  constant  of electron-external field interaction ($\lambda e_0$)  brings to the correct  combinatorial coefficient; see, for example, \cite{AGD}.) As regards the  ballistic contribution to the action, its calculatetion is well-known.

 One can see:
\begin{itemize}
\item
The ballistic part of the action ($\log \mathfrak{Det}[U]_{bal}$) appears
from the variation of the electron charge density under  influence of external fields. (It is the first term in r.h. Eq.(\ref{jf1})).
The result is
$$
\frac{i}{2\pi}
\int\frac{dkd \omega }{(2\pi)^2 }U(k,\omega)
U(-k,-\omega)\frac{k^2}{\omega^2 - k^2 + i\delta}.
$$
One should add to this expression  the weight-term arising from Hubbard transformation:
$$
\frac{i}{2}\int\frac{dk d\omega}{(2\pi)^2}\frac{U(k,\omega)
U(-k,-\omega)}{V_0(k)}.
$$
 As a result, the whole ballistic part of the action is
\begin{equation}
\log{\mathfrak{Det}[U]_{bal}} = \frac{i}{2}
\int\frac{dkd \omega }{(2\pi)^2 }U(k,\omega)
U(-k,-\omega)V^{-1}_0(k)\frac{\omega^2 - k^2 v_{c}^2(k)+ i\delta}{\omega^2 -
k^2 + i\delta} \label{b4a}.
\end{equation}
Eq.(\ref{b4a}) represents the Dzyaloshinsky-Larkin theorem \cite{Lar} in the form of a functional integral.
\item
According to Eqs.(\ref{final-rho},\ref{inlambda}), the effect of impurity  gives
additional term to the action:
\[
\log{\mathfrak{Det}}_{imp}=-i\int_0^1 d\lambda
\int\frac{dkd\omega}{(2\pi)^2}U(-k,-\omega)\frac{ik}{\omega^2-k^2+i\delta}{\mathfrak
D}([\lambda\alpha],\omega).
\]
It is essential for further consideration, this expression can be rewritten only as an $\alpha(\omega)$-functional:
\begin{equation}
\log{\mathfrak{Det}}_{imp}=-\frac{i}{2}\int_0^1
d\lambda\int\frac{d\omega}{(2\pi)}\alpha(-\omega)
{\mathfrak
D}([\lambda\alpha],\omega).
\label{det-good}
\end{equation}

The charge density in the external field is  variational derivative of the action over potential energy. Similar, the variational derivative of the $\log{\mathfrak{Det}}_{imp}$ in $\alpha(\tau)$ is the charge jump:
\begin{equation}
{\mathfrak D}[\alpha](\tau)=2i\frac{\delta}{\delta\alpha(\tau)}
\log{\mathfrak Det}_{imp}.
\label{jump-det}
\end{equation}
Indeed, after substitution (\ref{jump-det}) in (\ref{det-good}) one obtains
\[
\log{\mathfrak{Det}}_{imp}([\alpha])=\int_0^1\! d\lambda\int\! d\tau\;
\alpha(\tau)\frac{\delta}{\delta\lambda\alpha(\tau)}
\log{\mathfrak Det}_{imp}([\lambda\alpha])
=\int_0^1\! d\lambda\frac{d}{d\lambda}\log{\mathfrak
Det}_{imp}([\lambda\alpha]).
\]
In view of evident equality $\log{\mathfrak
Det}_{imp}[\lambda\alpha]|_{\lambda\to 0}=0$ one has proved the identity
(\ref{jump-det}).
\end{itemize}
As a result, the whole action ${\cal S}(\alpha)$  consists of
$\log{\mathfrak{Det}}_{bal}[\alpha]$ and $\log{\mathfrak{Det}}_{imp}[\alpha]$
(their analytical expressions are given by
Eqs.(\ref{b4a},\ref{det-good})).

\subsection{Linear response for the attracting LL.}
\label{det}
 We begin transition to the function variable $\alpha$ from the statistical sum in the Minkowski space:
\begin{equation}
{\cal Z}=\int {\cal D}U\exp\left\{\frac{i}{2}
\int\frac{dkd\omega}
{(2\pi)^{2}%
}\frac{U(-k,-\omega)U(k,\omega)}{V_0(k)}
\left[ \frac{\omega^2-v^2_c(k)k^2+i\delta}
{\omega^2-k^2+i\delta}\right]
\right\}
\mathfrak{Det}_{imp}([\alpha]).
\label{Z}
\end{equation}
To pass to variable $\alpha$, we will use the Faddeev-Popov
trick \cite{FP}. To that, we multiply the Eq.(\ref{Z}) on the factor
equals to $1$
$$
\int {\cal D}\alpha\delta\left\{\alpha+\int\frac{dk}{2\pi}U(\omega
,k)\frac{2ik}{\omega^2-k^2+i\delta}\right\}
$$
and represent the $\delta$-function as:
\begin{equation}
\delta\left\{\alpha(\omega)+\int\frac{dk}{2\pi}U(\omega,k)
\frac{2ik}{\omega^2-k^2+i\delta}\right\}=$$ $$=
\int\frac{{\cal D}\zeta(\omega)}{2\pi}\exp
\left\{-i\int\frac{d\omega
}{2\pi}\zeta(-\omega)\alpha(\omega)-
i\int\frac{d^{2}k}
{(2\pi)^{2}}%
U(k,\omega)\zeta(-\omega)
\frac{2ik}{\omega^2-k^2+i\delta}\right\}.
\end{equation}
The next step (integration over $U(x,t)$) is not difficult, as one has a Gaussian integral:
\begin{equation}
{\cal Z}=\int\! {\cal D}\alpha\;\mathfrak{Det}_{imp}[\alpha]\int
\frac{{\cal D}\zeta(\omega)}{2\pi}
\exp\left[-i\int\frac{d\omega}{2\pi}\zeta(-\omega)
\alpha(\omega)-
\frac{1}{2}\int\frac{d\omega}{2\pi}
\zeta(-\omega)W(\omega)\zeta(\omega)
\right]Z_{U},\label{ZU}
\end{equation}
where we have introduced "one-dimensional"\  renormalized potential:
\begin{equation}
W(\omega)=i\int\frac{dk}{2\pi} \frac{4k^{2}V_0(k)} {(\omega^2-k^2+i\delta)
(\omega^{2}-v^2_c(k)k^{2}+i\delta)}.%
\label{def-W}
\end{equation}
As the last step, one can integrate the Eq.(\ref{ZU})  in $\zeta(\omega)$:
\begin{equation}
{\cal Z}=Z_UZ_{\zeta}\int\! {\cal D}\alpha\;\mathfrak{Det}_{imp}[\alpha]
\exp{\left[ -\frac{1}{2}\int\frac{d\omega}{2\pi}
\frac{\alpha(-\omega)\alpha(\omega)}{W(\omega)}
\right]}.
\label{Zfinal}
\end{equation}So, we have obtained the "free part"\ of effective action \begin{equation}
S_{kin}([\alpha])= \frac{1}{2}\int\frac{d\omega}{2\pi}
\frac{\alpha(-\omega)\alpha(\omega)}{W(\omega)}.
\label{Skin}
\end{equation}
Here $Z_U$ and $Z_{\zeta}$ are the normalizing constants resulting from integration in $U$ and $\zeta$, they are cancelled from any observed value obtained in the same way. As regards the effective potential, it is proportional to $|\omega|^{-1}$\ by dimension, and
for the special case $\delta-$functional e-e interaction, one has
\begin{equation}
W(\omega) = \frac{2\pi}{|\omega|}\left[\frac{1}{v_c}-1\right].
\label{W}
\end{equation}
(For the point interaction limit, the integration region must be limited by $M$. The ultraviolet cutoff is determined by the nonlocality scale of the e-e interaction.)
So for the case point-like  interaction, we drive from initial problem to the effective field theory with
dimensionless variable $\alpha$ and statistical sum
\begin{equation}
{\cal Z}=\int\! {\cal D}\alpha\;\mathfrak{Det}_{imp}[\alpha]
\exp{\left[ -\frac{1}{4\pi\nu }\int^M_{-M}\frac{d\omega}{2\pi}
|\omega|\alpha(-\omega)\alpha(\omega)
\right]},
\label{Zdelta}
\end{equation}
where $\nu$ is a well-known parameter:
$
\nu=1/v_c-1.
$
The quantity $\nu$ plays a role of effective coupling constant. It tends to zero, if the strength of electron-electron interaction misses.

The iteration procedure of the functional integral (\ref{Zdelta}) is well-defined if $\nu>0$ only. This is true only for {\em attracting} potential. In the case
$\nu<0$ (repulsive potential) the direct expansion in $\nu$ diverges.  We will see later,   for
repulsive potential one can formulate the well-defined iteration procedure
starting from  weak permeable barrier (small ${\cal K}$) \cite{FK}. The procedure will be
formulated in terms of the new variable $\tilde\alpha$ with other “free part”\ of the action, $\tilde W(\omega)$ (see, Section \ref{LLrep}).

  Non-linear current
of non-interacting electrons, placed in an authentic external field, can be written  in the
form:
$$
j[U](x,t)=(\rho_R(U,x,t) - \rho_L(U,x,t))\exp{(\log{\mathfrak{Det}[U]})},
$$ here $\log{\mathfrak{Det}[U]}$ is the part of the action corresponding to
field $U$.
To calculate the linear response under the external field $\varphi\to 0$ applied to the channel, one has to substitute the total field in the form $U+\varphi$ and rewrite expression as
$$
j[U](x,t)=\int dx_1dt_1 \varphi (x_1,t_1)\frac{\delta}{\delta
U(x_1,t_1)}[(\rho_R(U,x,t) - \rho_L(U,x,t))\exp{(\log{\mathfrak{Det}[U]})}].
$$ To get  e-e
interaction we should average this expression over $U$:
$$
j(x,t)=\frac{1}{{\cal Z}}\int{\cal D}Uj[U](x,t)\exp{\left(
\frac{i}{2}\int\frac{dkd\omega}{(2\pi)^2}
\frac {U(k,\omega)
U(-k,-\omega)}{V_0(k)}\right)}.
 $$ After integration by parts, we  arrive to expression:
 \begin{equation}
j(x,t)=\frac{1}{{V_0\cal Z}}\int{\cal D}U\int dx_1dt_1 (-i\varphi
(x_1,t_1))U(x_1,t_1)(\rho_R(U,x,t) - \rho_L(U,x,t))
\mathfrak{Det}[U]_{bal}\mathfrak{Det}[\alpha] _{imp}
 \label{jfinal}
 \end{equation}

 Now, the Hubbard factor is hidden in $\mathfrak{Det}[U]_{bal}$. (Integration by parts
 is equivalent to using of a Ward's identity.) The final expression can be obtain  after  substitution here
 \begin{equation}
j(k,\omega) = j(k,\omega)_{bal} +
\frac{i\omega}{(\omega^2-k^2+i\delta)}{\mathfrak D}([\alpha],\omega); \quad j(k,\omega)_{bal}=\frac{i}{\pi}
\frac{\omega} {\omega ^2 -k^2 +i\delta}ikU(k,\omega).
 \label{jfinalA}
 \end{equation}(The ballistic part of the current is determined by the  Adler anomaly; see Appendix \ref{Adler}.)
 To express the integral in terms of variable $\alpha$, one should repeat the Faddeev-Popov trick described above. Let us discuss the integration in
$U$. There are two different terms in the pre-exponent. The first is related
to ballistic current, and it is quadratic in potential $U$. The second one is
the impurity-depended part of the current  multiplied by $U$. As a result, the latter
term is linear in $U$. So, we can repeat the Faddeev-Popov trick  with minimum  modification.  The impurity-depended part ($\delta j$) of the current is expressed by relation:
\begin{equation}
\delta j(\omega,k)=\frac{2i\omega}{(\omega^2-k^2+i\delta)}\int
\frac{dqd\omega^{'}}{(2\pi)^2}\frac{q\varphi
(q,\omega^{'})}{(\omega^{'2}-v^2_c(q)q^2+i\delta)}
(<\frac{i{\mathfrak D}([\alpha],\omega)\alpha (-\omega^{'})}{W(|\omega^{'}|)}>
+
\label{j1}
\end{equation}$$+
\frac{2}{\pi W(|\omega|)}\frac{k^2 V_0(k)}{(\omega^{2}-v_c^2(k)k^2+i\delta)}
 <2\pi\delta (\omega -\omega^{'} ) - \frac{\alpha (\omega)\alpha
 (-\omega^{'})}{W(|\omega^{'}|)}>);
$$
\begin{equation}
<...> =\frac{1}{{\cal Z}}\int{\cal D}\alpha
...\mathfrak{Det}[\alpha]_{imp}\mathfrak{Det}[\alpha]_{bal}=
 \frac{1}{{\cal Z}}\int{\cal
 D}\alpha...\exp{\left[ -\frac{1}{2}\int\frac{d\omega'}{2\pi}
\frac{\alpha(-\omega')\alpha(\omega')}{W(|\omega'|)}
\right]}\mathfrak{Det}[\alpha]_{imp}
\label{Hamilt}\end{equation}
 Further conversion consists in  applying identity  (\ref{jump-det}) and integrating by parts because
$$
<2\pi\delta (\omega -\omega^{'})  - \frac{\alpha (\omega^{'})\alpha
(-\omega)}{W(|\omega^{'}|)}>= \frac{1}{{\cal Z}}\int{\cal
D}\alpha\frac{\delta}{\delta \alpha (\omega)}\left[\alpha
(\omega^{'})\exp{(
-\frac{1}{2}\int\frac{d\omega'}{2\pi}
\frac{\alpha(-\omega')\alpha(\omega')}{W(|\omega'|)})}\right]
\mathfrak{Det}[\alpha]_{imp}.
$$
Take into account that after averaging $\omega=\omega'$, one can move Eq.(\ref{j1}) to
\begin{equation}
\delta j(\omega,k)= -\frac{2\omega |\omega |}{\omega^2-
v^2_c(k)k^2+i\delta}\int\frac{dq}{(2\pi)}
\frac{E(q,\omega)}
{\omega^2- v^2_c(q)q^2+i\delta}\frac{v_c(\omega)}{\pi}
{\cal R}_{\omega}^2.
\label{final-sigma-1}
\end{equation}
Expression for the {\em Feynman response (Eq.(\ref{final-sigma-1})) is  exact}. It completely determines dependence of conductivity on momenta $k$ and
$q$ (the problem with impurity lost a  translation invariance),  and contains only one
unknown function of $\omega$.  The dimensionless factor ${\cal R}_{\omega}$ can be called as a "renormalized reflection coefficient":
\begin{equation}
(2\pi)\delta(\omega-\omega'){\cal R}_{\omega}^2  = \frac{i\pi }{v_c(\omega
)|\omega|W(|\omega|)}\langle\mathfrak{D}(\omega)\alpha
(-\omega')\rangle,
\label{R-renorm}
\end{equation} here we have introduced the factor
$v_c(\omega)=\sqrt{1+V_0(k_0)/\pi}$, where $k_0$ is the root of the equation
$\omega=v_c(k)\cdot k$. (One should introduce the factor $1/v_c(\omega)$ here to secure relation ${\rm Re}{\cal R}_{\omega}^2\le 1$).
The r.h. of Eq.(\ref{R-renorm}) has to be calculated from microscopic theory.
Introduction of the ${\cal R}_{\omega}$ gives possibility to present the conductance (${\cal C}$) of a channel in a conventional form. It can be got as the  limit of ${\rm Re}\ {\cal R}^2_\omega$  at small frequencies, if the renormalized Fermi velocity is a smooth function. (Expression of charge-jump for the attracting problem are given in Appendix \ref{sumAttr}.)
To obtain  retarded response, one should calculate  the {\em Feynman one (not retarded)} response on the real $\omega$ axis and continue the resulting expression in accordance with $|\omega |
\to +\sqrt{\omega + i\delta}$. (For details, see \cite{Af}). As a result, one has
 \begin{equation}
 {\cal C}(\omega )=\frac{e^2_0}{2\pi v_c(\omega)}(1-{\rm Re}\ {\cal
 R}_{\omega}^2)|_{\omega\to 0}.
 \label{G}
 \end{equation}  One can define the renomalized
 transition coefficient ($ {\cal K}_{\omega}$) as: $1-{\rm Re}\ {\cal R}_{\omega}^2={\rm
 Re}\ {\cal K}_{\omega}^2$.

 It is useful to rewrite the expression of the exact reflection coefficient,  Eq.(\ref{R-renorm}), in terms
 Green's functions of the effective field theory. To this effect, it is
sufficient to
 use the identity Eq.(\ref{jump-det}), take  into account relation ${\mathfrak
 Det}_{imp}\delta\log{\mathfrak Det}_{imp}/\delta\alpha (\omega)=
 \delta{\mathfrak Det}_{imp}/\delta\alpha (\omega)$ and integrate received
 functional integral by parts. Variation the $\alpha (-\omega')$ gives a
 bare  Green's function ($ G_0(\omega)$) times on $W^{-1}(\omega)$, while
 variation the impurity part of action gives the exact Green's function $
 G(\tau-\tau')=\langle\alpha(\tau)\alpha(\tau')
\rangle $. At the end of this procedure we get:
\begin{equation}
{\cal R}_{\omega}^2=\frac{2\pi}{|\omega|W^2(\omega)v_c(\omega)}[
G_0(\omega)-G(\omega)].
\label{obs2}
\end{equation}

\subsection{LL with repulsive fermions.}
\label{LLrep}
To formulate
well-defined iteration procedure, one should expand the impurity action in
series in powers of $|{\cal K}|^2$ (The problem has to be formulated closely
to the split channel). The iteration procedure of Eq.(\ref{invers1}) over small
$|{\cal K}|^{2 n}$ will indicate the correct field variable replaced
the variable $\alpha$. It is expounded at Appendix \ref{K}. Here it is shown, the correct field variable is
\begin{equation}
\tilde\alpha(\omega)= -{\rm sign} (\omega)\alpha(\omega)
\quad {\rm or} \quad \tilde\alpha(\tau)=\alpha(\tau)_+-\alpha(\tau)_-,\quad{\rm while}\quad\alpha(\tau)=\alpha(\tau)_++\alpha(\tau)_-,
\label{tildalph}
\end{equation}
where $\alpha(\tau)_{\pm}$ is the part of the function analytical in
upper/lower semi-plane of $\tau$.
In term of $\tilde\alpha$ the "free part"\ of the total action can be written
 as
\begin{equation}
\log\widetilde{{\mathfrak Det}}[{\tilde\alpha}]_{ball}=-\frac{1}{2}\int
\frac{d\omega}{2\pi}\frac{{\tilde \alpha } (\omega){\tilde \alpha }
(-\omega)}{{\widetilde W (\omega)}},\quad {\rm where}\quad
\widetilde{W}^{-1}(\omega)=-W^{-1}
(\omega)-\frac{|\omega|}{2\pi}.
\label{Wdual}
\end{equation}For the case of point е-е interaction
$\widetilde W(\omega) = 2\pi\tilde \nu/|\omega|$,
 and $\tilde \nu=v_c-1$.

We will denote an average with the
action (\ref{Wdual}) and $\widetilde{{\mathfrak D}}([{\tilde\alpha}])_{imp}$
(charge jump for the repulsive interaction) as $<...>_K$. So,  transition to the $\tilde\alpha$-variable changes the incorrect signum of the "free part"\ of the action and produces the well-defined iteration procedure for the repulsive interaction.

It is very useful to rewrite the ${\cal C}(\omega)$ in terms of exact
transition coefficient ${\cal K}_{\omega}^2$. In Appendix \ref{DA} we have shown, the problems with attracting and repulsive e-e interaction are dual.
  It means,   results of one problem (i.e. $\K ^2_{\omega}$)  gets
 from other   (i.e. from $\R ^2_{\omega}$) by replacement ${\cal R}, \alpha, v_c   \leftrightarrow  {\cal K},\tilde\alpha, v_c^{-1} $. This property is exact for the point-like impurity and any e-e interaction  (provided the series for ${\mathfrak D}([{\alpha}],\omega)$ converge, even if asymptotically, \cite{As}).  At  first glance,  results of the problems with attracting and repulsive interactions cannot be obtained  one from the other,
because ${\mathfrak D}_{UV}$ should be proportional to $|\R |^2$ in  both cases (see Eqs.\ref{UV-rhoR}, \ref{UV-rhoL}).  This factor cannot  be changed  to the
$|{\cal K}|^2$,  because the hump of  electron density  in front of  impurity is determined by the reflection probability in both issues. However, duality takes place. To obtain duality, we must extract the ultraviolet part from the entire repulsive interaction charge jump (${\mathfrak
D}_{}([\tilde\alpha],\omega)$). This part is proportional to $|{\cal R}|^2$. Next, from the regular part, we must extract a term with the same base, but proportional to $|{\cal K}|^2$. The sum of these terms must be extracted from the entire charge jump. It means
\begin{equation}
\tilde\alpha(t)\widetilde{{\mathfrak D}}([\tilde\alpha],t)= \alpha (t) {\mathfrak
D}([\alpha],t)|_{{\cal R},\alpha \to{\cal K}, \tilde\alpha},\quad{\rm where}\quad{{\mathfrak D}}([\tilde\alpha],t)=-\frac{\tilde\alpha'(t)}{\pi}+
\widetilde{{\mathfrak D}}([\tilde\alpha],t)
\label{jumpDual}
\end{equation} is the total charge jump.
Extracting the first term
from the total charge jump is necessary to
eliminate the ballistic current from the expression for the total current. (It is absent in the split channel.) Note, the duality property should be formulated for the transition and reflection coefficient only, or for
$\alpha {\mathfrak
D}([\alpha],t)$ and $\tilde\alpha \tilde{\mathfrak
D}([\tilde\alpha],t)$. (These combinations determine expressions for the impurity-depending parts of action, currents etc.)
Due to duality, one can rewrite expression  for the ${\cal
 R}_{\omega}^2$ of the attracting problem to
\begin{equation}
(2\pi)\delta(\omega-\omega'){\cal K}_{\omega}^2  = \frac{i\pi {v_c(\omega)}
}{|\omega|{\widetilde W}(\omega)Z}
\int\!{\cal D}\widetilde{\alpha}\;\widetilde{{\mathfrak
D}}([{\tilde\alpha}],\omega)\widetilde{\alpha}(-\omega')
\widetilde{{\mathfrak Det}}[{\tilde\alpha}]_{imp}
\exp\left[-\frac{1}{2}\int\frac{d\omega}{2\pi}
\frac{\widetilde{\alpha}(-\omega)
\widetilde{\alpha}(\omega)}{\widetilde{W}(\omega)}
\right]
\label{Kk}
\end{equation}for the repulsion one. To make sure that derived in this way coefficient ${\cal K}_{\omega}^2$ holds expression $1-{\rm Re}\ {\cal R}_{\omega}^2={\rm
 Re}\ {\cal K}_{\omega}^2$ too,
  note that Eq.(\ref{R-renorm}) is correct in terms of both $\alpha$-variables.  Therefore to
 prove Eq.(\ref{Kk}) directly, it is enough to pass to the new $\tilde\alpha$ variable in
 Eq.(\ref{R-renorm}).
To that, let us substitute into
Eq.(\ref{R-renorm}) the Eq.(\ref{jumpDual}) and rewrite the
$\tilde\alpha$ from matrix element as $-\widetilde W(|\omega|)\delta\widetilde{{\mathfrak
Det}}[{\tilde\alpha}]_{ball}/\delta\tilde\alpha.$
After
integration by parts  and taking into account relation dual to
Eq.(\ref{jump-det}), one has
$$
2\pi\delta(\omega-\omega^{'}){\cal
R}_{\omega}^2=-2\pi\delta(\omega-\omega^{'})\frac{\tilde
W(\omega)}{v_c(\omega)W(\omega)}+\frac{\pi i}{|\omega
|v_c(\omega)W(\omega)}\left(1+\tilde W(\omega)\frac{|\omega|}{2\pi}
\right)<\tilde \alpha (-\omega^{'})\widetilde{{\mathfrak
D}}[\tilde\alpha](\omega)>_K.
$$ It remains to take into account relations
$v_c W(\omega)=-\tilde W(\omega);$   $1+\tilde
W(\omega)|\omega|/2\pi=v_c(\omega)$. This
allows us to write the expression for the conductivity in its usual form, showing that the exact coefficient ${\cal K}_{\omega}^2 $ in (\ref{Kk}) is entered correctly:
 \begin{equation}
 {\cal C}(\omega )=\frac{e^2_0}{2\pi v_c(\omega)}{\rm Re}\ {\cal
 K}_{\omega}^2|_{\omega\to 0},
 \label{GK}
 \end{equation}

\subsection{The final expression for electron-impurity part of the action.}
\label{main e-i}
  Our approach requires integration over the coupling constant to get an expression for electron-impurity action, $\log{\mathfrak{Det}}_{imp}$.  It is not a problem for the iterating procedure, but outside of it, a need of integrating in $\lambda$ leads to the more complicated calculations. Fortunately, in our problem one can integrate the impurity part of the action  over $\lambda$ in a general form. As a result of this operation, the action describing
e-i interaction  for the attracting e-e interaction is
\begin{equation}
\log\Det\!=\!\sum_{n=1}^\infty
\frac{(-1)^{n+1}}{n}\!\left(\!\frac{|\cal R|}{|{\cal K}|}
\!\right)^{2n}\!{\cal C}_{2n-1}
;
{\cal C}_{n}=\int\!\!\frac{d\tau_0.. d\tau_n}{(2\pi
i)^{n+1}}
\frac{1\!-\!\cos[\alpha(\tau_0)\!-\!\alpha(\tau_1)\!+
\!\ldots \alpha(\tau_n)]}
{(\tau_0-\tau_1-i\delta)(\tau_1-\tau_2-i\delta)..(\tau_n-\tau_0-i\delta)}
\label{det1text}
\end{equation}
(The path from initial e-i action to this one is expounded in Appendix \ref{K}.)

Consequently, we have got the non-local interaction. It is the payment for the
transition from the 1+1-dimension theory to 0+1 one. Nevertheless, this
Hamiltonian makes possible to study effects,  demanding summation of an
infinite number of diagrams.
For example, in \cite{CondReem} we have studied the effects associated with
the screening of  one-dimensional channel by the surrounding
three-dimensional environment. To this, we had to move beyond the  perturbation
theory both  the e-e and the e-i scattering \footnote{In particular, the paper shows that from expressions for the conductance  it follows: the  limit $k\to0$ in bare e-e interaction $V_0(k)$  should be understood
as $\lim V_0|_{k\sim 1/L}$ from the side of  1D
region. (Not as the value of e-e interaction in the 3D contact where  $V_0$
vanishes, and $v_c$ always equals to 1.) Therefore, the conductance is controlled not by the 3D contact region, but by the “bottleneck”. The role of which is played by 1D channel.
As a result, we have got the ordinary factor $v_c(k)|_{k\sim 1/L}$ in the conductance expression.
 We guess, the authors of the papers \cite{Steph},\cite{msl} came to the opposite conclusion, since they assumed in  mathematical model that 3D region can be described as the region with $v_c=1$ {\em in the 1D equation}. Thus, one does not account for,  the  wave packets with a linear spectrum  is not non-spreading in a 3D region.  This is why a three-dimensional region cannot be described by a 1D equation. If  $v_c=1$ at the edge of a one-dimensional region,  their result is correct,  but it is not always so.}.

\section{Reflection
coefficient in the lowest approximation.}
\label{AttrP1}
Let us calculate an expression  for reflection coefficient in the lowest approximation in $|{\cal R}|^2$  for the point-like attracting interaction.
First time, it was calculated in \cite{FK},\cite{Fur}. In this order we can
neglect corrections to the determinant and use expression of the UV-part of
charge density  $\mathfrak{D}(\omega)$ following from
Eqs.(\ref{UV-rhoR},\ref{UV-rhoL}).  According to Eq.(\ref{R-renorm}) we get expression:
\begin{equation}
2\pi\delta(\omega-\omega'){\cal R}_{\omega}^2=\frac{i|{\cal R}|^2}{4\pi
v_c|\omega|W(\omega)}\int\!\! d\tau dt e^{i\omega t}
\langle\big[\frac{\alpha(-\omega')}{(\tau-t-i\delta)^2}
+\frac{\alpha(-\omega')}{(\tau-t+i\delta)^2}\big]
\big[e^{i\alpha(\tau)-i\alpha(t)}
-e^{-i\alpha(\tau)+\alpha(t)
}\big]\rangle
\label{R1}
\end{equation}
Let us calculate the basic integral for one-dimensional theory:
\begin{equation}
\Xi(\tau_1-\tau_2)\equiv\langle e^{i\alpha(\tau_1)-i\alpha(\tau_2)}\rangle
=\frac{1}{{\cal Z}}\int {\cal D }\alpha
e^{i\alpha(\tau_1)-i\alpha(\tau_2)}\exp\left[-\frac{1}{2}
\int\frac{d\omega}
{2\pi}
\alpha(-\omega)W^{-1}(\omega)\alpha(\omega)\right].
\label{XiCommon}
\end{equation}
It is a Gaussian integral and saddle point $\alpha_0$ is:
$
\alpha_0(-\omega')=iW(|\omega'|)\left[e^{-i\omega' \tau_1}-e^{-i\omega'
\tau_2}\right].
$
 It leads to the following expression for the correlation function
\begin{equation}
\Xi(|\tau_1-\tau_2|)=\exp[-1/2\int\frac{d\omega}{2\pi}W(\omega)
\left|e^{i\omega \tau_1}-e^{i\omega \tau_2}\right|^2].
\label{propagator}
\end{equation}
This expression depends only on the difference $|\tau_1-\tau_2|$. For the
point-like potential, $V(k)=V_0$, the $W=2\pi\nu/|\omega|$ and integral
equals:
\begin{equation}
\Xi(|\tau_1-\tau_2|)=\left\{\begin{array}{cc}
1/(M|\xi|)^{2\nu}&|\xi|\gg 1/M\\
1&|\xi|\ll 1/M
\end{array}
\right.
\qquad \xi = \tau_1-\tau_2
\label{Xi}
\end{equation} A such  type of correlator  guaranties the absence of UV-divergence in an observed value, as it should be.
 Returning to the expression (\ref{R1}), we see - in our approximation it is a Gaussian-type integral and we arrive at:
\begin{equation}
2\pi\delta(\omega-\omega'){\cal R}^2_\omega=\frac{|{\cal R}|^2}{2\pi
v_c|\omega|}\int\!\! d\tau dt\; \Xi(|t-\tau|) e^{i\omega
t}\left[\frac{1}{(\tau-t-i\delta)^2}+\frac{1}{(\tau-t+i\delta)^2}
\right]\left[e^{-i\omega'
t}-e^{-i\omega'\tau}\right].
\end{equation}
Integration in center mass coordinate produces $\delta(\omega-\omega')$, and
one has:
\begin{equation}
{\cal R}^2_\omega=\frac{|{\cal R}|^2}{2\pi v_c|\omega|}\int\!\! d\xi\;
\left[\frac{1}{(\xi-i\delta)^2}+\frac{1
}{(\xi+i\delta)^2}\right]\left(1-e^{i\omega \xi}\right)\Xi(|\xi|)
\label{R2}
\end{equation}So, the real part of (\ref{R2}) is
\begin{equation}
{\rm Re}\ {\cal R}^2_\omega
=\frac{|{\cal R}|^2 M}{2\pi v_c|\omega|}\int_{0}^{\infty}d\tau
\frac{\Xi(\tau)}{(\tau)^2}\sin^2{(\frac{\omega}
{2 M} \tau)}.
\label{R2a}
\end{equation}
Calculating this integral for the point-like potential (\ref{Xi}), we obtain:
\begin{equation}
{\rm Re}\ {\cal R}^2_\omega=\frac{2}{v_c}\Gamma(-1-2\nu)
\frac{\sin\pi\nu}{\pi}|{\cal R}|^2\left(\frac{|\omega|}{M}\right)^{2\nu},
\label{R3}
\end{equation}while the ${\rm Im}\ $-part of (\ref{R2}) is zero due to the
oddness of the integrand (integration region in $\zeta$ is unlimited, and
$\Xi$ is depended on $|\xi|$). As it should be, expression Eq.(\ref{R2}) has
not any divergences at small $\xi$  for weak e-e interaction case (see Eqs.\ref{Xi},\ref{R2}). We have calculated the time-ordered response. To get  retarded one,  we should make the analytical continuation $|\omega|\to +\sqrt{\omega^2 +i\delta}$.

The frequency dependence of Eq.(\ref{R3}) is valid at  $ \nu < 1/2 $ and at $ \nu \ge 1/2 $ it has to be
slowly modified, because one cannot use  the asymptotic form $\Xi(|\xi|)$ for $|\xi|
\gg 1/M$. For the case, the region of small $\xi$ is emphasized. As one see
from (\ref{Xi}) the correlator $\Xi\to 1\ $ at $\tau_1\to\tau_2.$ In
this region, the frequency-dependence of the conductance becomes linear.
Indeed, in dimensionless variable $z=\omega \xi$ one can rewrite the integral
on r.h. of Eq.(\ref{R2}) as
$ \int_0^{\infty}dz\Xi(z M/\omega)(\sin z/z)^2 $. So, at $ \nu \ge 1/2 $ one
should use an opposite asymptotic form for $\Xi$ in the region $z_0\le
|\omega|/M$ (that is 1), and main contribution gives this small region. One can estimate the integral as $\int_0^{z_0}dz$. As a result we have:
\begin{equation}
{\rm Re}\ {\cal R}^2_{\omega}\cong c{|\cal R}|^2\frac{|\omega |}{M},
\label{R4a}
\end{equation} where "c"\ is a numerical coefficient of the order of 1. The result is valid provided ${|\cal R}|^2\to
0$  is the smallest parameter of the theory. Notice, the difference between the asymptotic forms at $\nu<1/2$ and $\nu\ge1/2$ results from the absence of a singularity in the properly regularized expression for the charge density in the UV region. We have seen, for $ \nu \ge 1/2 $ the coefficient in Eq.(\ref{R4a}) depends on the details of the e-i interaction at small distances. They determine value of "c".
Therefore scaling approach is valid only at $ \nu <1/2 $. The singularity of the $\Gamma$-function at $\nu=1/2$ likely arises from a change in the GS wave function. This may be an effect analogous to the phase-slip centre in superconductivity \cite{PS}.
 \subsection{Crossover region.}
 \label{smallK}
From  expression (\ref{R3}) for the first order correction in the
reflection/transition coefficient  one  see, the transport properties of
 a channel change drastically due to e-e interaction. However,  there should be a parameter's region where the channel still has a finite reflection/transition coefficients (the crossover region). Let us consider the case of repulsive interaction.
 For estimation  of the size  of crossover region, we will examine the domain of very  small bare reflection coefficient, i.e. we should describe the system by $\alpha$ -  variable (not $\tilde\alpha$). For this, we should expand Eq.(\ref{GK}) in
the power of $|{\cal R}|^2\left( {\cal M}/|\omega|
\right)^{2{|\nu|}}\ll 1$.
 It is legal only for a very  small $|{\cal R}|^2$.
In the case $v_c>1$ expansion of the $\mathfrak{Det}_{imp}[\alpha]$ in small
$|{\cal R}|^2_{}$  in the partition function (\ref{Zfinal}) is not well-defined, but we have understood,  the ratio of the functional integrals converges as a whole.  Therefore, to have an expansion of conductance expression in very small $|{\cal R}|^2$, one should  perform an analytical continuation procedure (transformation of the path of integration). For our
problem, it is equivalent to the replacement $\alpha (\pm \omega)\to i\alpha
(\pm \omega)$. It changes the  bare e-e interaction
(Eq.(\ref{W})) to
  \begin{equation}
 W_{cr}(\omega) = \frac{2\pi}{|\omega|}|\frac{1}{v_c}-1|.
\label{Wtrans}
\end{equation} As a result, the saddle point will change too:
$
\alpha_0(-\omega')= - W_{cr}(|\omega'|)\left[e^{-i\omega' \tau_1}-e^{-i\omega'
\tau_2}\right],
$ and the correlation function $\Xi$ will be $(M|\xi |)^{2|\nu|}$
at $M|\xi |\gg 1.$
So, for weak reflecting impurity and repulsive potential one has
\begin{equation}
G(\omega )=\frac{e^2}{2\pi v_c}\left( 1  -\frac{1}{\pi}|{\cal R
}|^2\left(\frac{M}{|\omega |}\right)^{2|\nu |}\int
d\zeta\frac{1-\cos{\zeta}}{\zeta^2}\zeta^{2|\nu|}\right),
\label{UR}
\end{equation}
and crossing  from the conducting mode to the split channel takes place when the second term is about the first one. It means,
transition from open to closed channel takes place at $|{\cal R }|^2 \cong
\left( |\omega |/M\right)^{2|\nu |}$. Note, the power is $2|\nu |$ not
the $2\tilde\nu$, as it seems at first glance.

\section{Renormalization group approach.}
\label{RGgrp}
In the previous section, we derived expression of the exact action of 1D channel with one
point-like impurity and e-e interaction.  Hard-to-use computation obtained (0+1)-dimension action follows from its nonlocality.  However,  nonlocality makes the theory convergent in the ultraviolet region. In this section we will show,  in frequency representation the Lagrangian of the problem can be transformed to the polynomial action with local non-trivial  vertices. (They  depend on bare reflection coefficient and frequency). Nevertheless, the cost of the step  is high. As usual, the long-wavelength expansion of a non-local Lagrangian brings to the UV divergencies in observed quantities.  Therefore a renormalization procedure is required.
Successful transition to the local action
is possible because the impurity part of the charge jump does not depend on $v_c$.
As a result, expansion of the non-trivial part of the action describing e-i
interaction  (Eq.\ref{det1text}), should be self-dual (i.e., the vertices of the problems should transform one to other under replacement  ${\cal K}\leftrightarrow{\cal R};-{\rm
sign}(\omega)\cdot\alpha(\omega)\leftrightarrow\alpha(\omega)$).  Otherwise, these problems would not be dual. So, duality determines the structure of the vertices. It is a powerful tool for the nonperturbative methods.
Exception is the trivial “free part” of the action. It depends on $\nu$ (i.e. $v_c$) directly.

As a first steep towards formulating  the renormalization  group (RG) approach,
we will expand the interacting part of renormalized Lagrangian \\$(
\mathfrak{Det}_{imp}([\alpha])=e^{-S_{r}([\alpha])}$)
in powers of $\alpha$. We will consider RG-approach in original
 Gell-Mann - Low formulation \cite{GML} for the attracting e-e interaction. Behaviour of conductance is determined by infrared divergences existing at small $\omega$.
  As a result, to calculate the conductance of the channel, one should sum the infrared logarithmically divergent terms. The RG-approach is a system tool for  solution the problem of a  such type.
 (To sum the items of the order of $\nu^{n}\log^{n-m}({\cal M}/|\omega|);\nu\ll1;$ here
 $m=0$ - leading logarithmic, or one-loop, approximation, $m=1$ - two-loop approximation,
 etc. ${\cal M}$ is an auxiliary quantity separating the low- and high-frequency regions. It is generally assumed that ${\cal M}$ is determined by the nonlocality of the effective Hamiltonian.) Sufficient conditions of the RG-method are
\begin{equation}
\nu\ln{({\cal M/|\omega |} })\ll 1;\quad \ln{({\cal M}/|\omega |})\gg 1.
\label{uneq}
\end{equation} In  certain problems, these conditions may be weakened up to
$\nu\ln{({\cal M/|\omega |} })\sim 1$, but it is not our case. Our
 observed value, effective reflection coefficient,  is directly related to the exact Green
function $G(\omega)$. For the point-like interaction, one can rewrite
(Eq.\ref{obs2})  in the form
\begin{equation}
{\rm }{\cal R}^2_{\omega}=-\frac{|\omega|(1+\nu)}
{2\pi\nu^{2}}[G(\omega)-G_{0}(\omega)].
\label{rc}
\end{equation}(Later we will denote ${\rm Re} {\cal R}_{\omega}^2$ as $\Ro.$)
In order for our calculations would be mathematical reasonable, they should be regularized in intermediate steps.
For that, we will use the Pauli-Villars regularization (one has to put $M_{P.V.}\to\infty$ at the calculation ending):
\begin{equation}
 G_{P.V.}(M_{PV},\omega)=\frac{2\pi\nu M_{P.V.}}{|\omega|(|\omega|+M_{P.V.})}.
\label{GPV}
\end{equation}
Whereby to Gell-Mann - Low approach, one has to add counter-terms in
Lagrangian  to compensate all ultraviolet divergence of all Green's
functions in each approximation. We will use a bit non-standard version
of renormalization procedure: we put $\alpha^2$ - power term, describing the
e-e interaction without impurity (Eq.\ref{W}), equals to the $S_{0}$ ("kinetic"\
energy"\ without e-i interaction). So, Z-factor renormalized of $\alpha$-fields is
1. As a result, the renormalized action  should have the form:
 \begin{eqnarray}
S_r([\alpha],\mu)=\sum_{n=1}g_{2n}(\mu
)\int\frac{d\omega_1...d\omega_{2n}}{(2\pi)^{2n}
}\frac{1}{(2n)!}\Gamma_{2n} (\omega_1...\omega_{2n})\cdot
\alpha (\omega_1)...\alpha (\omega_{2n})\cdot
2\pi\delta (\omega_1+...\omega_{2n}).\label{Gama}
\end{eqnarray}
Here $g_{2n}(\mu)$ are  the renormalized coupling constants at some (most convenient) point  $\mu$:
$
g_{2n}(\mu) = g^{(0)}_{2n}+\delta g_{2n}(\mu),
$ here $g^{(0)}_{2n}$ is bare coupling constant and $\delta
g_{2n}(\mu)$ is the sum  of counter-terms. The vertices $\Gamma_{2n}
(\omega_1...\omega_{2n})$  are completely symmetrical in rearrangement
$\omega_i\leftrightarrow\omega_{j}$. According to Eq.(\ref{det1text}) the renormalized
action in time representation is
\begin{equation}
S_{r}([\alpha],\mu)=\sum_{n=1}^{\infty}g_{2n}(\mu)\frac{(-1)^{n}}{n}
\left(  \frac
{|{\cal R}|}{|{\cal K}|}\right)  ^{2n}{\cal C}([\alpha])_{2n-1}
\label{rc11}
\end{equation}
The renormalized coupling  constants are normalized by the condition
$g_{2n}(\mu={\cal M})=1$. In the point, the action (\ref{rc11}) should
coincide with original one, Eq.(\ref{det1text}).  The action Eq.(\ref{Gama}) and Eq.(\ref{rc11})
have to coincide in both representations. It is the way to calculate all
$\Gamma_n$. Yet, we know some properties of the vertices without
calculations:
\begin{itemize}
\item All terms from Eq.(\ref{rc11}) are invariant with respect to
  replacement $\alpha(\tau)\to\alpha(\tau) + {\rm const}$. It means that
 $
    \Gamma_{2n} (\omega_1,..,\omega_{i}=0,...,\omega_{2n})=0
$ for any $\omega_{i}$. The property is a complete analogue of Goldstone theorem.
\item Self-duality imposes  strict limitations to the vertices. Indeed, all
    coefficients ${\cal C}([\alpha])_n$ in Eq.(\ref{rc11}) do not depend on
    $v_c$. Then the
    duality of the problems with repulsive and attracting fermions can exist
    only if
    $$\log {\mathfrak Det}_{imp}([\alpha ])=\log {\mathfrak Det}_{imp}({\rm
    [-sgn} \omega\cdot\alpha ])|_{\K<->\R} +
    \frac{1}{4\pi\tilde\nu}\int(d\omega)
    |\omega|\tilde\alpha(\omega)
    \tilde\alpha(-\omega)$$(the last term , - see Eq.(\ref{Wdual})). It means
    \begin{equation}
    \Gamma_{2n}{\rm sgn}(\omega_1)\cdots {\rm  sgn}(\omega_{2n})|_{\K<->\R}=
    \Gamma_{2n} (\omega_{1},\omega_2,...\omega_{2n})
    \label{GamaD}
    \end{equation} at $n>1,$ i.e. the symmetric under exchange
    $\R\leftrightarrow\K$   part  of $\Gamma_{2n}$ (we denote it as $ S_n$)
    has to be nonzero only in the frequency region where
    $\prod_i{\rm sgn}\omega_i>0$, and antisymmetric one ($ A_n$)  at
    $\prod_i{\rm sgn}\omega_i<0$. As a result, we have a vertex of the type:
    \begin{equation}
\Gamma_{2n} (\omega_1,\ldots)\!=\!\left[ S_n\theta(\prod_i{\rm sgn}\omega_i)
+A_n\theta(-\prod_i{\rm sgn}\omega_i)\right]\gamma_{2n}
(\omega_1,\ldots),
\label{GamaStruct}
\end{equation}  where $\gamma_{2n}$ is a continuous function of external
frequencies. The vertex $\Gamma_2$ is a special case. It is not invariant
under the dual transformation. Its expression is
\begin{equation}
\Gamma_2=\frac{1}{4\pi}|{\cal R}|^2\gamma_2(\omega_1,\omega_2)=\frac{|{\cal
R}|^2}{2\pi}|\omega_1|, \quad \omega_1=-\omega_2
\label{gama2}
\end{equation}

\item The coefficients $S_n$ and $A_n$ should be zero at the point $|{\cal
    R}|^2=0,$  and a vertex $\Gamma_{2n}$ has no powers higher than $|{\cal
    R}|^{2n}$. (The number of independent variables cannot increase after
    transition to other representation.)
\item     Taking into account, antisymmetric combination $|{\cal
    R}|^2-|{\cal K}|^2$ does not tend to zero at $|{\cal R}|^2\to 0$, we
    see
   $S_2\propto  |{\cal R}|^2|{\cal K}|^2,$ and $A_2=0.$

\end{itemize}
Other properties of the $\Gamma$-vertices are proved in Appendix \ref{ApGama}:
\begin{itemize}
\item the frequency dependance of the vertices is
\begin{eqnarray} \gamma
(\omega_1,..\omega_{2n})=\sum_i|\omega_i|-\sum_{i<j}
 |\omega_i+
\omega_j|+
\sum_{i<j<k}|\omega_i+\omega_j+\omega_k|-...,
\end{eqnarray}
\item for the frequency $\Omega\gg \omega_i$ (i=3,4...,2n) vertex
    $\Gamma_{2n}$ reduces to the previous one: \begin{equation}\Gamma_{2n}
    (\Omega,-\Omega,\omega_3,...,\omega_{2n})=-2x\partial_x
    \Gamma_{2n-2} (\omega_3,...,\omega_{2n}),
     \label{An}
     \end{equation} while
     $\gamma_{2n} (\Omega,-\Omega,\omega_3,...,\omega_{2n})=2\gamma_{2n-2}
     (\omega_3,...,\omega_{2n}),$\ $\gamma_2(\omega,-\omega)=2|\omega |$\
     (here $x=|\R |^2/|\K |^2$). For $n>2$ this property can be
     reformutated as a relation between $S_n$ and $A_n$ parts. They satisfy the relations:
\begin{equation}
S_n(x)=-\frac{\partial A_{n-1}(x)}{\partial \log{x }},
\quad
A_n(x)=-\frac{\partial S_{n-1}(x)}{\partial \log{x }}.\label{rec}
 \end{equation}
As a result, an antisymmetric combination cannot be constructed at $n < 3$.
Eq.(\ref{rec}) allows one to determine arbitrary $S_n,A_n$ starting from
$n=3$. The firsts vertices are
$$
S_1=\frac{1}{4\pi}\frac{x}{1+x}=\frac{1}{4\pi}|\R |^2;\quad S_2=-\frac{\partial
S_1}{\partial \log{x }}=-S_3;\quad S_3= \frac{1}{4\pi}\frac{x}{(1+x)^2}=\frac{1}{4\pi}|\R |^2|\K |^2; $$ $$A_{1(2)}=0,\quad A_3=\frac{1}{4\pi}x\partial_x \frac{x}{(1+x)^2}=\frac{1}{4\pi}|\R |^2|\K
|^2(|\K |^2-|\R |^2), \quad{\rm etc.}
 $$
\end{itemize}
In the following orders,  a number of invariants exists.
Therefore, the form of the higher
vertices cannot be determined from symmetry considerations and should be
calculated in the general way  formulated just now.
These properties of the vertices are enough for our RG-calculations. Note,
for $n>2$ there are two independent renormalized coupling constants in
the action ($g_{2n}^{s(a)}(\mu)$ - symmetric and antisymmetric).
\subsection{Calculation of the renormalized charges.}
\subsubsection{One-loop approximation.}
\label{oneloop}
The RG-approach is based on the assumption,  the original Hamiltonian of non-divergent theory  (usually unknown to us in UV-region and, probably, non-local there)  is equivalent at the large distances to our low-frequecy expansion  with a number of counter-terms. The latests are introduced to cancel the ultraviolet divergences in observed quantities.
 To this, one should calculate the divergent factors of the Green function existing in the non-renormalized problem and correct the vertices in a way to cancel the divergences. The counter-terms will depend on the normalization point $\mu$. This is an artificial parameter of the theory and observed values cannot depend on $\mu$ as well as on regularization method.

 In the subsection we will sum all terms of the order of $(\nu\log{{\cal M}/|\omega
 |)^n}$. We begin from the simplest case: one particle Green's function. In the principal order the logarithmically divergent  term of one-particle
 Green's function with $\Gamma_4$  vertex is the diagram depicted in (Fig.1A).
\begin{figure}
\begin{center}
\includegraphics[width=10 cm]{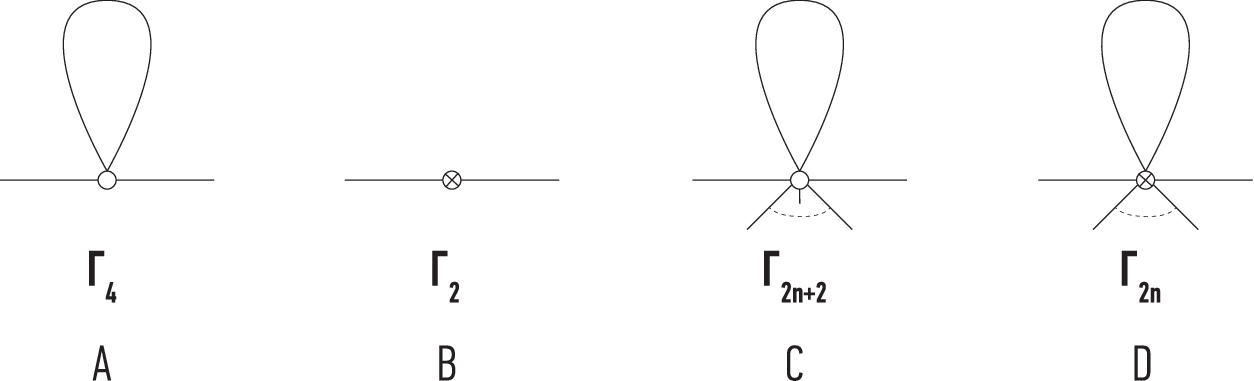}
\caption{ Renormalization of the  vertices in the  principal order: A)---the logarithmically divergent term of the simplest vertex, B)---its counter-term (the vertices with  crossed-out circles); C),D)--- the arbitrary vertices.}
\label{fig1}
\end{center}
\end{figure}
  In logarithmic approximation it equals:
 $$
 -\frac{4\times 3}{4!}(\frac{2\pi\nu}{|\omega
 |})^2g_4\int\frac{d\Omega}{2\pi}G_{PV}(M_{PV},\Omega
 )\Gamma_4(\Omega,-\Omega,\omega,-\omega)=-
 (\frac{2\pi\nu}{|\omega |})^2\nu\log{\frac{M_{PV}}{|\omega
 |}}g_4(\mu)S_2(x)\gamma_2(\omega,-\omega),
 $$ here the factor -1 arises from our definition of the "action,"\
 $4\times 3$ is the combinatorial symmetry factor \footnote{In counter-term the range of integration in $\Omega$ is limited by the condition $|\Omega|\ge|\omega|$, since at low frequencies the vertices of $\Gamma_{2n}$ tend to zero. Therefore,  the region does not contribute to the divergence.}. The divergence should be cancelled by  adding the counter-term $\delta g_2(\mu)$:  $$\delta g_2(\mu)=-
 2\nu\log{\frac{M_{PV}}{\mu}}g_4(\mu) S_1^{-1}(x)\frac{\partial S_1(x)}{\partial\log{ x}}.$$
It should be added to the coefficient near
$\Gamma_2(\omega,-\omega)$ vertex (Fig.1B). (In the figures renormalized charges are designated as vertices with crossed-out circles.)
So,  cancellation algorithm of the divergences in multi-particle Green's function is obvious. One should calculate an one-loop diagram Fig.1C with $\Gamma_{2(n+1)}$ vertex, extract the logarithmically divergent factor from it, divide it by the factor  depended on $|\R |^2/|\K |^2$ from $\Gamma_{2n}$ and multiply it by the same factor from $\Gamma_{2(n+1)}$ vertex.  The calculated in the way factor  has to be putted to the counter-term with $\Gamma_{2n}$ vertex (Fig.1D). It will cancel the diverging factor followed from the diagram with $\Gamma_{2(n+1)}$ vertex.  It is easy to make sure, the combinatorial factors at the definitions of $\Gamma_{2n}$ are chosen correctly: they reproduce
the correct combinatorial  coefficient of any diagram. For example:
$$\delta g_4(\mu)=-
 2\nu\log{\frac{M_{PV}}{\mu}}g_6^a(\mu) S_2^{-1}(x)\frac{\partial S_2(x)}{\partial\log{ x}}.$$
At the $n\ge 3$ there is
 one add-on: in the given order $n$ there are two
 constants $g^{(s)}_{2n}$ and $g^{(a)}_{2n}$ in front of symmetric and
 antisymmetric structures, which should be renormalized independently. So, to
 compensate divergencies in all Green's functions, one has to add to the action
the counter-terms
\begin{equation}
\delta g_{2n}^{(s)}(\mu )=-2\nu g_{2n+2}^{(a)}(\mu)
\log\left(\frac{M_{PV}}{\mu}\right)\frac{1}{S_{n}(x)}
\frac{\partial S_n(x)}{\partial \log{x }},
\label{Z2}
 \end{equation}
(and analogously for $\delta g_{2n}^{(a)}(\mu )$ with substitution $S_n\to A_n$ and indexes $a\leftrightarrows s $). To get the Gell-Mann - Low equation, one should take into account:
\begin{itemize}
\item the bare coupling constant in Eq.(\ref{Z2}) does not depend on $\mu$. It means \begin{equation}
\beta_{2n}(\mu)=\frac{\partial g_{2n}(\mu)}{\partial\log \mu}=\frac{\partial \delta g_{2n}(\mu)}{\partial\log \mu}\label{GLcom} \end{equation} It is the Gell-Mann - Low (GL) equation;
\item the renormalized coupling constants are depended on $\mu$. Yet,  differentiation in $\mu$ the coupling constant in equation of $\beta_{2n}(\mu)$-function is an over accuracy in the one-loop approximation (see, Eq.\ref{GL1}). (But it is not the case in  higher approximations. Here this dependency vanishes all terms about $(\log \mu/M_{PV})^n;n>1$ in $\beta$-function.)
\end{itemize}Accordingly,
in one-loop approximation we obtain the Gell-Mann - Low equation ($n\ge 3$):
\begin{equation}
\frac{\partial g_{2n}^s(\mu ) }{\partial \log{\mu }} =2\nu
g_{(2n+2)}^a(\mu) \frac{1}{S_{n}}
\frac{\partial S_n}{\partial \log{x }}.
\label{GL1}\end{equation}
It is useful to introduce the function ${\psi}(\mu,x)=g_2(\mu)S_1(x);\quad \psi({\mu=\cal M},x)=S_1(x)$. In term of $\psi-$function the GL-equation (\ref{GLcom}) for the charge $g_2(\mu)$ is $$
\frac{1}{2\nu}\frac{\partial \psi(\mu)}{\partial\log \mu}=g_4(\mu) \frac{\partial S_1(x)}{\partial\log{ x}}.
$$Taking into account  relation $S_2=-\partial S_1/\partial\log x$, one can rewrite the GL-equation for the charge $g_4(\mu)$ in the same form: $$
\left(\frac{\partial}{2\nu\partial\log \mu}
\right)^2\psi=g_6^a(\mu)\frac{\partial^2 }{(\partial\log{ x)^2}}S_1(x),$$ and the first equation of the Eqs.(\ref{GL1}) is $$
\left(\frac{\partial}{2\nu\partial\log \mu}
\right)^3\psi=g_6^a(\mu)\frac{\partial^3 }{(\partial\log{ x)^3}}S_1(x),
$$  etc. (Here we have used relation Eq.\ref{rec}.) These relations fix the functional dependency \\ $\psi(\mu,x)=\psi\left(x(\mu/{\cal M})^{2\nu}\right)$, and boundary condition tells as $$ \psi\left(x(\mu/{\cal M})^{2\nu}\right)= S_1(x(\mu/{\cal M})^{2\nu}).$$ In addition, we have known $S_1=|\R |^2/4\pi.$ It means
\begin{equation}
g_2^{(1)}(\mu/{\cal M})=\frac{(\mu/{\cal M})^{2\nu}}{|\K |^2 + |\R |^2(\mu/{\cal M})^{2\nu}};\quad g_4^{(1)}(\mu/{\cal M})=\frac{(\mu/{\cal M})^{2\nu}}{\left(|\K |^2 + |\R|^2(\mu/{\cal M})^{2\nu}\right)^2}\quad {\rm etc.}
\label{g2}
\end{equation} in one-loop approximation. The other charges can be obtained from it.
 \subsubsection{Two-loop approximation.}
 \label{twoloop}
 In the section, we will sum diagrams up to the order of
$\nu^{n+1}\log^n{(M/|\omega |)}$.
 \begin{figure}
\begin{center}
\includegraphics[width=10 cm]{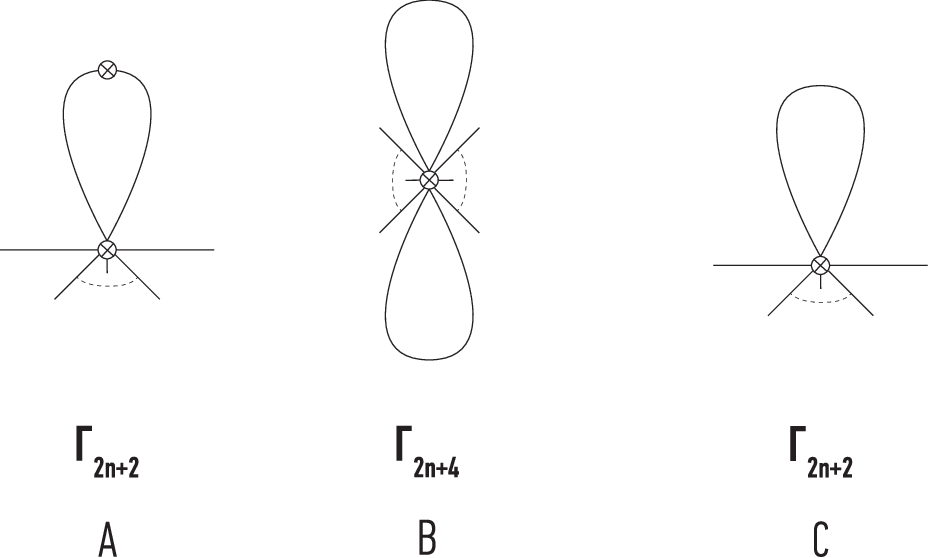}
\caption{A scheme of the cancellation the divergent contributions that would make the theory non-renormalizable in the two-loop approximation.}
\label{fig1}
\end{center}
\end{figure}
In this approximation, we should consider the
diagrams similar to Fig.(2A) and Fig.(2B).  The simplest first-type diagram is the next correction to one-loop
multiparticle Green's function. It has the divergent factor
$$
\frac{1}{2}g_2(\mu)g_4(\mu)\int\frac{d\Omega}{2\pi}G^2_{PV}
(M_{PV},\Omega
)\Gamma_4(\Omega,-\Omega,\omega,-\omega)\Gamma_2
(\Omega,-\Omega).
 $$ This expression differs from one-loop approximation by the factor $-\nu |\R
 |^2g_2(\mu)=-4\pi g_2(\mu)S_1$, because $\Gamma_2(\Omega,-\Omega)\propto |\Omega|$. Therefore, we have the same factor in
 counter-term $\delta g_2^{(2)}(\mu)$.

 The main difficulty, characteristic of the RG approach, is manifested by the diagram depicted in (Fig. 2B). Let us consider this diagram. One has a divergent factor in each  Green's function equals
 \begin{equation}
 -\frac{1}{8}\int_{-\infty}^{\infty}\frac{d\Omega_1d\Omega_2}{(2\pi
 )^2}G_{PV}(\Omega_1)G_{PV}(\Omega_2)
 g^i_{2n+4}(\mu)
 \Gamma^i_{2n+4}(\Omega_1,-\Omega_1,\Omega_2,-\Omega_2,
 \omega_1,\omega_2,...),\label{renormG}
 \end{equation}
 here sign is defined by expansion of $e^{-S_r}$, $1/8$ is the combinatorial factor ($1/(2n+4)! \times (2n+4)(2n + 3)...5$ - it is the ways to distribute 2n vertices and index $i$ is $a$ or $s$.
A direct  attempt  to compensate the full divergence of the diagram (Fig.2B) by subtracting from each loop the divergent part fails. The divergent term of each loop is proportional to  $\log{M_{PV}}$.  Immediate way to compensate the divergent term would  resulted  to the expression $(\log{M_{PV}/|\omega |})^2=
(\log{(M_{PV}/\mu}+\log{\mu/|\omega |})^2$. To cancel  the term \\$\log{(M_{PV}/\mu)}\log{(\mu/|\omega|)}$ one needs to introduce the counter-term depended on
$\omega$ (external Green's function frequency) into a Hamiltonian.
It is illegal for any problem. The  difference between the renormalizable  theories and others one is in  cancellation of the  such type terms. In renormalized  problems, the cancellations are realized  due to  non-trivial frequency properties of the renormalized vertices.

\centerline{{\bf Renormalizability of the problem in two-loop approximation.}}
\label{renorm2}

To prove the renormalizability of the problem,
we have developed a procedure similar to the decomposition of the divergent diagrams by cumulants. We will do this in the two-loop approximation. In the following approximations, the proof can be developed by induction.

The graph of the Green's function, excluding from the equation (\ref{renormG}) the terms linear in $\log{(M_{PV})}$, should be taken from the diagram Fig. 2C. Its "loop factor" is proportional to
  $$
 \frac{1}{2}\int_{-\infty}^{\infty}\frac{d\Omega}{2\pi}G_{PV}(M_{PV},\Omega
 )\Gamma^i_{2n+2}(\Omega,-\Omega,\omega_1,\cdots,
 \omega_{2n})
 $$ with counter-term  $\propto\nu\log{(M_{PV})/\mu}. $ (In fact, we have calculated the diagram in previous Section). Taking into
 account
 $$
\nu\log{\frac{M_{PV}}{\mu}}=\frac{1}{2}
 \int_{-\infty}^{\infty}\frac{d\Omega_1}{2\pi}\left[ G_{PV}(M_{PV},\Omega_1)
 -G_{PV}(\mu,\Omega_1) \right],$$
  one can rewrite the divergent at $M_{PV}\to \infty$ part of diagram (Fig.2C), originates from counter-term, in the form
$$
\int\limits_{-\infty}^{\infty}\frac
 {d\Omega_1d\Omega_2}{4(2\pi )^2}\left[
 G_{PV}(M_{PV},\Omega_1) -G_{PV}(\mu,\Omega_1) \right]
 G_{PV}(M_{PV},\Omega_2)\times$$ $$\times\lim_{|\Omega_1|\to\infty}
\Gamma^i_{2n+4}
 (\Omega_1,-\Omega_1,\Omega_2,-\Omega_2,\omega_1\cdots  \omega_{2n})
 g^i_{2n+4}(\mu)
$$
 Here we  take into account, the  $\lim_{|\Omega_1|\to\infty}\Gamma^i_{2n+4}
 (\Omega_1, -\Omega_1, \cdots,  \omega_{2n})$ does not depend on $\Omega_1$.
This part of diagram (2C) has to cancel the lineal in $\log M_{PV}$ summand from diagram (2B). Otherwise, the problem would be non-renormalizable.

 Now we are ready to calculate the sum of diagram  (Fig.2B) and counter-term.
 Let us rewrite the free Green's functions $G_{PV}(M_{PV},\Omega)$ at Eq.(\ref{renormG}) and counter-term
 in the form: $[G_{PV}(M_{PV},\Omega)
 -G_{PV}(\mu,\Omega)]+G_{PV}(\mu,\Omega)$.
 At present, we are interested  in the summands divergent in the limit $\log M_{PV}\to\infty$. The term $G_{PV}(\mu,\Omega_1)G_{PV}(\mu,\Omega_2)$ does not depend on $M_{PV}$ and hence should not require any counter-term. This sum has an interference term, which can be written as \begin{equation}-1/8\int_{-\infty}^{\infty}\frac{d\Omega_1d\Omega_2}{(2\pi )^2}2[G_{PV}(M_{PV},\Omega_1) -G_{PV}(\mu,\Omega_1)]
G_{PV}(\mu,\Omega_2)g^i_{2n+4}(\mu) \times\label{rc1}
\end{equation} $$\times\left[\Gamma^i_{2n+4}
 (\Omega_1,-\Omega_1,\Omega_2,-\Omega_2,\omega_1,..)-\lim_{|\Omega_1|\to\infty}\Gamma^i_{2n+4}
 (\Omega_1,-\Omega_1,\Omega_2,-\Omega_2,\omega_1,..)
\right].$$ In this expression  integral in $\Omega_2$  is convergent because at $\Omega _2\gg\mu$ the Green function $G_{PV}(\mu,\Omega_2)$ decreases as $1/\Omega _2^2$.
 Another integral in $\Omega_1$ is convergent too due to the difference of $\Gamma$-vertices in the square brackets. So, these terms should not require any counter-term as well.

 Thus, it is necessary to make regularization the terms with the factor\\
$[G_{PV}(M_{PV},\Omega_1) -G_{PV}(\mu,\Omega_1) ][G_{PV}(M_{PV},\Omega_2)
-G_{PV}(\mu,\Omega_2) ]$. The region $|\Omega_1|, |\Omega_2 |\gg\mu$ is
essential for the contribution. The term has the form
\begin{equation}-1/8\int_{-\infty}^{\infty}\frac{d\Omega_1d\Omega_2}{(2\pi )^2}[G_{PV}(M_{PV},\Omega_1) -G_{PV}(\mu,\Omega_1)]
[G_{PV}(M_{PV},\Omega_2) -G_{PV}(\mu,\Omega_2)]\times\label{rc1}
\end{equation} $$\times g^i_{2n+4}(\mu) \left[\Gamma^i_{2n+4}
 (\Omega_1,-\Omega_1,\Omega_2,-\Omega_2,\omega_1,)-
 2\lim_{|\Omega_1|\to\infty}\Gamma^i_{2n+4}
 (\Omega_1,-\Omega_1,\Omega_2,-\Omega_2,\omega_1,)
\right]$$

Let us add to the last bracket the term  $\lim_{|\Omega_1|,|\Omega_2|\to\infty}
\Gamma^i_{2n+4}(
 \Omega_1,-\Omega_1,\Omega_2,-\Omega_2,\omega_1,\cdots  )$ (the added term will be considered later) and consider expression is proportional to:
\begin{equation} \Gamma^i_{2n+4}
 (\Omega_1,-\Omega_1,\Omega_2,-\Omega_2,\omega_1,.)-
 (\lim_{|\Omega_1|\to\infty} + \lim_{|\Omega_2|\to\infty})
\Gamma^i_{2n+4}
 (\Omega_1,-\Omega_1,\Omega_2,-\Omega_2,\omega_1,.)+
 \label{ren2C}\end{equation}
 $$+\lim_{|\Omega_1|,|\Omega_2|\to\infty}
\Gamma^i_{2n+4}
 (\Omega_1,-\Omega_1,\Omega_2,-\Omega_2,\omega_1,\cdots  )
 $$One can see,
 \begin{itemize}
 \item if $\Omega_{1,2}\to\infty$ -  the whole sum with the factor (\ref{ren2C}) vanish;
 \item $\Omega_{1}\to\infty ;\,\, \Omega_{2}$ is finite - first and second
     terms are cancelled as well as the third and fourth terms;
 \item $\Omega_{2}\to\infty ;\,\, \Omega_{1}$ is finite - cancelation the
     first and the third terms, as well as the second and fourth ones.
 \end{itemize}
The added term is nothing else as the divergent multiplier of the diagram Fig.(2C). The diagram is proportional to the factor $\log (M_{PV}/|\omega|)$ (from the loop) and $\log (M_{PV}/\mu)$ from the vertex. (The last multiplier arises from the  first-order counter-term.) As a result, the diagram Fig.(2C) (vanishing the divergent part of multiplier of diagram Fig.(2B)) is proportional  to \\$\log (M_{PV}/|\omega|)\log (M_{PV}/\mu)$. This cancellation makes the problem renormalizable.

 So, the divergent part of  diagrams  Fig. (2b) and Fig. (2c) is proportional to the vertex $\lim_{|\Omega_i|\to\infty}
\Gamma^i_{2n+4}(
 \Omega_1,-\Omega_1,\Omega_2,-\Omega_2,\omega_1,..)$, and whole divergent factor equals:
\begin{equation}
\frac{1}{2}\nu^2\log^2{\frac{M_{PV}}{\mu}}\sum_{i} g^i_{2n+4}
(\mu)\lim\Gamma^i_{2n+4}(
 \Omega_1,-\Omega_1,\Omega_2,-\Omega_2,\omega_1,\cdots \omega_{2n} )|_{|\Omega_{1(2)}|\gg \mu , |\omega_j|},
\label{deltaG2countr}
\end{equation}where i=a or s. In this expression the vertices $\Gamma^i_{2n+4}(
 \Omega_1,-\Omega_1,\Omega_2,-\Omega_2,\omega_1,..\omega_{2n})$ is proportional to
$\Gamma^i_{2n}(\omega_1,\cdots \omega_{2n})$.

\centerline{{\bf Renormalized charges in two-loop approximation.}}

It is upshot, the divergence of the diagrams Fig.(2B) and Fig.(2C)   is cancelled by the counter-terms  with the vertex $\Gamma_{2n}$. Taking into account relation
$$
\lim_{|\Omega_{1,2}|\to \infty}\Gamma^s_{2n+4}(
 \Omega_1,-\Omega_1,\Omega_2,-\Omega_2,\omega_1,\cdots )=4\frac{\partial^2 S_n}{(\partial \log{x })^2}\gamma(\omega_1,..\omega_{2n})\theta({\rm sign}(
 \Pi_i\omega_i) )
 $$(and similar for $\Gamma^a_{2n+4}$), we see that Eq.(\ref{deltaG2countr}) is following from Lagrangian with counter-term
\begin{equation}
\delta g_{2n}^{s}(\mu)=-\frac{1}{2}(2\nu)^2 g_{2n+4}^s(\mu)
\log^2{\left(\frac{M_{PV}}{\mu}\right)}\frac{1}{S_{n}}
\frac{\partial^2 S_n}{(\partial \log{x })^2}.
\label{deltaG2countr1}
\end{equation} (Eq. for the antisymmetric counter-term is given by replacement
$S_{n}\to A_{n}$).)
As it should be, the counter-term is local (as it does not depend on incoming frequency). Besides, the form of  the vertex reproduces. So, we deal with renormalizable
theory.

 To get  Gell-Mann — Low equation, one should take into consideration  dependence  of  the charges on $\mu$  in one-loop approximation. Owing to this, all summands in $\beta$-function from Figs.(2B) end (2C),  proportional to $\log(M_{PV}/\mu) $, vanish in two-loop approximation, and $\beta$-function is determined only
 by the diagram, Fig.(2A). As a result, we get
$$
\beta(\mu,x) =
2\nu g_{2n+2}^{(a)}(\mu)\left(1-\nu g_2(\mu) |\R |^2\right)\frac{1}{S_n(x)}
\frac{\partial  S_n(x)}{\partial \log{x }}.
$$ We will replace $\left(1-\nu g_2(\mu) |\R
|^2\right) \to 1/\left(1+\nu g_2(\mu) |\R |^2\right)$. It corresponds to
summing  all diagrams with $g_2$-vertices in the loop. It is useful for
subsequent calculation. (Of course, we will expand the final expressions up to
appropriate order). So, GL-equations in two-loop approximation are$$
\frac{\partial g_{2}^{}(\mu,x ) }{\partial \log{\mu}}
=
2\nu g_{4}^{}(\mu,x )\frac{1}{\left(1+4\pi\nu g_2(\mu) S_1(x)\right)}\frac{1}{S_1(x)}
\frac{\partial S_1(x)}{\partial \log{x }}
$$
\begin{equation}
\frac{\partial g_{2n}^{(s)}(\mu,x ) }{\partial \log{\mu}} =
2\nu g_{2n+2}^{(a)}(\mu,x )\frac{1}{\left(1+4\pi\nu g_2(\mu) S_1(x)\right)}\frac{1}{S_n(x)}
\frac{\partial S_n(x)}{\partial \log{x }},\quad n>1.
\label{GL2} \end{equation}To get Eqs. for $g_{2n}^{(a)}$ one should replace $s,S_n \to a,A_n$

The system is divided into two systems for the sets
\begin{eqnarray}
h_{2n} &=&4\pi\left\{S_1g_2,\frac{\partial S_1}{\partial\log
x}g_4,\frac{\partial^2 S_1}{(\partial\log x)^2}g^{(a)}_6,\frac{\partial^3
S_1}{(\partial\log x)^3}g^{(s)}_8\ldots\right\}\nonumber\\
f_{2n} &=&4\pi\left\{\frac{\partial S_1}{\partial\log
x}g^{(s)}_6,\frac{\partial^2 S_1}{(\partial\log
x)^2}g^{(a)}_8,\frac{\partial^3 S_1}{(\partial\log
x)^3}g^{(s)}_{10}\ldots\right\},
\label{notation}
\end{eqnarray}here we have used the identity$$
\left(\frac{1}{A_{n+1}(x)}
\frac{\partial  A_{n+1}(x)}{\partial \log{x }}\right)
\left(\frac{1}{S_{n}(x)}
\frac{\partial  S_n(x)}{\partial \log{x }}\right)=\frac{1}{S_n(x)}\frac{\partial^2  S_n(x)}{(\partial \log{x })^2}
$$
The final GL-equations for these charges can be rewritten in the form
\begin{equation}
(1+\nu h_2(\mu,x))\frac{\partial h_{(2n)}(\mu , x ) }{\partial \log{\mu^{2\nu} }} =
h_{(2n+2)}(\mu , x )
\label{GL2final}\end{equation} and similar one for the $f_{2n};\, n\ge 3$.
It is useful to represent it in the form
 $$
h_4(\mu,x) \frac{ \partial h_{2n}(\mu,x)}{\partial
h_2(\mu,x)}=h_{2n+2}(\mu,x).
$$ and move to the new variable
$z=\log (h_2(\mu,x)/1-h_2(\mu,x));\quad
z |_{\mu={\cal M} }=\log x$. It ia easy to see
$$
\frac{\partial}{\partial h_2}= \frac{1}{h_2(1-h_2)}\frac{\partial}{\partial
z}\qquad\mbox{and}\qquad h_2(1-h_2)=4\pi\frac{\partial S_1(e^z)}{\partial z}.
$$ As a result, the system can be rewrite in the form:
\begin{equation} [4\pi \frac{\partial S_1 (e^z)  }{\partial z}]^{-1}h_4(\mu ,
x )\frac{\partial h_{2n}(\mu,x) }{\partial z} = h_{2n+2}(\mu,x),
\label{sysH}\end{equation} while the boundary condition at the point $\mu={\cal M}$
(following from Eq.(\ref{sysH}) and relation $h_4(\mu/{\cal M}=1,x)=4\pi\partial
S_1/\partial\log{x}$) are:
$$  \frac {\partial h_4(\mu,x)}{\partial
z}|_{\mu={\cal M}}=h_6(1,x)=4\pi\left(\frac{\partial
S_1(x)}{\partial\log{x}}\right)^2;\quad\frac {\partial h_6(\mu,x)}{\partial
z}|_{\mu={\cal M}}=h_8(1,x)=4\pi\left(\frac{\partial
S_1(x)}{\partial\log{x}}\right)^3, {\rm etc.}
$$
To determine these charges, it is sufficient to know only one charge: $g_4(\mu,x)$.
By representing the Gell-Mann–Low equations in the form Eq.(\ref{sysH}), it is easy to guess their solutions. Indeed, if one takes\\
$
h_4(\mu,x)=4\pi \partial_ z S_1 (e^z),$ then this function  will  satisfy the boundary condition, while the other functions
$$
h_{2n}(\mu,x) = 4\pi\frac{\partial^{n-1}S_1(e^z)}{(\partial
z)^{n-1}}=f_{2n+2}(\mu,x), \quad
n>2$$ will satisfy the GL-equation and automatically the boundary conditions because $z |_{\mu={\cal M} }=\log x$.

One thing remains:
determine $z$, or in other words, solve the GL equation for $h_2$.
Substituting the expression for $h_4$ in Eq.(\ref{GL2}; $n=1$), we have
 \begin{equation}
\frac{\partial h_{(2)}(\mu , x ) }{\partial \log{\mu^{2\nu} }} = \frac{h_2(\mu
, x )(1-h_2(\mu , x ))}{(1+\nu h_2(\mu , x ))};
h_2|_{\mu={\cal M}}=\frac{x}{1+x}\quad{\rm or}\quad
\frac{h_2(\mu , x )}{(1- h_2(\mu , x
))^{1+\nu}}=x(1+x)^{\nu}\left(\frac{\mu}{{\cal M}}\right)^{2\nu}
\label{H2eq}
\end{equation}
This algebraic equation can be solved iteratively.
We will assume:
$\nu\ll 1$, so that $\nu|\log(|\K|^2+(\mu/{\cal M})^2|\R |^{2\nu})|\ll1.$ Let us introduce  new function $xY(\mu , x )=h_2/1-h_2$ or $h_2 =
xY/1+xY$. After that, one can rewrite the algebraical equation in the form
\begin{equation}
Y(\mu , x )=\left[\frac{(\mu/{\cal M})^2}{|\K|^2 + |\R|^2Y(\mu , x )}\right]^\nu
\label{Y}
\end{equation}
The direct iteration gives
$$
Y=(\mu/{\cal M})^{2\nu}[1- \nu\log{(|\K|^2 + |\R|^2(\mu/{\cal M})^{2\nu})}+...],
$$ and renormalized charges in the second-loop approximation are
\begin{equation}
g^{(2)}_2(\mu)=\frac{(\mu/{\cal M})^{2\nu}}{|\K|^2 +
|\R|^2(\mu/{\cal M})^{2\nu}}-\frac{\nu(\mu/{\cal M})^{2\nu}|\K|^2\log{[|\K|^2 +
|\R|^2(\mu/{\cal M})^{2\nu}]}}{[|\K|^2 + |\R|^2(\mu/{\cal M})^{2\nu}]^2}
\label{g2two}
\end{equation}
\begin{equation}
g^{(2)}_4(\mu)=\frac{(\mu/{\cal M})^{2\nu}}{(|\K|^2 +
|\R|^2(\mu/{\cal M})^{2\nu})^2}\bigg[1-\nu\frac{|\K|^2-(\mu/{\cal M})^{2\nu}|\R|^2}{|\K|^2 + |\R|^2(\mu/{\cal M})^{2\nu}}
\log{(|\K|^2 +
|\R|^2(\mu/{\cal M})^{2\nu})}\bigg],\quad {\rm etc.}
\label{g4two}
\end{equation}
\subsection{Calculation of the reflection coefficient.}
\label{Robs}
To calculate the effective reflection coefficient, one should use
Eq.(\ref{rc}). The lowest diagrams of the Green function are presented  on
Fig.(3 A;B;D). (We will  calculate  it up to the terms of the order of $\nu^2$, i.e. one should calculate the diagram with $\Gamma_4$-vertex  (Fig.3D) only in one-loop approximation.) The renormalized Lagrangian depends on renormalized coupled constants and contains all counter-terms. (The vertices with renormalized coupling constancs are depicted  in Fig.3 as $\Gamma_{2n}(\mu)$.)
In the previous section,  we have vanished all divergences.
 \begin{figure}
\begin{center}
\includegraphics[width=10 cm]{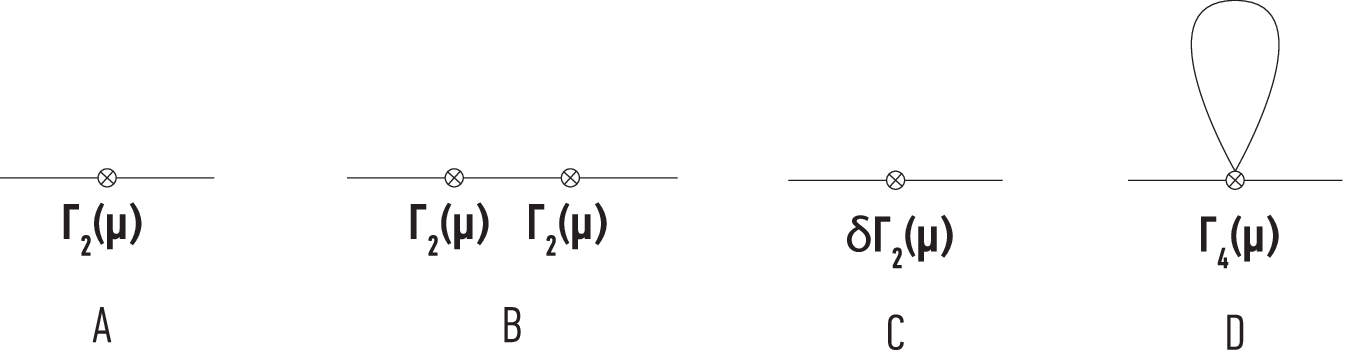}
\caption{  The lowest diagrams for the Green function. }
\label{fig1}
\end{center}
\end{figure}
In particular, the divergent part of diagram (3D) cancels by  counter-term (3C). It means, in diagram (3D) one has to consider  not only the big frequency region  (as for  the Gell-Mann - Low function), but a small one ($\cong \omega$). After that, one can put $M_{PV}\to\infty$.
The calculation gives
\begin{equation}
G_{\mu}(\omega)=\frac{2\pi\nu}{|\omega |}\left[1 - \nu g_2(\mu)|R|^2+\nu^2
g_2^2(\mu)|R|^4
+2\nu^2g_4(\mu)|R|^2|K|^2(\log{\frac{\mu}{|\omega|}} + 1) \right]+...
\label{Gz}\end{equation}
Yet, in our problem the Green function can be considered
as an observed quantity, i.e. it cannot depend on regularization point. It means, all terms depending on $\mu$  should vanish.
It is possible because our expansion is the expansion of the  $G_{\mu}(\omega)$ in $\nu g_{2n}\sim\nu\log{\mu/|\omega|}\ll 1$. As the $G_{\mu}(\omega)$ does not depend on $\mu$ - the real expansion of the $G_{\mu}(\omega)$ is the expansion in $\nu\ll 1$ and the big  logarithms should be cancelled by the next terms expansion of the $G_{\mu}(\omega)$ and $g_n(\mu)$. To see it directly, the series (\ref{Gz}) has to be rewritten as the series  of $\nu$ powers. It is a time-consuming calculation. Instead of this one can use relation  $d\Ro/d\mu=0$ is applied to the exact reflection coefficient (similar to derivation  Callan-Symanzik equation \cite{GML},\cite{CZ},\cite{ZJ}).
However, our problem is much  easier because it is sufficient to take the renormalization point $\mu=\omega$.  In this case, all logarithmic
terms  do not exist from the beginning, and expression
(\ref{Gz}) becomes regular expansion in $\nu\ll1$-powers.  As a result, all infrared
logs have been summed up by system of the Gell-Mann-Low equations and enter to renormalized charges do not depending on $\mu$ now. In two-loop approximation, we have
\begin{equation}
\Ro = |\R |^2[(1+\nu)g_2(\omega)-\nu |\R |^2 g_2^2(\omega)-2\nu g^{}_4(\omega)|\K
|^2].
\label{Ro2}
 \end{equation}

\subsection{Violation of the "poor man's" approach. }
\label{AndSec}
It is believed, "poor man's" approach \cite{And} is a simplified version of the Gell-Mann - Low one and it is valid in any "loop approximation".
In the case of our problem, one assumes \cite{FK},\cite{Fur},\cite{RG},\cite{aristov},\cite{RgM}:
\begin{itemize}
\item Lagrangian  depends on the only charge;
\item   this charge coincides with exact reflection coefficient.  This
    assumption is based on intuitive conviction, according to it the $\Ro $ (or conductance)
    is the only physical quantity that can define the low — frequency
    properties of the system. As a result of this, the observed conductance has to relate directly to renormalized   charge. Therefore, the conductance has to obey the Gell-Mann — Low equation itself.
\end{itemize}
 However, these assumptions cannot be correct in all “loop-approximation.”\ The point is,  beginning from some order in $\nu$  the GL-equation is depended on regularization scheme  always (i.e., on a calculation way).  The latter is an ancillary procedure, permitting one to give the mathematical meaning of the  divergent quantities. Therefore,  GL-equation cannot define the observed value in all orders in $\nu$.   Let us discuss the "poor man's" assumptions for the the Gell-Mann - Low approach in more detail. According to it one assumes, exact reflection coefficient coincides with the coupling charge $h_2$, and $\Ro$ is determined by the GL-equation Eq.(\ref{H2eq}). It is  true in the leading logarithm approximation (compare with \cite{RG},\cite{aristov}):
 \begin{equation}
 \frac{\partial \Ro }{\partial \log{\omega }} = -2\nu\Ro(1-\Ro),
\label{Rom}\end{equation}
  because $h_2(\mu=\omega) =\Ro=|\R |^2g_2(\omega)$ in accordance with Eq.(\ref{Ro2}). Yet, in the next approximation, our renormalized charges does not coincide with $\Ro$. Therefore, in our problem the observed quantity cannot be defined by the GL-equation already in two-loop approximation, Eq.(\ref{Gz}). From this approximation the renormalized charge  depends on the renormalization scheme. Indeed, transition from one scheme to another equals to the modification
$M_{PV}\to M_{PV}e^c$; $c$-is the number. This transition cannot change the leading logarithm approximation, but in the next loop it changes the counter term in Eq.(\ref{deltaG2countr1}). The extra counter term ($\propto c\log (M_{PV}/\mu)$) enters into GL-equation (Eq.\ref{GL2final}), that changes the vertices of the diagrams (3C) and (3D). The additional divergent part of the diagram (3D) vanishes by counter term from the vertex $\delta\Gamma_2$ (Fig.3C), and the finite contribution from small $\Omega\sim\omega$ will change the coefficient in front of charge $g_4$ in Eq.(\ref{Gz}). So, dependence of the GL-equation on the regularization scheme makes it impossible to use the GL-equation to determine the observed value in this order. As for the reflection coefficient, Eq.(\ref{Ro2}) does not change due to $h_2(\mu,x)$ (i.e., diagram (3A)) changing. This changing equals to $\delta h_2(\mu,x)=\nu ch_4^{}(\mu,x)$ entered into the expression for $\delta \Gamma_2(\mu)$ (Fig.3C). The modification cancels the extra factor before the charge $g^{}_4 (\omega)$ in Eq.(\ref{Ro2}). As a result, {\em $\Ro$ becomes independent on the regularization scheme, as it should be.}  Dependence of the renormalized charges on the scheme already in two-loop approximation is a bit unusually.  Typically,  such a dependence takes place in the three-loop
approximation. The reason is: usually, the logarithmically divergent factor is the loop with two vertices. This loop is proportional to the squared bare coupling constant. In our problem, the log-divergent loop has only one vertex. Therefore, in our case the scheme-dependency appears one step earlier: in the term proportional to $\nu g_4,$ Eq.(\ref{Ro2}).
Thus, dependence of  the $\beta -$function on the cut-off scheme, i.e. its “non-universality”  pointed in \cite{aristov}, is a common occurrence. The unobserved charges entering to the $\beta $-function can be scheme-dependence. Another matter, it does not mean a non-universality of conductance.

In the later article of the same authors  \cite{aristov2} the different version of the RG approach (Callan-Symanzik  approach, CS),  was used. The aim of the paper was to apply the “poor's man” assumption  to GS scheme. The GS version of RG investigates variation  any Green's function after a change of the ultraviolet regularization scale $\Lambda=L/a$, while conception of the regularization point does not exist here. Dependence of the  Green function on regularization scale appears from different places: directly  from divergence of the loop diagrams; implicitly  from  variation of the renormalized charges owing to a change of the regularization scale. It is an interaction effect.  If one derives a CS equation for an observed value, you can directly exploit  its  independence on $\Lambda$. For our problem,  CS equation for conductance has to have a form \cite{Ry}:
 \begin{equation}
 \left(\frac{\partial}{\partial \log\Lambda}+\sum_i\beta_{i} (\{g_i(\Lambda)\})\frac{\partial}{\partial g_{i}}
 \right){\cal C}(\omega)=0
 \label{CSQ}
  \end{equation} The $\beta-$function depends on the whole set of the charges $\{g_i(\Lambda)\}$ and never on the $\Lambda$ directly.
  Nature of the changing the $\beta$ function due to the  regularization scale is profound: variation of regularization scale changes a charge that “an observer sees from this scale.”\ The charge changes owing to vacuum  (ground state) polarization, and visible part of the polarization cloud is different  for a different regularization scale.
 The first term of equation (\ref{CSQ}) must be calculated directly from the original Hamiltonian, using any subtraction scheme to extract the log-divergent factor. Next, the $\beta$ function must be calculated too.
 The final step of calculation is solution Eq.(\ref{CSQ}). It is a time-consuming procedure. To simplify the problem, in the paper \cite{aristov2} the "poor man's" assumptions were adapted to the CS-scheme. To that, one calculates the function $Y$ directly connected with conductance: $ Y_{ren} =2{\cal C}-1$. According to "poor man's" assumptions it holds to Eq.:
  \begin{equation}
\frac{\partial Y_{ren}(g,Y_{b}, \Lambda)}{\partial \log\Lambda}=\beta (g,Y_{ren}(\Lambda))
\label{ArGL}
\end{equation}
(here $Y_{b}$ is determinate by the bare conductance, and $g$ is the bare e-e interaction charge).

We guess, the consequences of this step are clear.  Indeed,    Eq.(\ref{ArGL}) determinates the observed  quantity explicitly. Therefore, its  solution should be  independent of a calculation way. However, the authors of the article emphasize,  changing the normalization scale $\Lambda\to\Lambda e^c$ changes the $Y_{ren}-$function  and this is true.
 According to them, the problem is in the use different subtraction schemes.

 It is true, the different subtraction schemes define the $\Lambda$-independent parts of the log-divergent terms differently, but it does not mean that this leads to different results for the observed quantities. If the calculations are done correctly, the observed quantities calculated with different subtraction schemes should be the same.
The “correctly” for the conventional procedure means. i) One should determine relation between observed quantity and exact Green function exactly (not in logarithm approximation). The correct relation fixes the scale-independent factor in an observed value. (It is $1+\nu$ in our Eq.(\ref{rc}).)
ii) To calculate the observed value correctly, not only a region of large energy (about $\Lambda$) should be considered,  but a region of small energy (about $\omega$) should be taken into account.
We have pointed out the request at the beginning of the Section (\ref{Robs}).
 After this, $Y_{ren}$ for different subtraction schemes should be the same, {\em if the renormalized  charge can be identified with observed value}.

 To sidestep this question, in the paper \cite{aristov2} was used an unconventional way for the GS approach. As usual for CS-scheme, the function $Y_{ren}(Y_b)$ was inverted and  GS equation had been written down for the bare function
 $Y_b(Y_{ren})$ (Eq.(40) of the paper \cite{aristov2}):
 \begin{equation}
 0=\frac{d Y_b}{d\log \Lambda}=\frac{\partial Y_b}{\partial\log \Lambda}+\beta (g,Y_0)\frac{\partial Y_b}{\partial Y_{ren}}
 \label{ARd}
 \end{equation} with additional condition: Eq.(\ref{ArGL}), should be taken in the point  $\log\Lambda=0 .$   This condition determinate “the true $\beta(g,Y_0)$-function” in accordance to the terminology of the paper \cite{aristov2}. (Here $Y_0$ is $Y_b$ {\em plus the sum of all scale-independent contributions} in $Y_{ren}$.) At this step, a highlighted subtraction scheme is recorded.  This condition is equivalent to calculation of the $\beta$-function from renormalized Hamiltonian with some fixed subtraction scheme.    Indeed, under the changing subtraction scheme, the regular (i.e., does not depend on $\Lambda$) part  of renormalized charge will change too. Therefore, calculation of the $\beta$-function by computing the iterative sum of observed $Y(\Lambda)$ together with this condition, as it was done in \cite{aristov2}, is equivalent to fixation  one of the subtraction schemes.

 I guess, this  path was chosen to justify  existence  of the “single correct” subtraction scheme. The problem is:  existence of the highlighted subtraction scheme breaks the basic idea of renormalization group approach. It supposes, the observed quantity cannot depend on the renormalization scale or points, regularization or subtraction schemes, etc (in other words, from a calculation method in any RG-approach). In principle, one can demand independence of an observed value on  subtraction schemes to derive the CS equation, and this demand should not lead to the dependence of an observed value on the regularization scale, etc. There is the one exception: non-renormalizable theories. That is why we proved the renormalizability of our problem in Section \ref{renorm2}.
Therefore, the question about the scheme-dependence of the $\beta $-function (and as a result, the conductance), remains in this version of CS approach too.
The way out of the problem is simple: one should reject the “poor  man's” assumptions outside  the leading-log approximation, i.e. one should not associate the renormalized chargers, depending on a calculation method, with observed quantities.
We have seen that it is enough. I believe, dependence of an observed value on a subtraction scheme in the CS approach is no better than its dependence on the regularization point in the GL approach. In effect, together with correct calculation of an observed quantity, these approaches are the same. One just needs to accept the facts: if one changes a calculation scheme, then and a polarization cloud at a scale will change too, and different Hamiltonians will lead to the same observed quantity. Besides, transition to another Hamiltonian can change not only unobserved charges, but also a new diagram for an observed quantity can appear also.

\section{Conclusion}
Despite a rather long history, the problem of the LL remains relevant now. It turned out that LL is directly related to the problems of helical and chiral liquids. (Already in the first papers devoted to the topological insulators, it was pointed out  that the LL describes the low-energy properties of edge states \cite{Top}, \cite{TIsul}.)
However, dissemination of the 1D problem to these liquids  demands a reliable qualitative description of the phenomenon. That is why, one of the most essential aim of the paper was a qualitative discussion of the ground state of the LL (calculated in \cite{Ph}). It is done in Section “Overview of the Problem.” Here we  argue that the state with energy minimum  corresponds to the uncharged correlated state (the Kosterlitz-Thouless phase) not to the state with Peierls instability.
Description the ground state of the system as a state with exciton-like pairs  makes the break-off of a channel  with respect to direct current after implantation  point-like impurity understandable at the qualitative level. The effect appears due to  the appearance of a new “scattering” channel (not to an amplification of  e-i scattering amplitude). This new channel emerged due to formation  near impurity supplementary uncharged electron-hole pair. To conserve the electric charge of the entire system, the  process is accompanied  by creation of the electron moving in opposite direction in comparison with initial one. It is recorded as an electron reflected by impurity.

LL with point-like impurity is a problem enabling to trace the origin appearance of a non-local field theory from the problem  with local point-like interactions. The cause for the appearance of nonlocality in our case is the need to match solutions of the nonlinearized Schrödinger equation (described the interacting electrons)  at the impurity position point. An indemnity for the difficulties related to nonlocality is the absence of any ultraviolet divergences in the observed values.  It allowed to extend  domain of applicability theory from weak e-e interaction  to the strong one. In the strong interaction case ($\nu\ge 1/2$), behaviour of the conductance changes. It is proportional to $|\omega|$. The changing of frequency dependence of the conductance arises from the absence of the UV divergences in the problem.

 For the weak e-e interaction, expansion of the non-local effective action by  powers of the small frequency makes it possible to develop a new approach to the renormalization group method. We have compared results have been taken from our approach with “poor man's” one widely used in solid-state physics. The observed values differ in the second-loop approximation. The reason of this discrepancy is the dependence of unobservable renormalized charges on the regularization scheme already  in two-loop approximation.  It breaks assumptions of the “poor man's” approach is based on. This result is essential not only to the LL. The difference between a “standard” RG-approach and “poor man's”  one is principle from the viewpoint of theoretical physics. The first approach asserts, our lack of knowledge of the structure  of a Hamiltonian in the UV-region  does not affect in any observed value. The “poor man's” approach implies (outside the leading-log approximation) existence the only correct way of calculating the observed values, since changing this path changes them.

{\bf Acknowledgments.}
 \newline I am grateful to Natalia Belyaninova-Petrova for her assistance in preparing the manuscript, which was provided to me after Victor Petrov, my friend and co-author, passed away in 2021, and to Ya.M.Beltyukov for discussion of the problem.

\appendix
\section{Complete set of the wave functions.}
\label{funckSet}
 Let us discuss the meaning of the boundary condition of the Schr\"odinger equations.
Our Feynman Green function describes  transition
system of non-interacting electrons from the GS (with wave function $<F|$) at $t\to-\infty$ to the GS at $t\to\infty$ with
wave function $|F>$.  However, the
Schr\"odinger's equation is the first-order differential equation  in "t," and it is impossible to put two
boundary condition (at $t\to\pm\infty$) for one excited state $\psi_{\epsilon}(x,t)$.
To find a path out of the problem, let us represent a one-particle state at
$t\to\infty$ as $\hat {\varphi }(x)|F>$. Here the$$
\hat{\varphi}(x)=\int^\infty_{-\infty} \frac{d\varepsilon
}{2\pi}\hat{c}_{\epsilon}\psi_{\epsilon}(x)
$$ is the one-particle state. ($\hat{c}_{\epsilon}$ is the electron annihilation operator is defined
under empty state: $\hat{c}_{\epsilon}|0>=0$, and $\hat{c}_{\epsilon}^{\dag}|0>=|1>$)  The positive-energy part of electron wave function $\psi_{\epsilon}$ satisfies to the
relation: $\theta (\epsilon)\hat{c}_{\epsilon}\psi_{\epsilon}|F>=0,$  so corresponding part of the $\psi_{\epsilon}$ can be arbitrary.  In order for
the remaining part of the wave function with negative frequency does not destroy the GS
at $t\to\infty$, this part has to be forbidden. It is the required boundary condition
for $\psi_{\epsilon}$. Similar consideration for the  $<F|\hat {\varphi
}^{\dag}(x,t)$ at $t\to-\infty$ results to an arbitrary "hole-like"\ part of electron wave function and to the
prohibition of the "electron-like"\ state.
So, we have putted  one boundary condition for any state.  In such a way, one can prove all other boundary conditions.

To find Feynman Green function in the external time-dependent field $U(x,t)$, we should  find
8 solutions of the Schr\"odinger equation ($\hat{\psi}_{\epsilon}(x,t))$ with positive
and negative energies and corresponding $\hat{\tilde{ \psi}}(x,t)$. General
solution of the Schr\"odinger equation outside  impurity has a form:
\begin{equation}
\hat{\psi}(x,t)=\left(
\begin{array}{c}
\left[\mathfrak{c}(t-x)\theta(-x)+\mathfrak{d}(t-x)
\theta(x)\right]e^{i\gamma_R(x.t)}\\
\left[\mathfrak{e}(t+x)\theta(-x)+\mathfrak{f}
(t+x)\theta(x)\right]e^{i\gamma_L(x.t)}
\end{array}
\right)
\label{general-external}
\end{equation}
where $\mathfrak{c},\mathfrak{d},\mathfrak{e},\mathfrak{f}$ are unknown
functions of one variable.  They obey to the {\em second order}  Schr\"odinger equation
 with ${\cal H}_{e-i}(x)=g\delta (x)\Psi^{\dag}(x)\Psi(x),$ and $\Psi$ is a {\em total} electron wave function. Let us
integrate the equation around impurity position point:
$
\partial_x \Psi(+0) - \partial_x \Psi(-0) =2mg\Psi (0)
$
Here $\partial_x \Psi(\pm 0) =i p_F (\Psi(\pm 0)_R
-\Psi(\pm0))_L$ ($p_{F}$ is the biggest parameter of the problem). In view
of this  expression, one has
\begin{equation}
\Psi(+0)_R=\Psi(-0)_R+\frac{mg}{ip_F}(\Psi(+0)_R +\Psi(+0)_L);\quad
\Psi(+0)_L=\Psi(-0)_L-\frac{mg}{ip_F}(\Psi(+0)_R +\Psi(+0)_L)
\label{Sh1}
\end{equation}
After substitution Eq.(\ref{general-external}) to these expressions, we can  rewrite they in the form \begin{equation}
(1-\frac{mg}{ip_F})\mathfrak{d}(t)=\mathfrak{c}(t)
+\frac{mg}{ip_F}
\mathfrak{f}(t)\exp{(-\alpha(t))};\quad
(1+\frac{mg}{ip_F})\mathfrak{f}(t)=\mathfrak{e}(t)
-\frac{mg}{ip_F}
\mathfrak{d}(t)\exp{(\alpha(t))}
\label{Sh2}
\end{equation}
 Let us construct solutions obeying Feynman boundary conditions at
$t\to\pm\infty$. All $R$-particles at $t\to\infty$ should be located at
$x\to\infty$, and at $t\to-\infty$ location is
$x\to -\infty$. This means, the Feynman conditions for $R$-particles are
applied at $t\to -\infty$ {\em only}
for  $\mathfrak{c}$, and at $t\to \infty$ - for $\mathfrak{d}$. Analogously, for
$L$-particles at $t\to -\infty$ we have to apply boundary conditions for
$\mathfrak{f}$, and at $t\to\infty$ for $\mathfrak{e}$.

To illustrate the method of constructing a solution, consider the wave function $\psi^1_{\varepsilon}(x,t)$.
For the case, one allows the electron-type
solution $(\propto \exp{(-i\varepsilon t)})$ and only $R$-type wave can exist
at $x\to \infty $, i.e. $\mathfrak{d}(t-x) = \exp{(-i\varepsilon (t-x))};
\mathfrak{e}(t)= 0$. So, Eqs.(\ref{Sh2}) have to be rewritten in the form:
$$
\frac{ip_F +mg}{ip_F}\exp{(-i\varepsilon (t-x)) }=\mathfrak{c}(t) +
\frac{mg}{ip_F}\mathfrak{f}(t)\exp{(-i\alpha(t))};\quad
\frac{ip_F +mg}{ip_F}\mathfrak{f}(t)=-\frac{mg}{ip_F}
\exp{(-i\varepsilon t +i\alpha(t))}
$$
As a result, one has
$$
\mathfrak{c}(t)=\mathcal{K}^*\exp{(-i\varepsilon t)};\quad
\mathfrak{f}(t)=\mathcal{R}^*\exp{(-i\varepsilon t +i\alpha(t))},\quad{\rm where}\quad
\mathcal{K}=\frac{ip_F}{ip_F-mg};\quad \mathcal{R}=\frac{mg}{ip_F-mg}.
  $$So, we arrive to the Eq.(\ref{set1}). All  other functions can be calculated in the same way.

\section{ Adler anomaly.}
\label{Adler}

 As it was pointed above, the  expression for the ballistic current diverges because in our
approach the filled Fermi sphere is unlimited from below.  In fact, the charge density is  indeterminate. One should regularize its
expression. For the problem, the most convenient method  is
the symmetric argument splitting  method  \cite{reg}.
To renormalize a divergence, one should state  a physical principle that allows one to “calculate” an observed quantity.  We demand:
i)The gauge invariance of the problem (i.e., the fields depended  on time only  do not contribute to the observed);
ii) Conservation of the electric charge.
In order for gauge Hubbard fields do not enter into the observed value diverged in the UV range, we move to the new wave functions:
$\Psi_{R,L} (x,t) = \exp [-i\int^td\tau U(x,\tau)]\tilde\Psi_{R,L} (x,\tau)$.
If the initial wave functions obey the equation
$(i\partial_t\pm i\partial_x -U(x,t))\Psi_{R,L} (x,t)=0,$ the new wave functions
$\tilde\Psi_{R,L}$  obey  the expression$(i\partial_t\pm \partial_x \int^td\tau
U(x,\tau))\tilde\Psi_{R,L} (x,t)=0$. Thus, the gauge field cannot contribute to the observed values if we define expression for the electron density ($(\rho _{R,L}(x,t))_{reg}$) in terms of the new Green's function $\tilde{ G}_{R,L}$:
$$G_{R,L}(x-\delta x,t-\delta t|x+\delta x,t+\delta
t)\to \tilde{ G}_{R,L}(x-\delta x,t-\delta t|x+\delta x,t+\delta
t)e^{2i\delta tU(x,t)},$$ $\delta t\to +0;\delta x\to 0.$ So, one should
define the charge density as
$
(\rho _{R,L}(x,t))_{reg}=-< \tilde{G}_{R,L}(x-\delta x,t-\delta t||x+\delta
x,t+\delta t)>_{\delta t\to +0;\delta x\to 0}.$
Here, the angle  brackets label the regularization procedure,
describing below (see detailed discussion in
\cite{reg}). From Eq.(\ref{rhoR1}) with
${\cal K}=1$ and  expression for $ S_{as} $  (Eq. \ref{Sasympt}),  we have
$$
(\rho_{R,L})_{reg}(x,t)=\int\limits_{0}^{\infty}\frac{d\varepsilon_{1}
}{(2\pi)}e^{-2i\delta tU(x,t)}e^{-i\gamma_{R}(x+\delta x,t+\delta
t)+i\gamma_{R}(x-\delta x,t-\delta t)}e^{-2i\varepsilon_{1}(\delta t-\delta x)}|_
{\delta t\to +0;\delta x\to
0}=
$$
$$=\frac{1}{4\pi i}<\frac{\delta t\pm\delta x}{(\delta t)^2-(\delta
x)^2}\left(1-2i\delta t U(x,t) -2i( \delta t\partial_t +
\delta x\partial_x)\gamma_{R,L}(x,t) \right)>_{\delta t\to +0;\delta x\to
0}$$
 To obtain the continuity equation, it is necessary to determine the averaging of splits "in
direction:"
$$
<\delta t>=<\delta x>=0;\,\,\,\,\frac{(\delta t)^2}{(\delta t)^2-(\delta
x)^2}=1/2,\quad{\rm so \quad that}\quad \frac{(\delta x)^2}{(\delta t)^2-(\delta x)^2}= -1/2.
$$

It is easy to check the correctness of the approach on the example for
ballistic current. (Its expression diverges, too.)  Proceeding to Fourier transformation, we get:
\begin{equation}
(\rho _{R,L}(x,t))_{bal}=\frac{1}{4\pi}
\int\frac{d^2k}{(2\pi)^2}e^{ikx-i\omega t}\left[\frac{(\omega\pm
k)^2}{\omega^2-k^2+i\delta}-1\right]U(k,\omega).
\end{equation}
As a result, the ballistic current and electric charge density
($\rho=\rho_R+\rho_L$, $j=\rho_R-\rho_L$) equal:
\begin{equation}
(\rho(x,t))_{bal}=\frac{1}{\pi}\int\frac{d^2k}{(2\pi)^2}
\frac{e^{ikx-i\omega t}}{\omega^2-k^2+i\delta}k^2U(k,\omega);\quad (j(x,t))_{bal}=\frac{1}{\pi}\int\frac{d^2k}{(2\pi)^2}\frac{
e^{ikx-i\omega t}}{\omega^2-k^2+i\delta}\omega k U(k,\omega).\label{bal}
\end{equation}
One can check, our regularization leads to the conserved ballistic current.
However, corresponding chiral charge ($\rho_{cir}=\rho_R-\rho_L=j)$ is {\em
not conserved} due to {\em Adler anomaly}, in spite of the fact that
Hamiltonian is invariant under the chiral transformation:
\begin{equation}
\frac{\partial j}{\partial t}+\frac{\partial\rho}{\partial
x}=-\frac{1}{\pi}\int\frac{d^2k}{(2\pi)^2} ik U(k,\omega)e^{ikx-i\omega t}
=-\frac{1}{\pi} \partial_x U(x,t))
\label{A}
\end{equation}
 Also we see, the ballistic  current and charge density  (\ref{bal}) are expressed in terms of electric field of the Habbard potential
 $E(x,t)=-\partial_x U$, i.e. they are gauge invariant as it should be.
To understand the physical meaning of Adler's anomaly, consider the static limit of these equations:
$$
\frac{\partial}{\partial x}(\rho_R+\rho_L)=\frac{1}{\pi} E(x);\quad
\frac{\partial}{\partial x}(\rho_R-\rho_L)=0,
$$ and calculate the changes in the number of the $R$-electrons at the distance $|x_1-x_2|.$ It is
$$
N_R(x_1)-N_R(x_2)=\frac{L}{2\pi }(U(x_2)-U(x_1))
$$ Considering that $2\pi /L$ is the distance between energy levels ($v_F=1$), we see: the difference changes in the same way as if the electron distribution function were quasi-equilibrium (i.e., if it depends on the electrochemical potential). However, for that one needs:
$\omega\tau_{\epsilon} \ll 1,$ where $\tau_{\epsilon}-$ is the energy relaxation time, but our channel is ballistic. It means,  we have reflection  of slow electrons, existing very deep under Fermi level, from Habbard fields. (This process accompanies by creation of a number of L-electrons from the $R$ ones.) Our approach does not work here, but the conservation laws
define theirs quantities correctly. Therefore, chirality in LL does not conserve ever, and Eq.(\ref{A}) should hold for sufficiently large frequencies.

\section{Calculation of the charge jump. Integration in coupling constant.}
\label{K}
The charge jump iteration procedure has to make in a way, as to best emphasize  similarity between the theories with attracting and repulsive interactions. (To prove later
duality of the problems.) We achieve this in two stages. The
first step is transition in Eqs.(\ref{invers1}) from $S_{ik}(-\epsilon, -\epsilon_2$) to a more
convenient unknown function. To make this, let us rewrite these Eqs. in terms of theirs Fourier transforms $S_{ik}(\tau_,-\epsilon_2)$. They are defined as
$$
S_{ik}(-\epsilon, -\epsilon_2)=\int d\tau
e^{-i\epsilon\tau}S_{ik}(\tau,-\epsilon_2).$$
(It is convenient to define it with opposite sign in exponent; $\epsilon_2>0$)
As a result one has:
$$
{\mathcal{K}}S_{1,1}(\tau,-\epsilon_2) + {\mathcal{R}}\exp{(-i\alpha (\tau
))}\int \frac{d\tau_1}{2\pi
i}\frac{S_{2,1}(\tau_1,-\epsilon_2)}{\tau_1-\tau-i\delta}=
\exp{(i\varepsilon_2\tau)}
$$\begin{equation}
{\mathcal{K}}S_{2,1}(\tau,-\epsilon_2) + {\mathcal{R}}\exp{(i\alpha (\tau
))}\int \frac{d\tau_1}{2\pi
i}\frac{S_{1,1}(\tau_1,-\epsilon_2)}{\tau_1-\tau-i\delta} =0.
\label{invers2}
\end{equation}
  I.e., non-trivial parts of these equations are determined only by branches of the kernel that are analytic in the upper half-plane in $\tau$. We will denote theirs  as
 $\left[S_{ik}(\tau,-\epsilon_2)\right]_{+}$.
 As a result, one can rewrite Eqs. (\ref{invers2}) in the form:
\begin{equation}
{\cal K}S_{1,1}(\tau,-\epsilon_2)+
+{\cal R}e^{-i\alpha(\tau)}
\left[S_{2,1}(\tau,-\epsilon_2)\right]_
+=e^{i\epsilon_2\tau};\quad
{\cal K}S_{2,1}(\tau,-\epsilon_2)
+{\cal R}e^{i\alpha(\tau)}\left[
S_{1,1}(\tau,-\epsilon_2)\right]_
+=0
\label{main-eq}
\end{equation}

\subsection{Attracting interaction.}
\label{sumAttr}

 To calculate the charge jump for an attracting e-e interaction, one
should expand it in powers of $|{\cal R}|^2$ because the final expression of conductance corresponds to a picture close to the open channel. For the case, the lowest order expansion is proportional to $|{\cal R}|^2$. It is determined by the UV-charge jump (see Eqs. \ref{UV-rhoR},\ref{UV-rhoL}).
In higher orders, one should calculate only the regular part of the charge jump. It describes the laminar wake.
For it, one can use Eqs.(\ref{main-eq}). The direct iteration
Eq.(\ref{main-eq}) in $|{\cal R}|^2\ll 1$ gives
\begin{equation}
\left[S_{21}\right]_+=-\frac{{\cal R}}{{\cal
K}^2}\left[e^{i\alpha+i\varepsilon_2\tau}\right]_+
-\frac{{\cal R}^3}{{\cal K}^4}\left[e^{i\alpha}\left[e^{-i\alpha}\left[
e^{i\alpha+i\varepsilon_2\tau}\right]_+\right]_+\right]_+
-\ldots
\label{S1}
\end{equation}
It is very useful to parameterize ${\cal R}=i|{\cal R}|\exp{(i\chi)};\  {\cal
K}=\sqrt{1- |{\cal R}|^2}\exp{(i\chi)}$, so that $({\cal R}^2/{\cal
K}^2)^n=(-1)^n(|{\cal R}|^2/|{\cal K}|^2)^n$, etc...
To obtain the regular (convergent) part of the density one should
calculate the partial sum $ \R{\mathcal S_n} -{\cal R}\left[
S_{21}\right]_+^{(as)} $. (Here the $ {\mathcal S_n}$ is defined by the first
$n$-th terms of series (\ref{S1}), and ${\cal R}\left[
S_{21}\right]_+^{(as)} $ is defined by Eq.(\ref{Sasympt}).) To obtain
the charge jump, expression has to be integrated over energy. After this, the entire expression for the remaining charge jump (the total charge jump minus the divergent portion in the UV energy region) should converge.
However, this is not the case for each term in the sum. Indeed, let us consider the contribution from the
first term of the partial sum, $\R{\mathcal S_n}$. It is proportional to $|{\cal R}|^2/(1-|{\cal R}|^2)$. After integration by energy, the term proportional to $\left[e^{i\alpha+i\varepsilon_2\tau}\right]_+$ gives the divergent expression of the total electron concentration. The latter was equal to ${\cal R}\left[
S_{21}\right]_+^{(as)}\propto |{\cal R}|^2 $ at any $|{\cal R}|^2$. It has been regularized in expression for the total electron density (see Eqs.(\ref{UV-rhoR}-\ref{UV-rhoL}).  It is the  UV-part of the total charge jump.
 It means, at small $|{\cal R}|^2 $ the next term of expansion of the $1/(1-|{\cal R}|^2)$ (of the order of
$|{\cal R}|^4$) is divergent. This divergence can be vanished only by the lowest
term of  expansion of the second term in ${\mathcal S_n}$ (with is $|{\cal
R}|^4$), etc. So, the sum with convergent summands is the series in power  $|{\cal
R}|^{2m}$, not in $(|{\cal
R}|/|{\cal
K}|^{})^{2m}$. Direct rewriting the series in powers of small $|{\cal R}|^2 $ gives:
\begin{equation}
{\mathcal S_n}(t,\varepsilon_2)=\sum_{m=0}^n|{\cal
R}|^{2(m+1)}\sum_{k=0}^m(-1)^{k+1}J_{2k+1}(t,\varepsilon_2)
C^{k}_m
\end{equation}
where $C^{k}_m $ is a binomial coefficients and $J_k$ is $k$-fold analytical part:
\begin{equation}
J_n(t,\varepsilon_2)=\left[\cdots \exp{(i\alpha (t) )}\left[\exp{ (-i\alpha
(t) )}\left[\exp{ (i\alpha (t)+i\varepsilon_2t )}
\right]_+
\right]_+\cdots\right]_+
\end{equation}
So, expansion ${\cal R}\left[S_{21}\right]_+$ in power $|{\cal R}|^2$ is effected  by the term with coefficient \\$\sum_{k=0}^m(-1)^{k}C^{k}_m J_{2k+1}$ near the
$|{\cal R}|^{2(m+1)}$ (here $m\ge 1$). (We will label the $m-$th summand of this expansion as $s_{21}^{(m)}(t,\varepsilon_2)$).
After substituting the $n$-order series term into the regular part of the charge jump, it is rewritten as
\begin{equation}
{\mathfrak D}([\alpha],t)^{(n)}_{reg}=2\exp{\left(-i\alpha (t)\right)}{\cal
R}\int_0^{\infty}\frac{d\varepsilon_2}{2\pi}
\exp{(-i\varepsilon_2t)}
\left[s_{21}^{(n)}(t,
\varepsilon_2)\right]_+ - (\alpha \to -\alpha ).
\label{Dn}
\end{equation}
($n\ge 0$ here.)
In view of the identity$$
\int \frac{d\omega_i}{2\pi}...\frac{d\omega_{2n+1}}{2\pi}
\varphi^-(\omega_i)e^{-i\omega_it}
...\varphi^+(\omega_{2n+1})e^{-i\omega_{2n+1}t}=e^{-i\alpha (t)}e^{i\alpha (t)}...e^{-i\alpha (t)}e^{i\alpha
(t)}=1$$ ($\varphi^{\pm}(\omega)$ were defined earlier, Eq.(\ref{defphi})),  one can rewrite Eq.(\ref{Dn}) in the form $$
{\mathfrak D}([\alpha],\omega)_{reg}^{(n)}=2e^{-i\alpha
(t)}{|\cal R}|^{2(n+1)}\int_0^{\infty}\frac{d\varepsilon_2}
{2\pi}\int
\frac{d\omega_1...d\omega_{2n+1}}{(2\pi )^{2n+1}}\varphi_+(\omega_1)e^{-i\omega_1t}
\varphi_-(\omega_2)
e^{-i\omega_2t}...
$$ $$\times\varphi_+(\omega_{2n+1})e^{-i\omega_{2n+1}t}
\left[ C^0_n \theta (\varepsilon_2-\omega_1)- C^1_n \theta
(\varepsilon_2-\omega_1) \theta (\varepsilon_2-\omega_1-\omega_2)\theta
(\varepsilon_2-\omega_1-\omega_2-\omega_3)+...\right]-$$  $$- (\alpha\to
-\alpha)
=2e^{-i\alpha (t)}{|\cal
R}|^{2(n+1)}\int_0^{\infty}\frac{d\varepsilon_2}
{2\pi}\int
\frac{d\omega_1...
d\omega_{2n+1}}{(2\pi )^{2n+1}}d\tau_1...\tau_{2n+1}e^{i\alpha (\tau_1)-i\omega_1(t-\tau_1)}\times$$
\begin{equation}\times e^{i\alpha
(\tau_2)-i\omega_2(t-\tau_2)}\times ...
\times e^{i\alpha(\tau_{2n+1})-i\omega_{2n+1}(t-\tau_{2n+1})
}\left[ C^0_n \theta (\varepsilon_2-\omega_1)- \right.
\label{Dn2}
\end{equation}
 $$\left.-C^1_n \theta (\varepsilon_2-\omega_1) \theta
(\varepsilon_2-\omega_1-\omega_2)\theta
(\varepsilon_2-\omega_1-\omega_2-\omega_3)+...\right] - (\alpha\to -\alpha)
.$$
Let us calculate the auxiliary integral ($k\geq 2$):
$$
I_k(\tau_1\ldots \tau_k)=\int\frac{ d\omega_1\ldots d\omega_{k}}{(2\pi)^{k}}
\int_0^\infty\frac{d\varepsilon}{(2\pi)}e^{i\omega_1 \tau_1}\ldots
e^{i\omega_k \tau_k}\left[\theta(\varepsilon-\omega_1)\ldots\theta(\omega_1
+\ldots+\omega_{k-1}-\varepsilon)\right]=
$$
To uncouple the integrals, let us introduce the new variables
$$
\omega_1=\Omega_1, \qquad \omega_1+\omega_2=\Omega_2 \qquad \ldots \qquad
\omega_1+\ldots +\omega_k=\Omega_k.
$$
Then we arrive at:
$$
=\frac{1}{2\pi i (-t_1-i\delta)}\cdot\frac{1}{2\pi i (t_1-t_2-i\delta)}\ldots
\frac{1}{2\pi i (t_{k-1}-t_k-i\delta)}\cdot\frac{1}{2\pi i (-t_k-i\delta)}
$$
Let us rewrite  Eq.(\ref{Dn2}) in terms of auxiliary integrals. It is easy to see:
only a term
with $\theta (\varepsilon_2-\omega_1)$ remains divergent, but the coefficient
in front of divergent term equals $\sum_{k=0}^n(-1)^kC_n^k=0$. So, we have achieved
the aim: each term of the expansion in $|{\cal R}|^2$ is convergent. The
partial sum of charge jump Eq.(\ref{Dn}) can be rewritten in terms of $I_k$:
$$
{\mathfrak D}_{reg}^{(n)}(t)=2|{\cal R}_0|^{2(n+1)}\exp{(-i\alpha(t))}
\int dt_1.. dt_{2n+1}\exp{(i\alpha(t_1))}\exp{(-i\alpha(t_2))} ..
\exp{(i\alpha(t_{2n+1}))}\times$$
$$\left\{C^0_{n-1}
\left[I_2(\tilde{t_1},
\tilde{t}_2)\times\right.
\Delta({\tilde t}_3,..,{\tilde t}_{2n+1})
+I_3(\tilde{t_1},\tilde{t}_2,\tilde{t}_3)
\Delta({\tilde t}_4,..,{\tilde t}_{2n+1})
\right]-C^1_{n-1}\left[I_4(\tilde{t_1},
\tilde{t}_2,\tilde{t}_3,
\tilde{t}_4)\Delta({\tilde t}_5,..,{\tilde t}_{2n+1})\right.$$
$$\left.+I_5(\tilde{t_1},\tilde{t}_2,\tilde{t}_3,\tilde{t}_4,
\tilde{t}_5)
\Delta({\tilde t}_6,..,{\tilde t}_{2n+1})    \right]+..\big\}
$$
 here $\tilde{t}_i=t_i-t$ and
$
\Delta({\tilde t}_k,.. {\tilde t}_{2n+1})=\delta(t_k-t)\delta(t_{k+1}-t)
\delta(t_{2n+1}-t).$
After integration over the times, are not entering into $I_k$, we arrive at
\begin{equation}
{\mathfrak D}_{reg}^{(n)}(t)=|{\cal R}|^{2(n+1)}\left( {\mathfrak
B}_2(t)+{\mathfrak B}_3(t)-
C^1_{n-1} \left( {\mathfrak B}_4(t)+{\mathfrak B}_5(t)\right)+ C^2_{n-1} \left( {\mathfrak B}_6(t)+{\mathfrak B}_7(t)\right)-\ldots
\right)
\end{equation}
and ${\mathfrak B}_i$ is defined by Eqs. (\ref{defB}).

Now one can calculate the regular part of the density
jump, ${\mathfrak D}(t)_{reg}=\sum _n {\mathfrak D}^{(n)}(t)$:
\begin{equation}
{\mathfrak D}([\alpha],t)_{reg}=|{\cal R}|^{2}\left( \frac{|{\cal
R}|^{2}}{|{\cal K}|^{2}} \left( {\mathfrak B}_2(t)+{\mathfrak B}_3(t)\right)-
\left(\frac{|{\cal R}|^{2}}{|{\cal K}|^{2}}\right)^2\left( {\mathfrak
B}_4(t)+{\mathfrak B}_5(t)\right)+\ldots\right)
\label{Dfin}
\end{equation}
  Here we have used the identity:$$
\sum_{n=1}^{\infty}|{\cal R}|^{2(n+1)}C^k_{n-1}=|{\cal R}|^2\left(\frac{|{\cal
R}|^2}{1-|{\cal R}|^2}\right)^k.
$$ To calculate the full charge jump, one should add to the ${\mathfrak D}(t)_{reg}$ the UV-part. So,  expansion of the regular part of the charge jump begins from $|{\cal R}|^4$, and the lowest  term of expansion gives the UV-part. This expression is {\em exact}, but  to  calculate  reflection coefficient
one should perform a functional integration in $\alpha$.
Direct calculation gives the explicit expression for UV-part of the charge jump. It follows from Eqs.(\ref{UV-rhoR},\ref{UV-rhoL}). In terms of ${\mathfrak
B}_i(t)$ functions it is
\begin{equation}
{\mathfrak D}_{UV}([\alpha],t)=- |{\cal R}|^{2}{\mathfrak B}_1(t);\quad
{\mathfrak D}_{}([\alpha],t)=- |{\cal R}|^{2}{\mathfrak B}_1(t)+
{\mathfrak D}_{reg}([\alpha],t).
\label{D-fullfin}
\end{equation}Here ${\mathfrak D}_{}([\alpha],t)$ is the total charge jump, while
 coefficients ${\mathfrak B}_i(t)$ are given by the expressions:
\[
{\mathfrak B}_1(t)=\frac{1}{\pi}\int\frac{d\tau}{(2\pi
i)}\left[\frac{1}{(\tau-t+i\delta)^2}+\frac{1}{(\tau-t-i\delta)^2}
\right]
\sin[\alpha(\tau)-\alpha(t)]
\]
\[
{\mathfrak B}_2(t)=\frac{2}{\pi}\int\frac{d\tau_1d\tau_2}{(2\pi
i)^2}\frac{\sin[\alpha(\tau_1)-\alpha(\tau_2)]}{(\tau_1-t+i\delta)
(\tau_1-\tau_2-i\delta)(\tau_2-t+i\delta)}
\]
\[
{\mathfrak B}_3(t)=\frac{2}{\pi}\int\frac{d\tau_1d\tau_2d\tau_3}{(2\pi
i)^3}\frac{\sin[\alpha(\tau_1)-\alpha(\tau_2)+\alpha(\tau_3)-
\alpha(t)]}{(\tau_1-t+i\delta)(\tau_1-\tau_2-i\delta)
(\tau_2-\tau_3-i\delta)(\tau_3-t+i\delta)}
\]
\begin{equation}
{\mathfrak B}_4(t)=\frac{2}{\pi}\int\frac{d\tau_1d\tau_2d\tau_3d\tau_4}{(2\pi
i)^4}\frac{\sin[\alpha(\tau_1)-\alpha(\tau_2)+\alpha(\tau_3)-
\alpha(\tau_4)]}{(\tau_1-t+i\delta)(\tau_1-\tau_2-i\delta)
(\tau_2-\tau_3-i\delta)}\frac{1}{(\tau_3-\tau_4-i\delta)
(\tau_4-t+i\delta)}\quad
{\rm etc}.
\label{defB}
\end{equation}

\subsection{Repulsive interaction.}
\label{KlR}
In this section, we will assume that  the transition coefficient is small. The point is, from the final expression of conductivity we can make sure that   the channel will be close to shutting down.
It means, a well-defined iteration procedure exists only at ${\cal K}\ll1$. In addition, the resulting expression of the charge jump has to be reduced to the form detecting duality of the problems with repulsive and attracting electrons.
 To expand the charge jump in $|{\cal K}|^{2n}$ series,
 let us introduce the new functions:
\begin{equation}
[\sigma_{11}]_+=[S_{2,1}]_+e^{-i\alpha_+}, \qquad
[\sigma_{21}]_+=[S_{1,1}]_+e^{i\alpha_+}.
\label{dual-var}
\end{equation}We will assume, the functions $\sigma_{ik},S_{ik},\alpha_{}$ are rapidly
decreasing  at $\tau\to\pm\infty$. Therefore, they can be represented as a sum of two branches, analytical in the upper/lower semiplane ($\alpha_\pm(\tau)$).

In term of the functions (\ref{dual-var}) Eq.(\ref{main-eq}) can be rewritten in the form
$$
{\cal R}[\sigma_{11}(\tau,-\varepsilon_2) +{\cal K}e^{-i\tilde\alpha
(\tau)}[\sigma_{21}(\tau,-\varepsilon_2)]_+
]_+ =e^{i\varepsilon_2 \tau+i\alpha_-};
{\cal R}[\sigma_{21}(\tau,-\varepsilon_2)+{\cal
K}e^{i\tilde\alpha(\tau)}[\sigma_{11}
(\tau,-\varepsilon_2)]_+
]_+ =0,
\label{}
$$
(here we have taken into account identity $\left[\exp(\pm
i\alpha_-(\tau))[\sigma_{ik}(\tau,-\varepsilon_2)]
_-\right]_+=0$.)
At this step, we have introduced a new
field, closely related to $\alpha(\tau)$. It is
the dual field:
$ \widetilde{\alpha}(\tau)=\alpha_+(\tau)-\alpha_-(\tau).
$ As a second step, we will transfer the Eqs. to the {\em dual form.} To this, one can "solve"\ the first equation:
$$
{\cal R}\left[\sigma_{11}\right]_+=e^{i\alpha_- +
i\varepsilon_2 \tau}-{\cal K}[\sigma_{21}]_+e^{-i\tilde\alpha}
+ f_-(\tau),
$$
where $f_-(\tau)$ is an arbitrary function analytical in the lower semi-plane.
This function should be chosen from the requirement: the l.h.s. of the expression is analytical
in upper semi-plane function. This leads to the expression:
$$
f_-(\tau)=-\left[e^{i\alpha_- +
i\varepsilon_2 \tau}-{\cal K}[\sigma_{21}]_+e^{-i(\alpha_+
-\alpha_-)}\right]_-\quad{\rm and}\quad
{\cal R}\left[\sigma_{11}\right]_+=\left[e^{i\alpha_- +
i\varepsilon_2 \tau}-{\cal K}[\sigma_{21}]_+e^{-i(\alpha_+
-\alpha_-)}\right]_+.
$$
The second equation can be obtained by the same manipulations. As a result, one
has
\begin{equation}
{\cal R}\left[\sigma_{11}\right]_+
+{\cal K}\left[e^{-i\widetilde{\alpha}(\tau)}\left[\sigma_{21}
\right]_
+\right]_+
=\left[e^{i\varepsilon_2\tau+i\alpha_-(\tau)}\right]_+
\quad{\rm and}\quad
{\cal R}\left[\sigma_{21}\right]_+
+{\cal
K}\left[e^{i\widetilde{\alpha}(\tau)}\left[\sigma_{11}\right]_+\right]_+
=0.
\label{dual-eq}
\end{equation}

 We see now, transition to the {\em dual variables} convert Eq.(\ref{main-eq})
to the  Eq.(\ref{dual-eq}). Indeed, after transformation
$ {\cal R}\rightarrow {\cal K},
\qquad
{\cal K}\rightarrow {\cal R},
\qquad
\alpha(\tau)\rightarrow \widetilde{\alpha}(\tau),
\qquad
e^{i\varepsilon_2\tau}\rightarrow
\left[e^{i\varepsilon_2\tau+i\alpha_-(\tau)}\right]_+
$ the equations  move one to other, i.e. $
S_{11}\rightarrow \sigma_{1,1}, \ S_{21}\rightarrow \sigma_{2,1}.
$
Hence, the solution of (\ref{main-eq}) should obey these symmetry requirements too.
So, an asymptotic solution of dual equation Eq.(\ref{dual-eq}) for
$\sigma_{11}(\tau,-\varepsilon_2)_{as}$ can be taken from
$S_{11}(\tau,-\varepsilon_2)_{as}={\cal K}^* e^{i\varepsilon_2\tau}$ by same substitution.  As a result, the  asymptotic solution is
\begin{equation}
\left(\sigma_{11}(\tau)\right)_{as}={\cal R}^*\left[
e^{i\varepsilon_2\tau+i\alpha_-(\tau)}
\right]_+.
\label{sigma-as}
\end{equation}
The Eqs.(\ref{dual-eq}) is easy to iteration in small $\mathcal{K}$. Indeed,
it is clear from the dual equations that
\begin{equation}
\sigma_{11}(\tau)= \frac{1}{{\cal
R}}\left[e^{i\varepsilon_2\tau+i\alpha_-(\tau)}\right]_+
+
\frac{{\cal K}^2}{{\cal R}^3}\left[e^{-i\widetilde{\alpha}(\tau)}\left[
e^{i\widetilde{\alpha}(\tau)}\left[e^{i\varepsilon_2
\tau+i\alpha_-(\tau)}\right]_+\right]_+\right]_++\ldots
\label{alphaIter}
\end{equation}
and analogously for $\sigma_{21}$.

From Eq.(\ref{dual-var}) we can restore quantities $S_{2,1}(\tau)$ and
$S_{1,1}(\tau)$. One can directly substitute theirs into  Eqs. for
electron density. As regards asymptotic solutions (Eqs.(\ref{Sasympt})), they
are valid at all ${\cal K}$ and, broadly speaking, one can use their "as is".\
Yet,  to emphasize  the dual symmetry between the cases with small
${\cal K}$ and small ${\cal R}$, we will define $\Pi_{2,1}$ not in the form
Eq.(\ref{defPi}), but as
\begin{equation}
\widetilde{\Pi}_{21}(\tau)=\int_0^{\infty}\frac{d\varepsilon_1d\varepsilon_2}
{(2\pi)^2}\int d\tau_1
e^{i(\varepsilon_1-\varepsilon_2)\tau}e^{-i\varepsilon_1\tau_1}
e^{i\alpha_+(\tau_1)}\left[
[\sigma_{11}(\tau_1)]_+-\left(\sigma_{11}(\tau_1)\right)_{as}
\right] .
\label{tildPi}
\end{equation}We have done this step to remove the ballistic current
from the expression of linear response. Thereafter, the  charge jumps for the
attracting and repulsive problems should be dual.
Seeking duality, we have subtracted to  Eq.(\ref{}) not $(S_{2,1})_{as}$, as it should be for the correct
calculation the regular part of charge jump, but $(\sigma_{1,1})_ {as}$. We should
take  into account this operation and to redefine the charge jump:
\begin{equation}
\frac{{\mathfrak D}_{reg}(\tau)-{\widetilde{\mathfrak D}}_{reg}(\tau)}{2|{\cal
R}|^2e^{-i\alpha(\tau)}}=\int\limits_0^\infty\frac{d\varepsilon_1d
\varepsilon_2}{(2\pi)^2}d\tau_1
e^{i(\varepsilon_1-\varepsilon_2)\tau-i\varepsilon_1\tau_1}
\{
[ e^{i\varepsilon_2\tau_1+i\alpha_-(\tau_1)}
]_+e^{i\alpha_+(\tau_1)}-
[e^{i\varepsilon_2\tau_1+i\alpha(\tau_1)} ]_+\}-..
\label{dualCh}
\end{equation}
here and below the symbol
($\pm\ldots$)  means $\pm$ term with substitution $\alpha\to -\alpha$ and
\begin{equation}
{\widetilde{\mathfrak D}}_{reg}(\tau)=2{\cal R}
\left[e^{-i\alpha(\tau)}\widetilde{\Pi}_{21}(\tau)
-e^{i\alpha(\tau)}\widetilde{\Pi}_{12}(\tau)
\right]
\label{}
\end{equation}
is a {\em dual} regular charge jump. It has a property dual to the property of
${\mathfrak D}_{reg}(\tau)$: the ${\mathfrak D}_{reg}(\tau)$ at small
reflection coefficients has  expansion starting from $|{\cal R}_|^4$ and
${\widetilde{\mathfrak D}}_{reg}(\tau)$ at small transition
coefficients  has expansion  starting from $|{\cal K}_|^4$. However, there is the price  we have to pay for such definition of ${\widetilde{\mathfrak D}}_{reg}(\tau)$: one should
change the expression of the charge jump ( Eqs.(\ref{dualCh})). For that, we have inserted  the second and third terms to this expression.
The second term  cancels $(\sigma_{1,1})_{ as}$ from the expression for $\widetilde{\Pi}_{21}$,
while the third one is the correct subtraction equals to $(S_{2,1})_{as}$. Let us
rewrite the  integral term at Eq.(\ref{dualCh}). One can easily integrate the terms in $\varepsilon_1$ and then in $\tau_1$ using the analyticity of integrand in the upper semi-plane. It allows rewriting  the latter term in the form  $\left\{e^{-i\alpha
(\tau)} F(\tau) -\ldots \right\}$,
where
\begin{equation}
F(\tau)= 2|{\cal
R}|^2\int_0^\infty\frac{
d\varepsilon_2}{(2\pi)}
e^{-i\varepsilon_2\tau}
\left\{
\left[ e^{i\varepsilon_2\tau+i\alpha_-(\tau)}
\right]_+e^{i\alpha_+(\tau)}
 -
\left[e^{i\varepsilon_2\tau+i\alpha(\tau)} \right]_+ \right\}-\ldots.
\label{F}
\end{equation}
It is convenient to introduce Fourier  transform of the functions
\begin{equation}
e^{i\alpha_\pm(\tau)}= \int \frac{d\omega}{2\pi}e^{-i\omega
\tau}\varphi_+^{(\pm)}(\omega)\quad {\rm or}\quad \varphi_+^{(\pm)}(\omega)= \int d\tau e^{i\omega
\tau +i\alpha_{\pm}(\tau)}.
\label{anphi}
\end{equation}(These expressions are  full analogue of the definition Eq.(\ref{defphi}).)
Let us note, the $\varphi_+^{(+)}(\omega)$ is non-zero only at $\omega<0$,
while $\varphi_+^{(-)}(\omega)$  is non-zero at $\omega>0$, i.e. $\varphi_+^{\pm}(\omega)=\theta(\mp\omega)\varphi (\omega)$.  The similar property is valid
for any function. These functions allow rewriting Eq.(\ref{F}) as
\begin{equation}
F(\omega)=2|{\cal R}|^2
\int_0^\infty\frac{d\varepsilon_2}{(2\pi)}\int
\frac{d\omega_1d
\omega_2}{(2\pi)^2}
2\pi\delta(\omega-\omega_1-\omega_2)\times\varphi_+^{(-)}(\omega_1)\varphi_+^{(+)}(\omega_2)
\left[\theta(\varepsilon_2-\omega_1)-
\theta(\varepsilon_2-\omega_1-\omega_2)
\right].
\label{}
\end{equation}According to our definition, the integrant is non-zero only if $\omega_1>0$, and $\omega_2<0$. After integration in $\varepsilon_2$, we have
\begin{equation}
 F(\omega)=\frac{|{\cal R}|^2}{\pi}\int(d\omega_1d\omega_2)2\pi\delta(\omega-
 \omega_1-\omega_2)\varphi_+^{-}(\omega_1)\varphi_+^{+}
 (\omega_2)[\omega_2\theta(\omega)-\theta(-\omega )\omega_1)].
\label{F}
\end{equation}
Let us make the inverse Fourier-transform. As a result, the curly  bracket in Eq.(\ref{dualCh}) is
\begin{equation}
\left\{ F(t)e^{-i\alpha(t)}-[\alpha\to -\alpha]\right\}=\frac{i|{\cal
R}|^2}{\pi^2 }
\int\! d\tau\;\cos[\alpha(\tau)-\alpha(t)]\left[
 \frac{\alpha'_+(\tau)}{t-\tau-i\delta}+\frac{\alpha'_-(\tau)}{t-\tau+
 i\delta}
\right].
 \label{jump-reg1}
\end{equation}
In addition, we have another term of order unity — the ultraviolet charge jump (Eq.(\ref{UV-rhoR}). It equals
$$
{\mathfrak D}_{UV}(t)=-\frac{i|{\cal R}|^2}{2\pi^2 }
\int\! d\tau\;\alpha' (\tau)\cos[\alpha(\tau)-\alpha(t)]\left[
 \frac{1}{t-\tau-i\delta}+\frac{1}{t-\tau+
 i\delta}\right] .
 $$
 All terms of the order of unit have to be extracted from the charge jump to have a well-defined iteration procedure.
 The sum of these two quantities $\big($we will call it as a total "ultraviolet"\
part  $
 \widetilde{{\mathfrak D}}_{UV}(t)=\left\{
 F(z)e^{-i\alpha(\tau)}-\ldots\right\}+ {\mathfrak D}_{UV}(t)\big)
 $
 can be represented in the simple form because this sum is proportion to $\partial\tilde\alpha^{}(\tau)\delta(t-\tau)$
\begin{equation}
\widetilde{{\mathfrak D}}_{UV}([\tilde\alpha],t)=-\frac{|{\cal
R}|^2}{\pi}\left[\alpha_+'(t)-\alpha'_-(t)\right] .
\label{jump-reg5}
\end{equation}

Adding the regular part, we obtain expression for the total charge jump:
\begin{equation}
{\mathfrak D}([\tilde\alpha]t) =-\frac{|{\cal R}|^2}{\pi}\tilde{\alpha}'(t)+
{\widetilde{\mathfrak D}}_{reg}(t) .
\label{jump-reg6}
\end{equation}
The first term in r.h. of the expression violates duality for the full charge
jump, yet it has to  exist.  The summand with $\tilde{\alpha}'(t)$ should cancel the ballistic current, existing in response and, what is more important, it should renormalize the "free part"\ of the action (Eq.\ref{Wdual}).
To cancel the ballistic current, one should replace $|{\cal R}|^2\to
1$ in Eq.(\ref{jump-reg6}). For that, we should extract the item $-|{\cal K}|^2\tilde{\alpha}'(t)/\pi$ from
the regular part of the charge jump.
To take the term, we are interested in, one should iterate $\sigma_{11}$ in Eq.(\ref{tildPi}) up to the next order of $|{\cal K}|^2$. It equals:
$$ {\tilde{\mathfrak D}}(t)^{(0)}_{reg}=
-2\exp{\left(-i\alpha_-
(t)\right)}{|\cal K}|^2\int_0^{\infty}\frac {d\varepsilon_2}{2\pi}\int
\frac{d\omega_1.. d\omega_3}{(2\pi)^3}e^{-i\varepsilon_2t}
\left[e^{-i\tilde\alpha(t)}
\left[e^{i\tilde\alpha(t)}\left[e^{i\varepsilon_2t
+i\alpha_-(t)}
\right]_+
\right]_+
\right]_+-...
$$
$$
{\tilde{\mathfrak D}}([\alpha],t)^{(0)}_{reg}=-2\exp{\left(-i\alpha_-
(t)\right)}{|\cal K}|^2\int_0^{\infty}\frac {d\varepsilon_2}{2\pi}\int
\frac{d\omega_1d\omega_2
d\omega_3}{(2\pi)^3}\varphi_+^{(-)}
(\omega_3)\exp{(-i\omega_3t)}\times$$
$$\times
{\tilde
\varphi}_+(\omega_2)\exp{(-i\omega_2t)}
{\tilde\varphi}_-(\omega_{1})
\exp{(-i\omega_1t)} \theta (\varepsilon_2-\omega_3)\theta
(\varepsilon_2-\omega_2-\omega_3)
(\varepsilon_2-\omega_1-\omega_2-\omega_3)
-\ldots$$ where $\varphi_+^{(-)}$ is defined at Eq.(\ref{anphi}) while
${\tilde\varphi}_+$ differ from  Eq.(\ref{defphi}) by replacement
$\alpha\to{\tilde\alpha}.$ (The index "tilde"  replaces the upper indexes of $\varphi^{\pm}_{\pm}$, where they marked  the analytical branches of $\alpha$.)   This expression has to be calculated more accurately than the previous one. Let us proceed to the Fourier transformation  of this expression.  One can integrate it in $\varepsilon_2$ (using condition $\omega_3>0$):
$${\tilde{\mathfrak D}}([\alpha],t)^{(0)}_{reg}=
-2\exp{\left(-i\alpha_-
(t)\right)}{|\cal K}|^2\int
\frac{d\omega d\omega_1d\omega_2
d\omega_3}{(2\pi)^4}2\pi\delta(\omega-\omega_1-\omega_2-
\omega_3)e^{-i\omega t}\times$$
$$\times\varphi^-_+(\omega_3)\tilde\varphi_+(\omega_2)
\tilde\varphi_-(\omega_1)\left\{(\omega_1+\omega_2)\theta
(\omega_1+\omega_2)\theta(\omega_1)+\omega_2\theta
(-\omega_1)\theta(\omega_2)\right\}-\cdots$$It is important, expression in parentheses does not depend on $\omega_3$.  For this reason, one
can integrate back in $\omega$ and $\omega_3$ and return to the $e^{i\alpha(t)}$, which are cancelling out. The remaining expression depends only on $\tilde\alpha$:
$$
{\tilde{\mathfrak D}}(t)^{(0)}_{reg}=
-\frac{2{|\cal K}|^2}{\pi}\int
\frac{d\omega_1d\omega_2
}{(2\pi)^2}\tilde\varphi_+(\omega_2)
\tilde\varphi_-(\omega_1)\left\{(\omega_1+\omega_2)\theta
(\omega_1+\omega_2)\theta(\omega_1)+\omega_2\theta
(-\omega_1)\theta(\omega_2)\right\}e^{-it(\omega_1+
\omega_2)}
$$For calculation Fourier transformation of $
{\tilde{\mathfrak D}}([\alpha],\omega)^{(0)}_{reg},$  we will use the following integrals:$$
I_1=\omega\theta(\omega)\int\frac{d\omega_1d\omega_2
}{(2\pi)}\delta(\omega-\omega_1-\omega_2)e^{i\omega_1+
i\omega_2v}\theta(\omega_1)=
\frac{\omega\theta(\omega)e^{i\omega v}}{2\pi i(u-v-i\delta)}
$$
$$
I_2=\int\frac{d\omega_1d\omega_2
}{(2\pi)}\delta(\omega-\omega_1-\omega_2)e^{i\omega_1u+
i\omega_2v}\omega_2\theta(-\omega_1)\theta(\omega_2)=\frac
{\partial}{i\partial v}
\frac{1}{2\pi i(u-v-i\delta)}\times$$ $$\times[\theta(-\omega)e^{i\omega u}+\theta(\omega)e^{i\omega v}]=-\frac{e^{i\omega u}\theta(-\omega)+e^{i\omega v}\theta(\omega)}{2\pi (u-v-i\delta)^2}+\frac{\omega e^{i\omega v}\theta(\omega)}{2\pi i(u-v-i\delta)}
$$In the sum $I_1$ and $I_2$ (which determines $
{\tilde{\mathfrak D}}([\alpha],\omega)^{(0)}_{reg}$) the $\delta$-functions appear, but they are vanished due to relation $\omega\delta(\omega)\theta(\omega)=0$:
$$
{\tilde{\mathfrak D}}([\alpha],\omega)^{(0)}_{reg}=
-\frac{2{|\cal K}|^2}{\pi}\int dudve^{i\tilde\alpha(u)-i\tilde\alpha(v)}\{\delta(u-v)
\omega\theta(\omega)e^{i\omega v}-\frac{e^{i\omega u}\theta(-\omega)+e^{i\omega v}\theta(\omega)}{2\pi (u-v-i\delta)^2}-\cdots
\}
$$  As to the second term, after  adding terms with replacement $\alpha\to -\alpha$ it produces expression:
\begin{equation}
\tilde{\mathfrak D}([\alpha],\omega)^{(0)}_{reg}=
-\frac{2{|\cal K}|^2}{i\pi^2}\int dudv\frac{\sin
[\tilde\alpha(u)-\tilde\alpha(v)]}{(u-v-i\delta)^2}
[e^{i\omega u}\theta(-\omega)+e^{i\omega v}\theta(\omega)].
\label{tildD}
\end{equation} After substitution $u\to v$ in the second term, we have
\begin{equation}
\tilde{2\mathfrak D}([\alpha],\omega)^{(0)}_{reg}=-
\frac{2{|\cal K}|^2}{i\pi^2}\int dudv \sin[\tilde\alpha(u)-\tilde\alpha(v)]e^{i\omega u}\left\{\frac{\theta(-\omega)
}{(u-v-i\delta)^2} -\frac{\theta(\omega)
}{(u-v+i\delta)^2}
\right\}.
\label{tildD1}
\end{equation} Now, let us note that:
$$
\frac{1}{(u-v-i\delta)^2}=\frac{1}{2}\left[ \frac{1}{(u-v-i\delta)^2}+\frac{1}{(u-v+i\delta)^2}
\right]-i\pi\frac{\partial}{\partial v}\delta(v-u);$$
$$
\frac{1}{(u-v+i\delta)^2}=\frac{1}{2}\left[ \frac{1}{(u-v-i\delta)^2}+\frac{1}{(u-v+i\delta)^2}
\right]+i\pi\frac{\partial}{\partial v}\delta(v-u).$$
Hence:
\begin{equation}
\tilde{\mathfrak D}([\alpha],\omega)^{(0)}_{reg}=
\frac{{|\cal K}|^2}{i\pi^2}\rm{sign}(\omega)\int dv du \sin[\tilde\alpha(u)-\tilde\alpha(v)]\left\{\frac{1
}{(u-v-i\delta)^2} +
\frac{1
}{(u-v+i\delta)^2}
\right\}-\frac{1}{\pi}|\K|^2\tilde\alpha^{'}(\omega).
\label{tildD2}
\end{equation}
The later term of this expression serves for cancellation of the ballistic current and for transition to the new "free part"\ of the action (after consolidation with first term of Eq.(\ref{jump-reg6}). The first one, enters to the dual charge jump. It is exactly $\rm{sign}(\omega)\tilde{\cal B}_1(\omega)$, as it should be:
\begin{equation}
\tilde\alpha (-\omega)|\K|^2\tilde{\cal B}_1(\omega)|_{\K\to\R;\tilde\alpha\to\alpha}\to
\alpha (-\omega)|\R|^2{\cal B}_1(\omega),
\label{K2dual}
\end{equation}here and later $\tilde{\cal B}_i(\omega)$ is the
 ${\cal B}_i(\omega)$ with replacement $\alpha\to\tilde\alpha$ (see Eq.(\ref{defB})).

Let us calculate the new "free part"\ of the action for repulsive interaction. After consolidation with first term of Eq.\ref{jump-reg6}, the term is proportional to $|\K|^2$ gives the addition to the action Eq.(\ref{det-good}). It is proportional to
$$
\frac{1}{2\pi}\int_0^1\lambda d\lambda\int(d\omega)\rm{sign}(\omega)\tilde\alpha(-\omega)
\omega\tilde\alpha(\omega)=\frac{1}{4\pi}\int(d\omega)|\omega|
\tilde\alpha(\omega)\tilde\alpha(-\omega)$$ According to our definition of the action, transition to the variable $\tilde\alpha$ gives
\begin{equation}
S_{kin}([\tilde\alpha])= \frac{1}{2}\int\frac{d\omega}{2\pi}
\frac{\tilde\alpha(-\omega)\tilde\alpha(\omega)}{\tilde W(|\omega|)}
\label{SkinD}
\end{equation}So, we have arrived to the Eq.(\ref{Wdual}).

 It remains to calculate the remainder part of the dual charge jump. The sum for $\sigma_{11}$, Eq.(\ref{alphaIter}), is the expansion in $(|\K|/|\R|)^{2n}$. As well as for attraction problem, one should  rewrite the series to the $(|\K|)^{2n}$ powers instead of $(|\K|/|\R|)^{2n}$. The problem is simplified  by the fact,  cancellation of the
divergent parts occur in each  n-th term separately. One can express  the n-th order term as $$
|\R|\sigma_{11}^n=-\frac{1}{|\R|}(|\K|)^{2(n+1)}\sum_{k=0}^n
(-)^k \tilde J_{2k+3}C_n^k,
$$where $\tilde J_n$ is n-fold analytical part:
$$
\tilde J_n=\left[e^{i\tilde\alpha(t)}\left[e^{-i\tilde\alpha(t)}...
\left[e^{i\varepsilon_2t+ i\tilde\alpha_-(t)}
\right]_+\right]_+\right]_+,$$and for the
charge jump
$$
{\tilde{\mathfrak D}}([\alpha],t)^{(n)}=-2
\exp{\left(-i\alpha (t)_-\right)}{|\cal K}|^{2(n+1)}\int_0^{\infty}\frac
{d\varepsilon_2}{2\pi}\int
\frac{d\omega_1...d\omega_{2n+3}}{(2\pi
)^{2n+3}}\varphi_+^{(-)}(\omega_1)\exp{(-i\omega_1t)}
{\tilde\varphi}_+(\omega_2)
...
$$ $$\times{\tilde\varphi}_-(\omega_{2n+3})\exp{(-i\omega_2t)}...\exp{(-i\omega_{2n+1}t)}
\left[ C^0_n \theta (\varepsilon_2-\omega_1)\theta
(\varepsilon_2-\sum_1^2\omega_i)
(\varepsilon_2-\sum_1^3\omega_i)- C^1_n \theta (\varepsilon_2-\omega_1)\times
\right.
$$
$$\left.\times\theta (\varepsilon_2-\sum_1^2\omega_i)\theta
(\varepsilon_2-\sum_1^3\omega_i)\theta (\varepsilon_2-\sum_1^4\omega_i)\theta
(\varepsilon_2-\sum_1^5\omega_i)+...\right] - (\alpha\to -\alpha).$$
Let us transform the expression to the form with $I(t_1,...)$,  as it was
described in previous section. Keeping in the memory identity
$\sum_{k=m}^n(-1)^kC^k_n=(-1)^mC^{m-1}_{n-1}$, one has
$$
 \widetilde{{\mathfrak D}}^{(n)}([\tilde\alpha],t)=-2|{\cal
 K}|^{2(n+1)}\exp{(-i\alpha_-(t))}
\int dt_0\ldots dt_{2n}
e^{i\alpha_-(t_0)}e^{i\widetilde{\alpha}(t_1)}..
e^{-i\widetilde{\alpha}(t_{2(n+1)})}\left\{C^0_{n-1}
\left[I_4(\tilde{t}
_0,
.. \tilde{t}_3)\times\right.\right.
$$
\begin{equation}
\left.\left.\times
\Delta(t_4,\ldots,t_{2n})+I_5(\tilde{t_0},\ldots
\tilde{t}_4)\cdot\Delta(t_5,\ldots,t_{2n})
\right]
\ldots
+\right.\end{equation}
$$\left.+C^{n-1}_{n-1}\left[I_{2n+2}(\tilde{t}_0,\ldots
\tilde{t}_{2n+2})\Delta(t_{2n+3})+
I_{2n+3}(\tilde{t}_0,\ldots \tilde{t}_{2n+3})
\right]
\right\}.$$
After performing all possible integration in $t_i$, one can convert the series
beginning with $|{\cal K}|^4$ as
\begin{equation}
\delta_4{\tilde{\mathfrak D}}_{reg}([\tilde\alpha],t)=-|{\cal K}|^2\big\{
\frac{|{\cal K}|^2}{|{\cal R}|^2}[\widetilde{\mathfrak
B}_3+\widetilde{\mathfrak B}_4]-|{\cal K}|^2\left(\frac{|{\cal K}|^2}{|{\cal
R}|^2}\right)^2[\widetilde{\mathfrak B}_5+\widetilde{\mathfrak B}_6]
+|{\cal K}|^2\left(\frac{|{\cal K}|^2}{|{\cal
R}|^2}\right)^3[\widetilde{\mathfrak B}_7+\widetilde{\mathfrak B}_8]-
\ldots \big\} .
\label{DR4}
\end{equation}
 To
 have the total dual charge jump, one should add to Eq.(\ref{DR4}) the
 $\tilde{\mathfrak D}^{(0)}([\tilde\alpha ],t)$. As a result, we have got an expression for the dual part of charge jump ($\widetilde{{\mathfrak D}}([\tilde\alpha],\omega)$). It is the $\delta_4{\tilde{\mathfrak D}}_{reg}([\tilde\alpha],t) $ and the first term from Eq.(\ref{tildD2}). The explicit expression for the dual part of charge jump is
\begin{equation}
\tilde{{\mathfrak
D}}([\tilde\alpha],t)= |\K|^2\tilde{\cal B}_1(t)+\delta_4{\tilde{\mathfrak D}}_{reg}([\tilde\alpha],t).
\label{B1}\end{equation} At this step, we have not yet seen the duality of the repulsion and attraction problems, since the indices of
${\cal B}_i$ in (\ref{DR4}) and (\ref{Dfin}) are shifted by one.

\subsection{Duality of the problems.}
\label{DA}
As we have pointed out, the transition  coefficient for the repulsive
interaction (Eq.\ref{Kk}) can be obtained from reflection one (Eq.\ref{R-renorm})  calculated for the attracting interaction.
We will see here, the dual transformation
\begin{equation}
\widetilde{{\mathfrak D}}([\tilde\alpha],\omega)=\rm{ sign}(\omega) {\mathfrak
D}([\alpha],\omega)|_{{\cal R},\alpha \leftrightarrow{\cal K}, \tilde\alpha };
\ W(\omega) \leftrightarrow \tilde W(\omega) ({\rm or }\  v_c\rightarrow
1/v_c\quad\rm{for\quad point-like\quad imteraction})
\label{FDual2}
\end{equation}
is exact for the arbitrary e-e interaction. (As a result, the matrixes elements
$\langle\mathfrak{D}([\alpha],\omega)\alpha
(-\omega')\rangle$ and
$\langle\widetilde{\mathfrak{D}}([{\tilde\alpha}],\omega)
{\tilde\alpha }
(-\omega')\rangle_K$ will be equal.)

Let us compare the series (\ref{defB}) and (\ref{DR4}). To begin the proof, we show that the higher-order even coefficients of ${\mathfrak B}_i$
are not independent and can be expressed in terms of the odd ones:
\begin{equation}
{\mathfrak B}_{2n}=-\theta (\omega)\left( 2{\mathfrak B}_{2n-1}+{\mathfrak
B}_{2(n-1)}
\right)\,\,\,n>2
\label{odd-even}
\end{equation}
$\rhd$
Fourier's representation of the even ${\mathfrak B}_n$  may be expressed in the
 form:
\begin{equation}
{\mathfrak B}_n(\omega)=\frac{2}{\pi}\theta (\omega)\int
\frac{d\tau_1...d\tau_n}{(2\pi i)^{n-1}} \frac{\exp {(i\omega\tau_1 )}-\exp
{(i\omega\tau_n )}}{\tau_1-\tau_n - i\delta}\cdot\frac{\sin{\left(\alpha (\tau_1)-...-\alpha (\tau_n)
\right)}}{(\tau_1-\tau_2-i\delta)....
(\tau_{n-1}-\tau_{n}-i\delta)}.
\label{B2n}
\end{equation}The first term at the Eq.(\ref{B2n}) differs from the odd one (with the
index $n-1$) only by the factor $\theta (\omega)$. The second term (after
renaming $\tau_n \to t$) differs from a ${\mathfrak B}_i$ by the signs of
$i\delta$. The sign can be changed by extracting $-2\pi i\delta (t-\tau)$. Now, the term (up to the sign) coincides with the antecedent even
 coefficient ${\mathfrak B}_{n-2}$. So, we  proves identity
 (\ref{odd-even}).
 $\lhd$

As a result, one can represent the even ${\mathfrak B}$ via
 odd ones
$$
{\mathfrak B}_{2n}(\omega)=2\theta
(\omega)(-1)^{n+1}\sum_{k=1}^n(-1)^k{\mathfrak B}_{2k-1}(\omega),\quad{\rm i.e.}
$$ $$
{\mathfrak B}_{2n-1}(\omega) +{\mathfrak B}_{2(n-1)}(\omega)=- {\rm
sign}(\omega )\left( {\mathfrak B}_{2n}(\omega) +{\mathfrak
B}_{2n-1}(\omega)\right).
$$
The last identity just is the evidence of the duality in the meaning discussed earlier. Indeed, if one takes  Eq.(\ref{defB}) (the charge jump for the attracting problem) and changes the
$\alpha\to{\tilde\alpha};\,\,{\cal R}\to{\cal K}$, then one will have the charge
jump for repulsive interaction (Eq.\ref{DR4})  with extra factor ${\rm sign}(\omega).$ The latter factor is needed   to receive the necessary matrix elements. Expression for the action (Eq.(\ref{det-good}) as well as a definition
of $|{\cal R}_{\omega}|^2$ (Eq.\ref{R-renorm})  contains as a factor the $\alpha
(-\omega)$. The factor ${\rm sign}(\omega )$ will change $\alpha$ to the $\tilde\alpha$. It
replaces the
$\langle{\mathfrak D}[\alpha(\omega)]\alpha
(-\omega')\rangle$ with
$\langle\tilde{\mathfrak D}[\tilde\alpha(\omega)]
\tilde\alpha (-\omega')\rangle_K$.
Also, one needs to change
$W(\omega)\to \tilde W(\omega)$ in the "free part"\ of the action.
If each of the sum for the charge jump is convergent (at least asymptotically) the duality property is exact.
\subsection{Action expansion: exact integration over coupling constant.}
\label{DAction}
In this section, we will integrate the action over the electron coupling constant $\lambda e_0$ (see Eq.\ref{det-good}).  This point is important for the problem, especially outside the iteration procedure.  It allows to work with an action  depending on the actual coupling constant.
Otherwise, if we tried to simplify the problem (say, by considering the e-e interaction to be strong)
 then we would not be able to do this before integrating over $\alpha$. This would  be possible at the final calculations  stage only.

Let us begin from symmetrization the series for the charge jump. (Now one can
consider only one type of interaction,  let's say - attracting.) To produce
this, we change the sign of the image part in the  pole $t=\tau_n$ by eliminating the $\delta$-function from expression. Then: ${\mathfrak T}_1(t) ={\mathfrak B}_1(t)$,  and \[
{\mathfrak T}_2([\alpha],t)=\frac{2}{\pi}\int\frac{d\tau_1d\tau_2}{(2\pi
i)^2}\frac{\sin[\alpha(\tau_1)-\alpha(\tau_2)]}{(\tau_1-t+i\delta)
(\tau_1-\tau_2-i\delta)(\tau_2-t-i\delta)},
\]
\[
{\mathfrak
T}_3([\alpha],t)=\frac{2}{\pi}\int\frac{d\tau_1d\tau_2d\tau_3}{(2\pi
i)^3}\frac{\sin[\alpha(\tau_1)-\alpha(\tau_2)+\alpha(\tau_3)-
\alpha(t)]}{(\tau_1-t+i\delta)(\tau_1-\tau_2-i\delta)
(\tau_2-\tau_3-i\delta)(\tau_3-t-i\delta)},
\]
\begin{equation}
{\mathfrak
T}_4([\alpha],t)=\frac{2}{\pi}\int\frac{d\tau_1d\tau_2d\tau_3d\tau_4}{(2\pi
i)^4}\frac{\sin[\alpha(\tau_1)-\alpha(\tau_2)+\alpha(\tau_3)-
\alpha(\tau_4)]}{(\tau_1-t+i\delta)(\tau_1-\tau_2-i\delta)
(\tau_2-\tau_3-i\delta)
(\tau_3-\tau_4-i\delta)(\tau_4-t-i\delta)},
\label{defT}
\end{equation}
 etc. One can see easily: ${\mathfrak B}_n={\mathfrak T}_n-{\mathfrak T}_{n-1}$ or
${\mathfrak B}_{2n}+{\mathfrak B}_{2n+1}={\mathfrak T}_{2n+1}-{\mathfrak
T}_{2n-1}$ and
\begin{equation}
{\mathfrak D}([\alpha],t)= -\frac{|{\cal R}|^2}{|{\cal K}|^2}{\mathfrak
T}_1(t)+
\left(\frac{|{\cal R}|^2}{|{\cal K}|^2}\right)^2{\mathfrak T}_3(t)-
\left(\frac{|{\cal R}|^2}{|{\cal K}|^2}\right)^3
{\mathfrak T}_5(t)
+\ldots
\label{jumpT-exact}
\end{equation} The sum has to be substituted in
the relation $\log{\mathfrak{Det}}_{imp}=$ $=-i/2\int_0^1
d\lambda\int dt\alpha (t){\mathfrak D}[\lambda\alpha](t)$. Notice,
$$
\partial_{\lambda}\int\frac{dt... d\tau_n}{(2\pi
i)^n}\frac{1-\cos{\lambda (\alpha (\tau_1) .. -\alpha
(t))}}{(t-\tau_1-i\delta)\cdots (\tau_n-t-i\delta)}=(-1)^{n+2}(n+1)\int\frac{dt..
d\tau_n}{(2\pi i)^n}\frac{\alpha
(t)\sin{\lambda (\alpha (\tau_1) .. -\alpha
(t))}}{(t-\tau_1-i\delta)\cdots (\tau_n-t-i\delta)}
$$ Here (before differentiation) we have made the cyclic permutation of the integration
variables. So, the sums of multiloop
diagrams, describing interaction in the effective theory, reduces to the
action:
\begin{equation}
\log\Det\!=\!\sum_{n=1}^\infty
\frac{(-1)^{n+1}}{n}\!\left(\!\frac{|\cal R|}{|{\cal K}|}
\!\right)^{2n}\!{\cal C}_{2n-1}
;
{\cal C}_{n}=\int\!\!\frac{d\tau_0.. d\tau_n}{(2\pi
i)^{n+1}}
\frac{1\!-\!\cos[\alpha(\tau_0)\!-\!\alpha(\tau_1)\!+
\!\ldots \alpha(\tau_n)]}
{(\tau_0-\tau_1-i\delta)(\tau_1-\tau_2-i\delta)..
(\tau_n-\tau_0-i\delta)}
\label{det1}
\end{equation}

 \section{Properties of  $\Gamma_{2n}$ vertices.}
\label{ApGama}
 Obtaining a general expression for $\Gamma_{2n}$ for any $n$ is a bit cumbersome. We will make it in few stages. To this, one has to expand all cosines
  in Eq.(\ref{det1}) in  Taylor's series and collect the terms with same
power of $\alpha$. The general expression for the k-th contribution from
${\cal C}_{2k-1};\,\,(k\le n )$ to the vertex $\Gamma_{2n}$ is
\begin{equation}\Gamma_{2n}^k= \frac{(-1)^{k+n+1}}{k}\left(\frac{|{\cal
R}|^2}
{|{\cal K}|^2}\right)^{k}\int \frac{d\tau_1.. d\tau_{2k-1}}{(2\pi i)^{2k}}
\frac{(1-e^{-i\omega_1\tau_1}+e^{-i\omega_1
\tau_2}-..
e^{-i\omega_1\tau_{2k-1}})}
{(\tau_1+i\delta)(\tau_1-\tau_2-i\delta)}\times
\label{GammaEx}\end{equation} $$\ldots\times
\frac{(1-e^{-i\omega_{2n}\tau_1}+
e^{-i\omega_{2n}\tau_2}-\cdots
e^{-i\omega_{2n}\tau_{2k-1}})}{(\tau_{2k-2}-\tau_{2k-1}-i\delta)
(\tau_{2k-1}-i\delta)}.
$$ This expression has been got from Eq.(\ref{det1}) after transition  to the variables $\tau_i
-\tau_0$ and integration in $\tau_0$ a
summand of Taylor's series with the same $\alpha^{2n}$ factor.

To calculate the common expression,
let us calculate the contribution from ${\cal C}_1$ to the all vertices
$\Gamma_{2n}$. The simplest expression, giving the $\Gamma_{2}^1$, results
from the summand proportional to
$$
\int (d\omega_1d\omega_2)d\tau\alpha(\omega_1)\alpha(\omega_2)2\pi\delta(
\omega_1-\omega_2)\frac{(1-e^{-i\omega_1\tau})
(1-e^{-i\omega_2\tau})}{(\tau+i\delta)(\tau-i\delta)}.
$$So, we have
$$
\Gamma^1_2=\frac{1}{4\pi}\frac{|\R |^2}{|\K|^2}\gamma(\omega_1,\omega_2).
$$ The next term of the cosine expansion, generating the $\alpha^4$ vertex,
will have the factor $\Pi_{i=1}^{i=4}(1-e^{-i\omega_i\tau})$ in the numerator, etc. As a result one has $$
\Gamma^1_{2n}(\omega_1,...,\omega_{2n})=\frac{(-1)^{n+1}}{4\pi}\frac{|\R
|^2}{|\K|^2}\gamma(\omega_1,...,\omega_{2n}).
$$
For an arbitrary $k$, the frequency dependence of the $\Gamma_{2n}^k$ follows from the product $$
(1-e^{i\omega_1\tau_1}+e^{i\omega_1\tau_2}
-..^{i\omega_1\tau_{2k-1}})(1-e^{i\omega_2\tau_1}+
e^{i\omega_2\tau_2}-..e^{i\omega_2\tau_{2k-1}}).. \times (1-e^{i\omega_{2n}\tau_1}+
e^{i\omega_{2n}\tau_2}-.. e^{i\omega_{2n}\tau_{2k-1}})$$ It means,  $\Gamma_{2n}^k$
will have the same factor
\begin{eqnarray} \gamma (\omega_1,..\omega_{2n})=\sum_i|\omega_i|-\sum_{i<j}
 |\omega_i+
\omega_j|+
\sum_{i<j<k}|\omega_i+\omega_j+\omega_k|-...
\end{eqnarray} The single problem is  dependency of the vertices on
$|\R|/|\K|$.
To uniquely determine all vertices, we will derive a recurrent
relation for vertices in the particular case of $\omega$. It will give us opportunity to calculate dependence of any vertex on $|\R|/|\K|$. Namely, let us consider the
contribution to the $\Gamma_{2n}^k(\Omega, -\Omega,\omega_3,...,\omega_{2n})$
assuming $|\Omega|\gg|\omega_i|$ is the biggest parameter of the problem, so
\begin{equation}
\Gamma_{2n}^k(2\pi)\delta(\sum \omega)= \frac{(-1)^{k+n}}{k}\left(\frac{|{\cal
R}|^2}
{|{\cal K}|^2}\right)^{k}\int \frac{d\tau_0d\tau_1.. d\tau_{2k-1}}{(2\pi
i)^{2k}}(e^{-i\Omega\tau_0}-e^{-i\Omega\tau_1}
-e^{-i\Omega\tau_{2k-1}})
\times
\label{GammaAsimp1}
\end{equation}
$$\times(e^{i\Omega\tau_0}-e^{i\Omega\tau_1}+..
-e^{i\Omega\tau_{2k-1}})\cdot\frac{\prod^{(k)}_{2n-2}
(\tau_0,\tau_1,\ldots\tau_{2k-1};\omega_3,
\ldots,\omega_{2n}) }
{(\tau_0-\tau_1-i\delta)(\tau_1-\tau_2-i\delta)\ldots
(\tau_{2k-2}-\tau_{2k-1}-i\delta)(\tau_{2k-1}-
\tau_0-i\delta)}
$$ We have denoted as $\Pi$ the product of all other parenthesis, depending on
all other $2(n-1)$ frequencies. It equals
\begin{equation}
\Pi^{(k)}_{2n-2}(\tau_0,..;\omega_3,
..,\omega_{2n})=(e^{-i\omega_3\tau_0}
-e^{-i\omega_3\tau_1}+..-
e^{-i\omega_{3}\tau_{2k-1}})\cdot.
..(e^{-i\omega_{2n}\tau_0}-e^{-i\omega_{2n}\tau_1}+..-
e^{-i\omega_{2n}\tau_{2k-1}})
\label{PiAs}
 \end{equation}
One can integrate Eq.(\ref{GammaAsimp1}) neglecting dependence $\Pi$ on
$\tau_i$ if a closing  contour is controlled by the factor with frequency $\Omega$.
Expanding parenthesis, we obtain a number of
integrals ($i\leq j$)
\begin{equation}
I_{ij}^{\pm}=\frac{(-1)^{k+n}}{k}\left(\frac{|R|^2}{|K|^2}\right)^k
\int\frac{d\tau_0d\tau_1\ldots d\tau_{2k-1}}{(2\pi i)^{2k}}
\frac{e^{\pm i|\Omega| (\tau_i-\tau_j)}
\Pi^{(k)}_{2(n-1)}(\tau_0\ldots\tau_{2k-1})}
{(\tau_0-\tau_1-i\delta)
\ldots(\tau_{2k-1}-\tau_{0}-i\delta)},
\label{Ias}
\end{equation}
and $
I_{jj}^{\pm}=-\Gamma^k_{2(n-1)}(\omega_3,
\ldots,\omega_{2n})
$.
  Making cyclic redefinition of the variables, we can always put $i=0$. Therefore, this
integral depends on difference $j-i$. (In other words, the cyclic
redefinition $i\to i+1;2k-1\to 0$ shows, the set of the points $\{ i,j\}$
has not a distinguished point.)

 Expression for $I_1^+$; $j-i=1$.

 In the case, one can integrate in $\tau_0$
 using the pole $\tau_0=\tau_1+i\delta$ in the upper semi-plane. Then  all exponential function with $\tau_1$ in
 $\Pi$   cancel out. Taking into account the coefficients in
 Eqs.(\ref{GammaAsimp1}) and (\ref{Ias}), we have:
\begin{equation}
I_1^{(+)}=\frac{k-1}{k}\frac{|R|^2}{|K|^2}\Gamma^{(k-1)}_
{2n-2}(\omega_3,\ldots,\omega_{2n}).
\end{equation}
At $j=2$ we integrate first in $\tau_0$ (using the pole in upper semi-plane) and
then in $\tau_1$ using also the pole in upper semi-plane as dictated by factor
$e^{i\Omega\tau_1}$. Anyway, again $\tau_0=\tau_1, \tau_1=\tau_2$ and we are obtaining
the result is identical to the previous one. (The $e^{-i\omega_{i}\tau_2}$ in
$\Pi$ are not vanish.)
This is, in fact, a general case
\begin{equation}
I^{(+)}_{2j-1}=\frac{k-j}{k}(-1)^{2j+1}\left(\frac{|R|^2}{|K|^2}\right)^j
\Gamma^{(k-j)}_{2n-2}(\omega_3,\ldots,\omega_{2n}),\quad{\rm and}\quad I^{(+)}_{2j-1}=I^{(+)}_{2j}.
\label{Ij}
\end{equation}
However, the last possible integral at $I^+_{2k-1}$ is zero:
\[
I^+_{2k-1}=\frac{(-1)^{k+n}}{k}\left(\frac{|R|^2}{|K|^2}\right)^k
\int\frac{d\tau_{2k-2}d\tau_{2k-1}}{(2\pi i)^{2}}
\frac{e^{\pm i|\Omega| (\tau_{2k-2}-\tau_{2k-1})}
}
{(\tau_{2k-2}-\tau_{2k-1}-i\delta)}\times
\frac{\Pi^{(1)}_{2(n-1)}(\tau_{2k-2},\tau_{2k-1})}
{(\tau_{2k-1}-\tau_{2k-2}-i\delta)},
\]
as it does not contain any poles in $\tau_{2k-2}$ (poles of denominator are
canceled by $\Pi$).

Integrals $I^{-} $ may be considered in the same way, but one should make integration in
opposite direction. (In the case, the poles are in the lower
semiplane: $\tau_j=\tau_{j-1}-i\delta,..$.) As a result, after  "j"\
integrations we will have expression coinciding with Eq.(\ref{Ij}). However,
integrals $I^{\pm }_{2j-1}$ and $I^{\pm }_{2j}$ enter to the
Eq.(\ref{GammaAsimp1}) with the same footing but different signs. For this reason they cancel each other out and we are left with $I_{jj}^{\pm}$ only.
There are exactly $2k$  integrals with $i=j$ in Eq.(\ref{GammaAsimp1}). Therefore
\begin{equation}
\Gamma_{2n}^{(k)}(\Omega,-\Omega,\omega_3,\ldots
\omega_{2n})=-2k\Gamma_{2n-2}^{(k)}(\omega_3\ldots\omega_{2n})
\label{recurNK}
\end{equation}
This is the recurrent relation, we are looking for.

One can turn relation Eq.(\ref{recurNK}) into the relation between full
vertices. After introducing  $x=|R|^2/|K|^2$ one can rewrite
Eq.(\ref{recurNK}) in the form
\[
\Gamma_{2n}^{(k)}(\Omega,-\Omega,\omega_3,\ldots
\omega_{2n})=-2x\frac{\partial}{\partial
x}\Gamma_{2n-2}^{(k)}(\omega_3\ldots\omega_{2n})
\]
Now we sum up this relation in $k$ and arrive finally at:
\begin{equation}
\Gamma_{2n}(\Omega,-\Omega,\omega_3,\ldots
\omega_{2n})=-2x\frac{\partial}{\partial
x}\Gamma_{2n-2}(\omega_3\ldots\omega_{2n}).
\label{recurn}
\end{equation}(Let us note, automatically
$\gamma_{2n}(\Omega,-\Omega,\omega_3,\ldots
\omega_{2n})=2\gamma_{2n-2}(\omega_3\ldots\omega_{2n})$
for sufficiently large $\Omega$.)
Also, Eq.(\ref{recurn}) can be formulated as a relation between $S_n$ and
$A_n$.  From the other hand due to $\theta$-function
$A$ and $S$-structures exchange. In other words:
\begin{equation}
S_{n}=-x\frac{\partial}{\partial x}A_{n-1}, \qquad
A_{n}=-x\frac{\partial}{\partial x}S_{n-1}
\label{recurAS}
\end{equation}


\begin{thebibliography}{9}
\bibitem{T}
S.Tomonaga, "Remarks on Bloch's Method of Sound Waves applied to Many-Fermion
Problems," \  Prog. Theor. Phys., 5 (1950), 544.
\bibitem{L}
J. M. Luttinger, "An Exactly Solvable Model of a Many-Fermion System,"\  J. Math. Phys., (1963),  4 1154, doi:10.1063/1.1704046
\bibitem{M}
D.C.Mattis and E.H.Lieb, , "Exact Solution of a Many-Fermion System and Its Associated Boson Field,"\ J. Math. Phys., 6  (1965), 304.
\bibitem{Hold} F.D.M. Haldane, "Properties of the Luttinger model and their extension to the general 1D interacting
spinless Fermi gas,"\ J. Phys. {\bf C}, 14  (1981), 2585.
\bibitem{Top} C. Wu, B.A. Bervering, and S.C. Zhang, Phys.Rev.Let, "Helical Liquid and the Edge of Quantum Spin Hall Systems,"\ 96 (2006), 106401.
 \bibitem{E}
    V.J.Emery, in ``Highly Conducting One-Dimensional Solids,"\ ``Highly Conducting One-Dimensional Solids,"
    p.327 (Plenum Press, New York, 1979).
\bibitem{Gim}
T.Giamachi, "Quantum Physics in One Dimension,"\ Clarendon Press, Oxford 2003.
\bibitem{Gr}G.Gruner,  "The dynamics of charge-density waves,"\ Rev.Mod.Phys., 60 (1988), 1129.
\bibitem{Land} L.D.Landau, Sov.Phys.JETP, "On the theory of phase transitions,"\ 7 (1937), 19.
\bibitem{Af} V.V.Afonin, "Luttinger liquid  with attacting interaction and one  impurity: exact solution,"\  JETP, 163 (2023), 238.
\bibitem{Ph} V.V. Afonin and V.Yu. Petrov, "BKT Phase in Systems of Spinless  Strongly Interacting One-Dimensional Fermions,"\ JETP, 107 (2008), 542.
 \bibitem{NT}
 V.V.Afonin, V.L.Gurevich, V.Yu.Petrov, JETP,"\ Spontaneous Symmetry Breaking in a System of Strongly Interactiong Multicomponent Fermions (Electron with Spin and Conducting Nanotubes), 108 (2009), 845.
\bibitem{BKT} V.L.Berezinskii, "Destruction of long-range order in one-dimensional and
two-dimensional systems having a continuous symmetry group,"\ Sov.Phys.JETP, 32 (1971), 493.
\bibitem{KT} J.M. Kosterlitz and D.J. Thouless,  "Ordering, metastability and phase transitions in two-dementional system,"\  J.Phys.C, 6 (1973), 1181.
\bibitem{AGD}
Abrikosov A.A., Gorkov L.P., Dzyalosliinski I.E.  "Methods of quantum fields
theory in statistical physics."\  Prentice - Hall, Englewood Cliffs, 1963.
\bibitem{next} V.V.Afonin and V.Yu.Petrov, "Is the Luttinger liquid a new state of matter?"\    Found.Phys., 40 (2010), 190.
 \bibitem{Pham} Pham K.V., Gabay M., Lederer P., "Fractional excitations in the Luttinger liquid,"\ Phys. Rev. B, 61 (2000),  1637.
\bibitem{FK}
Kane C.L. and Fisher M.P.A., "Transmission through barriers and resonant tunnelingin an interacting one-dimensitional electron gas,"\ Phys.Rev.B, 46 (1992), 15233.
\bibitem{Fur}
A.Furusaki  and N.Nagaosa, "Single-barrier problem and Anderson localization
in a one-dimentional interacting electron system,"\ Phys.Rev.{\bf B}, 47 (1993),  4631.
\bibitem{BeteF}P.Fendley, A.W.W. Ludwig and H.Saleur, "Exact nonequilibrium transport through point contacts in quantum wires and fractional quantum
    Hall devices,"\  Phys.Rev.B, 52 (1995), 8934.
\bibitem{sup}
A.F.Andreev, "Thermal conductivity of the intermediate state of superconductors,"\ Sov.Phys.JETP, 46 (1964),  1823.
\bibitem{ExRep}V.V.Afonin, V.Yu.Petrov,
  "On the Exact Solution for a Luttinger Liquid
with Repulsion and a Single Point Impurity,"\  JETP, 137 (2023), 384.
\bibitem{B} J.Bardeen,  "Tunnelling from a Many-Particle Point of View,"\ Phys.Rev.Lett, 6 (1961), 57.
\bibitem{Hub} J.Hubbard,  "Calculation of Partition Functions,"\ Phys. Rev.Let. 3  (1959), 77.
\bibitem{aristov2} D. N. Aristov and P. Woelfle, "Conductance through a potential barrier embedded in a Luttinger liquid: Nonuniversal scaling at strong coupling,"\
 Phys. Rev. B,  80  (2009), 045109.
\bibitem{EqTh}V.V.Afonin, V.Yu.Petrov, "Luttinger liquid with one impurity: equivalent field theory and duality,"\ Pis’ma v ZhETF,  97   (2013), 587.
\bibitem{RG}
  L.I.Glazman,  K.A.Matveev and  D.Yue, "Conduction of a weakly interacting one-dimesional electron gas througt a single barrir,"\ Phys. Rev. {\bf B}, 49 (1994), 1966.
\bibitem{And}
P. W. Anderson,  "A poor man's derivation of scaling laws for the Kondo problem,"\ J. Phys. C: Solid St. Phys.,    3 (1970),  2436.
\bibitem{Collins} J.Collins, "Renormalization,"\ Cambridge, Cambridge University Press 1998.
\bibitem{aristov} D. N. Aristov and P. Woelfle,
"Transport of interacting electrons through a potential barrier: nonperturbative RG approach,"\
EPL, 82  (2008), 27001.
\bibitem{AP}    V.V. Afonin and V.Yu.Petrov,  "Breaking
an one-parameter "poor man’s" scaling approach in the Luttinger liquid,"\
J.Phys.: Condens. Matter,  30  (2018), 355601.
\bibitem{AD} S.L.Adler, R.F.Dashen "Current Aldebras and Applications to Particle
 Physics,"\  W.A.Benjamin Inc, N.Y.-Amsterdam 1968.
\bibitem{reg}
M. Peskin, D. Schroeder, "An introduction to quantum field theory,"\
Addison-Wesley, pp.654-656, 1996.
\bibitem{Lar}
I.E.Dzyaloshinskii and A.I. Larkin, "Correlation function for a one-dimensional Fermy system with long-range interaction (Tomonaga model),"\ JETP, 38  (1974), 202.
\bibitem{FP}
 L.D. Faddeev and V.N. Popov,  "Feynman diagrams for the Yang-Mills field,"\  Phys.Lett.{\bf B},  25  (1967), 29.
\bibitem{As} G.H. Hardy, "Divergent Series,"\ Oxford, 1949.
\bibitem{CondReem} V.V. Afonin and V.Yu.Petrov, "One Dimensional Strongly Interacting Electrons
with a Single Impurity: Conductance Reemergence,"\
JETP Letters, 101  (2015), 622.
\bibitem{Steph} D.L.Maslov and M.Stone, "Conductance of Luttinger-Liquid Wires Connected to Reservoirs,"\  Phys. Rev.{\bf B},  52 (1995), R5539.
\bibitem{msl} D.L.Maslov, "Transport through dirty Luttinger liquids connected to reservoirs," Phys.Rev.{\bf B}, 52  (1995), R14368.
\bibitem{PS} B.I. Ivlev and N.B. Kopnin, "Theory of current states in narrow superconducting channels,"  Adv. Phys., 33 (1984), 80.
\bibitem{GML}
Murray Gell-Mann, F.E. Low, "Quantum Electrodynamics at Small Distances," Phys.Rev., 95 (1954), 1300.
\bibitem{CZ} K. Symanzik, Com.Mat.Phys., "Small-distance-behaviour analysis and Wilson expansions,"  23 (1971), 49.
\bibitem{ZJ} J.Zinn-Justin "Quantum Field Theory and Critical Phenomena."\ Clarendon Press, Oxford, 1996.
\bibitem{RgM}
L. I. Glazman, K. A. Matveev, and D. Yue,"Tunneling in one-dimensional non-Luttinger electron liquid,"\
Phys. Rev. Lett.,  71 (1993),3351.
\bibitem{Ry} Lewis H. Ryder, "Quantum Field Theory,"\ Cambridge University Press, 1996.
 \bibitem{TIsul} C. Xu and J.E. Moore, "Stability of the quantum spin Hall effect: Effects of interactions, disorder, and $Z_2$ topology,"\ Phys.Rev.{\bf B}, 73  (2006),
 045322.
 \end{thebibliography}
\end{document}